\documentclass[a4paper,11pt]{article}
\usepackage[utf8]{inputenc}
\usepackage[]{units}
\usepackage[margin = 2.5cm]{geometry}
\usepackage{mathrsfs}
\usepackage{multirow}
\usepackage{multicol}
\usepackage{colortbl}
\usepackage{comment}
\usepackage{amsmath}
\usepackage{extarrows}
\usepackage{mathtools}
\usepackage{amsfonts}
\usepackage{amssymb}
\usepackage{enumerate}
\usepackage{diagbox}
\usepackage{amsthm}
\usepackage{bm}
\usepackage{tabularx}
\usepackage{float}
\usepackage{threeparttable}
\usepackage{graphicx}
\usepackage{stackengine}
\usepackage{enumitem}
\usepackage{cancel}
\usepackage{hyperref}

\usepackage[svgnames]{xcolor}
\usepackage[labelfont=bf]{caption}
\usepackage{subcaption}
\usepackage{amsmath}
\usepackage{blindtext}
\usepackage{tikz}

\usepackage{geometry}
\usepackage{rotating}
\usepackage{multirow}
\usepackage{blindtext}
\usepackage{wrapfig}
\usepackage{rotating}
\usepackage{booktabs}
\usepackage{caption}
\usepackage{subcaption}
\usepackage{booktabs}
\usepackage{xltabular}
\usepackage{titling}
\usepackage{pdflscape}
\predate{}
\postdate{}
\usepackage{authblk}
\usepackage{booktabs, makecell, longtable}
\usepackage{multirow}
\usepackage{float}

\hypersetup{
    colorlinks=true,
    linktocpage=true,
    pdfstartpage=1,
    pdfstartview=FitV,
    breaklinks=true,
    pdfpagemode=UseNone,
    pageanchor=true,
    pdfpagemode=UseOutlines,
    plainpages=false,
    bookmarksnumbered,
    bookmarksopen=true,
    bookmarksopenlevel=1,
    hypertexnames=true,
    pdfhighlight=/O,
    linkcolor=red,  
    citecolor=blue,
    urlcolor=blue,
    pdfsubject={},
    pdfkeywords={},
    pdfcreator={pdfLaTeX}
}

\newtheorem{remark}{Remark}

\usepackage{natbib}

\title{\textbf{Space time modeling for drought classification and prediction}}
\author[1]{Touqeer Ahmad\thanks{Corresponding author: \textcolor{blue}{\textit{touqeer.ahmad@univ-orleans.fr}}}}
\author[2]{Taha Hasan}

\affil[1]{Institute Denis Poisson,
Université d'Orléans, France. }
\affil[2]{Department of Statistics, Islamabad Model
College for Boys, F-10/4, Islamabad,
Pakistan}

\date{} 

\begin{document}
\maketitle
\begin{abstract}
The Standardized Precipitation Index (SPI) is a critical tool for monitoring drought conditions, typically relying on normalized accumulated precipitation. While longer historical records of precipitation yield more accurate parameter estimates of marginal distribution, they often reflect nonstationary influences such as anthropogenic climate change and multidecadal natural variability. Traditional approaches either overlook this nonstationarity or address it with quasi-stationary reference periods. This study introduces a novel nonstationary SPI framework that utilizes generalized additive models (GAMs) to flexibly model the spatiotemporal variability inherent in drought processes. GAMs are employed to estimate parameters of the Gamma distribution, while dual extreme tails flexible models are integrated to robustly capture the probabilistic risk of extreme drought events. Future drought and wet extremes events in terms of return levels are calculated using an extended generalized Pareto distribution, which offers flexibility in modeling the entire distribution of the data while bypassing the threshold selection step. Results demonstrate that the proposed nonstationary SPI model is both stable and capable of reproducing known nonstationary drought patterns, while also providing new insights into the evolving dynamics of drought. This approach represents a significant advancement in drought modeling under changing climatic conditions. 
\end{abstract}

\textbf{\textit{Keywords:}} Precipitation, Drought risk, Extreme value theory, Extended generalized Pareto distribution, Spatiotemporal modeling

\section{Introduction}\label{sec:1}
Drought is a growing global concern, particularly in semi-arid and Mediterranean regions where climatic variability, anthropogenic pressures, and increasing water demands converge to intensify water scarcity. Among these vulnerable areas, southern Europe has emerged as a hotspot for hydroclimatic extremes, with notable increases in drought frequency, duration, and severity over the past few decades~\citep{fontana2015early, ceglar2019observed}. Due to the higher vulnerability of the Mediterranean region to droughts, the study focused on the Puglia (Apulia) region in southeastern Italy. This region is challenged by the coexistence of scarce rainfalls and intensive agricultural activities, which often results in water scarcity through either groundwater abstraction~\citep{lopez2008planning}. 
Puglia is highly susceptible to both meteorological and hydrological droughts due to its fragile hydrological balance, limited surface water availability, and reliance on groundwater and interbasin water transfers~\citep{becherini2024multi}. A major agricultural producer and affected by climate change, Puglia faces challenges to meet its water needs, resulting in a water deficit situation~\citep{ellena2025influence}.  
Understanding drought dynamics in this region is thus essential for developing adaptive water management strategies and for informing regional resilience planning under future climate scenarios.

 In drought modeling, the normalized drought indices are considered key tools for drought monitoring and classification. In particular, the SPI, as introduced by \citet{mckee1993relationship} and further examined by \citet{guttman1999accepting} and \citet{lloyd2002drought}, is a widely used tool for quantifying meteorological drought. It assesses drought severity by comparing accumulated precipitation over a specified time period to a long-term climate reference using probability-based methods. The SPI converts this accumulated precipitation into a standard normal distribution, allowing the resulting values to be interpreted in terms of statistical probability or return periods \citep{guttman1999accepting}. For instance, an SPI value of $-2$ interprets that accumulated precipitation is two standard deviations drier than a typical year on this date, equivalent to approximately a 2.3\% probability of occurring in any given year, a 44-year event~\citep{stagge2022nonstationary}. Therefore, the primary advantage of the SPI is its normalization, which allows for consistent comparison of drought conditions across regions with varying climates and precipitation patterns. Recognized for its statistical reliability and ease of use, the SPI is recommended by the World Meteorological Organization \citep{WMO2006} and serves as a key tool for the United States drought monitoring.

The SPI quantifies precipitation anomalies over a defined $m\in (1,3,6,12, \dots)$  accumulation period using a backward-looking
rolling window. The resulting values are transformed to follow a standard normal distribution with mean zero and unit variance \citep{guttman1999accepting}. For instance, SPI-3 reflects precipitation anomalies over the preceding three months. By varying the accumulation period $m$, the SPI can approximate drought impacts across different components of the hydrological cycle, including soil moisture deficits \citep{lloyd2002drought} and hydrological drought conditions \citep{van2015hydrological}.

Seasonal accumulated precipitation data often exhibit positive skewness, making the two-parameter Gamma distribution a suitable choice for estimating the SPI due to its flexibility in modeling skewed distributions and its ability to approximate normality through shape parameter adjustments \citep{stagge2015candidate}. Although alternative distributions were evaluated in \citep{stagge2015candidate} to potentially improve SPI fitting, none demonstrated consistently superior performance over the Gamma distribution. Therefore, this paper will consider the Gamma distribution in a space-time modeling context.

The parameter estimates of the Gamma distribution used in constructing SPI benefit from longer precipitation records, as they reduce estimation uncertainty by providing more data for calibration \citep{van2015hydrological, carbone2018estimating}. However, while longer records are often preferred, using precipitation data spanning several decades or centuries can introduce errors due to nonstationarity in the climate. SPI calibration assumes a stationary climate by fitting 365 daily distributions with fixed parameters across all years. When applied to long-term records influenced by climate change or natural variability, this assumption can lead to biased estimates and misrepresentation of drought conditions. To address nonstationary climate variability in calibrating the SPI, two main approaches prevail: (1) using a quasi-stationary reference period such as the WMO 30-year reference period, which enhances comparability but increases parameters uncertainty due to shorter data records \citep{dubrovsky2009application, russo2013projection, carbone2018estimating}; and (2) using the full record precipitation, which reduces uncertainty but neglects climate trends, potentially biasing SPI values over time and making cross study comparison challenging.

Recent studies have proposed nonstationary approaches to construct SPI by allowing precipitation distribution parameters to vary over time. For instance, \citet{russo2013projection} applied a generalized linear model with a Gamma distribution, where the scale parameter changed linearly with year, capturing trends in mean precipitation while the keeps shape parameter remained constant. \citet{shiau2020effects} advanced \citet{russo2013projection} approach by using a generalized additive model for location, scale, and shape (GAMLSS), allowing both shape and scale parameters to vary nonlinearly using spline functions. \citet{ahmad2025estimating} also used GAMLSS in different period to observe trend in both parameters and estimate extreme drought risk by extreme value Bayesian model. \citet{stagge2022nonstationary} proposed a novel Bayesian framework using cyclic splines to model both stationary and nonstationary Gamma distribution parameters, including zero inflation, within a single unified model. In the nonstationary case, each parameter is allowed to vary over time through Bayesian spline functions, enabling more flexible and accurate representation of temporal changes in the precipitation distribution.

This work proposes a robust framework for modeling variation in the marginal distribution of non-zero accumulated precipitation. GAMs \citep{chavez2005generalized} are identified as suitable tools for capturing such marginal variability. By combining GAMs with the restricted maximum likelihood (REML) inference approach \citep{wood2011fast, wood2016smoothing, youngman2019generalized}, the framework enables objective and computationally efficient estimation of flexible spatio-temporal structures. We demonstrate how GAMs can effectively capture marginal spatio-temporal variation in the Gamma-distributed precipitation variable with applicability extending to a wide range of continuous covariates. The GAM framework here is implemented using regression splines, including thin plate splines for smooth two-dimensional effects, natural cubic splines and P-splines for one-dimensional effects, and tensor product smooths for smoothly varying interactions. The resulting estimates from the fitted spatio-temporal Gamma model are used to construct the SPI. This approach benefits from reduced parameter uncertainty and increased estimate stability by pooling information across locations and multi-decadal periods. An additional advantage is the enhanced flexibility to simultaneously model both the scale and shape parameters of the distribution, which is crucial for capturing realistic nonstationarity in space and time—particularly relevant for analyzing climate change projections.

This work further introduces flexible two-tailed models to estimate extreme drought risk using drought indices derived from a fitted space-time Gamma model. At first, a mixture model \citep{zhao2014extreme} is employed by combining generalized Pareto distributions (GPD) for the dry and wet tails with a normal distribution for the central part between the two thresholds. As this model requires pre-specified upper and lower thresholds, poor threshold selection or an inappropriate model for the non-extreme region can produce bias in tail estimates. The normal distribution is adopted for the central range because drought indices typically exhibit unimodal and symmetric behavior at each given location. Selecting optimal thresholds for modeling the lower and upper tails can be challenging and may introduce bias. To address this limitation, we adopt the \textit{bulk-and-tails} (BATs) model \citep{stein2021parametric}, which flexibly captures the entire distribution of drought index without requiring pre-specified thresholds. This approach enables a unified treatment of both central and extreme values. \citet{KROCK2022100438} applied the BATs model to nonstationary historical records of daily mean temperature across diverse climatic regions in the United States, demonstrating its adaptability to varying local conditions.

One of the objectives of this work is to predict future drought events in terms of return levels. While it is possible to model the entire distribution of drought index using BATs and capture both dry and wet conditions jointly, this approach makes it difficult to clearly distinguish the future behavior of each extreme condition. Therefore, we model the distribution of dry and wet events separately by fitting the extended GPD (EGPD) \citep{naveau2016modeling} to the respective tails. Since the EGPD is defined for positive values and traditionally assumes a positive shape parameter, it is commonly used to model the entire distribution of extreme events with the upper tail. However, in drought analysis, both tails of the distribution may be relevant and potentially bounded, particularly for variables such as drought duration or severity, which cannot grow indefinitely due to seasonal or physical constraints. To account for this, we allowed the shape parameter to take negative values, enabling the model to capture bounded upper tails characteristic of certain drought features.
To ensure compatibility with the EGPD framework, we inverted the dry events data to make it positive prior to model fitting. This transformation made the data compatible with the distribution support while preserving the interpretability of the fitted parameters. After estimation, the return levels were transformed back to their original signs for each location to accurately represent drought conditions.

In the remainder of the paper, the following structure is followed. Section~\ref{method} gives the details of GAMs for the Gamma model, construction of SPI, model for extremal probabilistic drought risk, and methodology for estimation of future return levels. Section~\ref{result-disc} provides the data description taken from Puglia, Italy, and details the discussion on the application of the proposed methodology. Section~\ref{sec:concl} concludes this work and provides some future direction to extend this research.

\section{Methodology}\label{method}


   



\subsection{Spatio-temporal Gamma model for drought indices}\label{sec: space-time-gamma}

Let $X(s, t)$ be an observed non-zero precipitation process at location $s$ for time $t$. In comprehensive probabilistic modeling across space and time, we assume that the process $X(s, t)$ follows a Gamma distribution. For each location $s \in S$ and time point $t \in T = \{1, \ldots, T\}$, we assume conditional independence
\begin{equation}\label{space-time-gamma}
X(s, t) \perp X(s^*, t^*) \mid \theta(s, t), \theta(s^*, t^*), \quad \forall (s, t), (s^*, t^*) \in S \times T, 
\end{equation}
where $\theta(s, t)$ represent an arbitrary parameter.

For full distribution of $X(s, t)$ we assume
\[
X(s, t) \sim \text{Gamma}(\alpha(s, t), \psi(s, t)),
\]
with density function:
\[
h(x; \alpha(s, t), \psi(s, t)) = \frac{\psi(s, t)^{-\alpha(s, t)} x^{\alpha(s, t) - 1} \exp(-x/\psi(s, t))}{\Gamma(\alpha(s, t))}, \quad x > 0,
\]
where $\theta(s, t) = \{\alpha(s, t)>0, \psi(s, t)>0\}$ denotes the Gamma distribution shape and scale parameters.

To model the spatial and temporal dynamics of the process $X(s, t)$, we embed the distribution parameters $\alpha(s, t)$ and $\psi(s, t)$ in a GAM framework. 
Specifically, we use
\[
\log \alpha(s, t) = g_{\alpha}(s, t), \quad \log \psi(s, t) = g_{\psi}(s, t),
\]
where $g_{\alpha}$ and $g_{\psi}$ are smooth functions of spatial and/or temporal covariates, each with a finite-rank basis representation. For example, following the general GAM representation:
\begin{equation}\label{basis}
g_{\ast}(s, t) = \beta_0^{(\ast)} + \sum_{k=1}^{K} \sum_{d=1}^{D_k} \beta_{kd}^{(\ast)} b_{kd}(s, t), \quad \ast \in \{\alpha, \psi\},
\end{equation}
where $b_{kd}(s, t)$ are pre-specified basis functions, and $\beta_{kd}^{(\ast)}$ are the associated coefficients, see for instance \citep{youngman2019generalized} for detail. This may be compactly written as
$
g_{\ast}(s, t) = \mathbf{w}_{\ast}(s, t)^{\top} \boldsymbol{\beta}_{\ast},
$
with $\mathbf{w}_{\ast}$ a row design matrix $\mathbf{W}_{\ast}$ with elements determined by $b_{kd}$.
Inference for the above Gamma model is carried out by adopting the REML based on penalized maximum likelihood, which was initially developed for the GAM forms~\citep{wood2011fast, wood2016smoothing, youngman2019generalized}.

Let \( X(s_j, t) \sim \text{Gamma}(\alpha(s_j, t), \psi(s_j, t)) \) for \( j = 1, \dots, J \), \( t = 1, \dots, T \), and their realization data vector is \( \mathbf{x} = \{ x(s_j, t) \} \). Assuming independence across \( s_j \) and \( t \), the log-likelihood has the form \(h(\mathbf{x}, \boldsymbol{\beta})=\prod_{j=1}^J\prod_{t=1}^T  h(x(s_j, t), \boldsymbol{\beta})\).  The penalized likelihood \citep{silverman1985some} approach is used to estimate the coefficients \( \boldsymbol{\beta} \) of the smooth terms by maximizing:
\begin{equation}
\ell_p(\mathbf{x}; \boldsymbol{\beta}, \boldsymbol{\lambda}) = \log h(\mathbf{x}; \boldsymbol{\beta}) - \frac{1}{2} \boldsymbol{\beta}^\top \mathbf{S}_\lambda \boldsymbol{\beta},
\end{equation}
where \(\log h(\mathbf{x}; \boldsymbol{\beta})= \sum_{j=1}^{J} \sum_{t=1}^{T}\log h(x(s_j, t); \alpha(s_j, t), \psi(s_j, t)) \), \( \mathbf{S}_\lambda \) is the block-diagonal penalty matrix constructed from the basis-specific smoothing matrices, scaled by smoothing parameters \( \boldsymbol{\lambda} \) which control the smoothness of $g$.
By following \citet{wood2011fast}, the smoothing parameters \( \boldsymbol{\lambda} \) are estimated via REML, using the Laplace approximation to integrate over \( \boldsymbol{\beta} \). This yields the restricted log-likelihood:
\begin{equation}\label{REML}
R(\mathbf{x}; \boldsymbol{\lambda}) = \ell_p(\mathbf{x}; \widehat{\boldsymbol{\beta}}_{\lambda}, \boldsymbol{\lambda}) + \frac{1}{2} \log |\mathbf{S}_\lambda|^+ - \frac{1}{2} \log | \mathbf{H}(\widehat{\boldsymbol{\beta}}_{\lambda}) |^+ + \text{const},
\end{equation}
where \( \widehat{\boldsymbol{\beta}}_{\lambda} \) maximizes the penalized log-likelihood for given \( \boldsymbol{\lambda} \), \( \mathbf{H} \) is the Hessian of \( -\ell_p(\mathbf{x}; \boldsymbol{\beta}, \boldsymbol{\lambda}) \), and \( | \cdot |^+ \) denotes the product of positive eigenvalues \citep{youngman2019generalized}. 

\subsection{Multi-scale Spatio-temporal Drought Index}\label{sec:drought-index}

To capture drought severity over different temporal scales, we define an accumulated precipitation process over \(m\)-month windows as
\begin{equation}
X^{(m)}(s, t) = \sum_{i=0}^{m-1} X(s, t - i), \quad m \in \{1, 3, 6, \dots\},
\end{equation}
where \(X(s, t)\) is the observed non-zero precipitation at spatial location \(s\) and time \(t\). This accumulation enables the analysis of drought at various timescales linked with short-term, seasonal, medium-term, and long-term hydrological conditions. 


For each temporal scale \(m\), we assume a spatio-temporal Gamma distribution for the accumulated precipitation
\begin{equation}
X^{(m)}(s, t) \sim \text{Gamma}(\alpha^{(m)}(s, t), \psi^{(m)}(s, t)),
\end{equation}
and modeled using the framework proposed in Section~\ref{space-time-gamma}.

To obtain a standardized drought index that is comparable across locations and times, we transform the fitted Gamma distribution using the probability integral transform, followed by the inverse standard normal cumulative distribution function (cdf). The cumulative probability under the fitted Gamma model is
\begin{eqnarray}
H(x^{(m)}(s, t)) &=& \mathbb{P}(X^{(m)}(s, t) \leq x^{(m)}(s, t)) = \int_{0}^{x^{(m)}(s, t)} h(u; \alpha^{(m)}(s, t), \psi^{(m)}(s, t)) \, \mathrm{d}u \nonumber\\ 
&=& \frac{1}{\Gamma(\alpha^{(m)}(s, t))} \, \gamma\left( \alpha^{(m)}(s, t), \frac{x^{(m)}(s, t)}{\psi^{(m)}(s, t)} \right),
\end{eqnarray}
where \( \Gamma(\cdot) \) is the complete Gamma function and \( \gamma(\cdot, \cdot) \) is the lower incomplete Gamma function.




The standardized multi-scale spatio-temporal drought index is then defined as
\begin{equation}
D^{(m)}(s, t) = \Phi^{-1} \left( H(x^{(m)}(s, t)) \right),
\end{equation}
where \(\Phi^{-1}(\cdot)\) denotes the inverse cdf of the standard normal distribution. Thus, $D^{(m)}(s, t)\sim \mathcal{N}(0, 1),$ Table~\ref{tab:drought-catagories} interpret the conditions of drought at different values of $D^{(m)}(s, t)$. It is noteworthy that zero accumulated precipitation, especially for short accumulation periods ($m=1$). The $D^{(1)}(s, t)$ can be effectively adjusted to account for the likelihood of zero precipitation by utilizing Weibull non-exceedance probabilities, which are averaged to determine the center of probability mass for multiple zero values. For further details, see, e.g., \citet{stagge2015candidate}. Our study focuses solely on non-zero precipitation in order to avoid the zero problem in the short accumulation period, i.e., $m=1$.  
\begin{table}[htbp]
\centering
\caption{Classifcation of drought levels based on $D^{(m)}(s, t)$ values.}
\label{tab:drought-catagories}
\begin{tabular}{@{}cc@{}}
\hline
$D^{(m)}(s, t)$ value & Drought category \\ \hline
$D^{(m)}(s, t) \geq 2.00$ & Extremely wer \\
$1.50 < D^{(m)}(s, t) < 2.00$ & Very wet \\
$1.00 < D^{(m)}(s, t) \leq 1.50$ & Wet \\
$-1.00 \leq D^{(m)}(s, t) \leq 1.00$ & Normal \\
$-1.50 \leq D^{(m)}(s, t) < -1.00$ & Dry \\
$-2.00 < D^{(m)}(s, t) < -1.50$ & Very dry \\
$D^{(m)}(s, t) \leq -2.00$ & Extremely dry \\ \hline
\end{tabular}
\end{table}

\subsection{Extremal Drought Risk Assessment}\label{sec:droughtrisk}

Let \( Y^{(m)}(s) = D^{(m)}(s, t) \) be an independent variable taking estimated drought index values independently at location \( s \in S \) over an aggregation temporal scale \( m \). The subscript \( t \) is omitted as it remains constant across all locations. To assess drought risk, we define a critical threshold \( u_c \) indicating extreme drought conditions. The probabilistic drought risk at location \( s \) is then defined as the probability that \( Y^{(m)}(s) \) falls below this threshold:
\begin{equation}
    \rho(\mathcal{F}, u_c) = P(Y^{(m)}(s) < u_c) = \int_{-\infty}^{u_c} f_{Y^{(m)}(s)}(y) \, dy, \quad s \in S,\quad m \in \{1, 3, 6, \dots\},
\end{equation}
where \( \mathcal{F}(.) \) and \( f(.) \) are the cdf and pdf of \( Y^{(m)}(s) \), respectively. For notational simplicity, we omit the temporal scale \( m \) in subsequent sections.

This probabilistic framework provides a robust method for quantifying the likelihood of drought events and evaluating their potential impact. Furthermore, a model-based risk measure is defined as:
\begin{equation} \label{cdfrisk}
    \mathcal{R}_\rho (\mathcal{F}(.), u_c) = \rho,
\end{equation}
where \( \mathcal{R}_\rho \) quantifies the probability that the random variable \( Y(s) \) falls below the critical threshold \( u_c \), and \( \mathcal{F}(.) \) is the cdf of the underlying distribution. Given that \( Y(s) \) exhibits symmetric behavior with heavy tails, we model \( \mathcal{F}(.) \) using flexible extreme value distributions that capture lower and upper tail behavior accurately. Detailed construction of this approach is presented in the following sections.

\subsubsection{Two tails mixture generalized Pareto distribution}\label{sec:gng}

For estimating drought risk at each location, we first replace $\mathcal{F}(.)$ in \eqref{cdfrisk} with the cdf of a two-tailed mixture GPD   that marginally models the lower and upper tails separately beyond each threshold using a GPD \citep{zhao2014extreme}. The normal distribution is chosen as a suitable candidate for modeling the portion of the data between the left and right thresholds $d_l$ and $d_r$. 
For the remainder of the paper, we will refer to the two-tailed GPD-Normal-GPD mixture model as the (GP-N-GP). 
The CDF of this mixture model is defined as
\begin{equation}\label{gng}
\mathcal{F}(y\mid \theta) = 
\begin{cases} 
\Phi(d_l(s) \mid \mu(s), \sigma(s))\left[1 - \mathcal{G}(-y(s) \mid \xi_l(s), \beta_l(s), -d_l(s))\right] & \text{for } -\infty <y(s) \leq d_l(s) \\
\Phi(y(s) \mid \mu(s), \sigma(s)) & \text{for } d_l(s) < y(s) < d_r(s) \\
\Phi(d_r(s) \mid \mu(s), \sigma(s)) + \left[1 - \Phi(d_r(s) \mid \mu(s), \sigma(s))\right]\times\\\mathcal{G}(y(s) \mid \xi_r(s), \beta_r(s), d_r(s)) & \text{for } d_r(s) \leq y(s) < +\infty,
\end{cases}
\end{equation}
where \( \Phi(.) \) is the cdf of a Gaussian distribution with mean \( \mu(s) \) and standard deviation \( \sigma(s) \), \( \mathcal{G}(.) \) is the cdf of the GPD with shape parameter \( \xi(s) \), scale parameter \( \beta(s) \), and $d_l(s)$ and $d_u(s)$ denote the left and right thresholds.
To estimate the parameters of the above model, we define a log-likelihood function as
\begin{eqnarray}
\ell(\theta) &=& \sum_{i=1}^{n} \Big[
\log\left(\Phi(d_l | \mu, \sigma) \times g(-x_i | \xi_l, \beta_l, -d_l)\right) \cdot \mathbf{1}_{\{x_i \leq d_l\}}
+ \log \phi(x_i | \mu, \sigma) \cdot \mathbf{1}_{\{d_l < x_i < d_r\}}\nonumber\\ 
&+& \log\left(\left[1 - \Phi(d_r | \mu, \sigma)\right] \times g(x_i | \xi_r, \beta_r, d_r)\right) \cdot \mathbf{1}_{\{x_i \geq d_r\}}\Big]
\end{eqnarray}
where $1_{\{.\}}$ represents the indicator function. 

This approach is particularly well-suited for hydrological and climate time series, where extreme droughts or intense rainfall events exhibit heavy-tailed behavior, while the central part of the distribution remains approximately symmetric and unimodal. By separating the modeling of the tails from that of the center, the GP-N-GP model offers a robust and flexible framework for capturing and quantifying extremes in drought behavior.

\begin{remark}\label{remark1}
    A key limitation of the GP-N-GP mixture model lies in the selection of the thresholds \( d_l \) and \( d_r \), which delineate the boundary between the central region modeled by a normal distribution and the tails modeled by the GPD. The accuracy and robustness of the model heavily depend on the appropriate choice of the thresholds. If the thresholds are set too close to the center, the GPD components may be applied to regions that do not exhibit extreme behavior, leading to overfitting or distortion of the tail properties. Conversely, if the thresholds are set too far into the tails, insufficient data may be available to reliably estimate the GPD parameters, resulting in high variance and instability in the tail estimates.
\end{remark}
To address this constraint, the following section introduces an alternative approach that retains the flexibility of GPD-based tail modeling without requiring predefined thresholds.

\subsubsection{Bulk-and-Tails Model}\label{sec:bats}

To avoid threshold pre-selection, we adopt the  BATs distribution proposed by \citet{stein2021parametric} as an alternative to the GP-N-GP model. The BATs model offers flexible tail behavior and a well-defined central region for modeling $Y(s)$. The acronym “BATs” here explicitly reflects the modeling of dual-tail formulation. The BATs model introduces a parametric family of distributions capable of capturing both heavy and bounded tails, offering improvements over classical bulk-and-tail models that often lack flexibility in the lower tail. 
The  BATs model is specified by its cdf as
\begin{equation}\label{bat-cdf}
    \mathcal{F}_{\theta}(y) = \mathcal{T}_\nu(\mathcal{H}_\theta(y(s))),
\end{equation}
where $\mathcal{T}_\nu$ is the cdf of the Student's $t$ distribution with $\nu > 0$ degrees of freedom, and $\mathcal{H}_\theta(.)$ is a monotonic transformation determined by the parameter vector $\theta$. To define a proper model, we suppose that strictly increasing function \( \Upsilon: \mathbb{R} \rightarrow \mathbb{R} \), which satisfies:\[
\lim_{y(s) \to -\infty} \Upsilon(y(s)) = 0, \quad \text{and} \quad \lim_{y(s) \to \infty} \left[\Upsilon(y(s)) - y(s)\right] = 0. \] By using cumulative distribution function $\mathcal{G}$ with an analytic, positive density and finite mean, \citet{stein2021parametric} define $\Upsilon$ function which satisfying the above limiting conditions as 
$\Upsilon(x(s)) = \int_{-\infty}^{y(s)} \mathcal{G}(x(s))dx(s).$ 
Adopting the standard logistic distribution $
\mathcal{G}(y(s)) = {e^{y(s)}}/{1 + e^{y(s)}},$ we get $\Upsilon(y(s)) = \log(1 + e^{y(s)})$ and define as a CDF $\mathcal{H}_\theta(y)$ as
\begin{equation}\label{inner-cdf}
    \mathcal{H}_\theta(y(s)) = \left[1 + \gamma_2(s) \Upsilon\left(\frac{y(s) - \alpha_2(s)}{\beta_2(s)}\right)\right]^{1/\gamma_2(s)} - \left[1 + \gamma_1(s) \Upsilon\left(\frac{\alpha_1(s) - y(s)}{\beta_1(s)}\right)\right]^{1/\gamma_1(s)},
\end{equation}
where $\theta = (\alpha_i(s) , \beta_i(s)>0, \gamma_i(s))$ are the  location, scale, shape parameters
for $i = 1$ (lower tail) and $i = 2$ (upper tail). In analogy with the other extreme value distributions, the tail index parameter $\gamma_i(s), i\in(1,2)$ determines the tail's behavior: positive values indicate a heavy-tailed distribution with infinite support, whereas negative values correspond to a light-tailed distribution with bounded support in the tail. In the special case where $\gamma_i(s) =0 $ for $i\in(1,2)$, the $\mathcal{H}_\theta(y(s))$ simplifies to
\begin{equation*}
    \mathcal{H}_\theta(y(s)) = \exp\left(\frac{y(s) - \alpha_2(s)}{\beta_2(s)}\right) - \exp\left(\frac{\alpha_1(s) - y(s)}{\beta_1(s)}\right).
\end{equation*}

Let \( \Upsilon^{-1} \) denote the inverse of \( \Upsilon \). The distribution $ \mathcal{F}_{\theta}(y)$ given in \eqref{bat-cdf} has interior support $(L, U)$ with 
\begin{align}
    L(s) &= 
    \begin{cases}
        -\infty, & \text{if } \gamma_1(s) \geq 0, \\
        \alpha_1(s) - \beta_1(s) \Upsilon^{-1}(-1/\gamma_1(s)), & \text{if } \gamma_1(s) < 0, \nonumber
    \end{cases} \quad\\
    U(s) &= 
    \begin{cases}
        \infty, & \text{if } \gamma_2(s) \geq 0, \\
        \alpha_2(s) + \beta_2(s) \Upsilon^{-1}(-1/\gamma_2(s)), & \text{if } \gamma_2(s) < 0. \nonumber
    \end{cases}
\end{align}
To ensure the function is well-defined, we require \( L < U \). In the case where both \( \gamma_2(s) < 0 \) and \( \gamma_2(s) < 0 \), this becomes the condition $
\alpha_2(s) - \alpha_1(s) + \beta_2(s) \Upsilon^{-1}\left(-{1}/{\gamma_2(s)}\right) + \beta_1(s) \Upsilon^{-1}\left(-{1}/{\gamma_1(s)}\right) > 0 $ \citep{stein2021parametric}.
Differentiating the expression \eqref{bat-cdf} yields the probability density function of the distribution as
\begin{equation}\label{dens-bat}
    f(y|\theta) = t_\nu(\mathcal{H}_\theta(y(s))) \cdot \mathcal{H}_\theta'(y(s)),
\end{equation}
where $t_\nu$ is the Student's $t$ density with $\nu$ degrees of freedom and \begin{align}
\mathcal{H'}_\theta(y(s)) = & \ \frac{1}{\beta_2(s)} \left[1 + \gamma_2(s) \Upsilon\left(\frac{y(s) - \alpha_2(s)}{\beta_2(s)}\right)\right]^{\frac{1}{\gamma_2(s)} - 1} \Upsilon'\left(\frac{y(s) - \alpha_2(s)}{\beta_2(s)}\right) \nonumber  \\ &+ \frac{1}{\beta_1(s)} \left[1 + \gamma_1(s) \Upsilon\left(\frac{\alpha_0(s) - y(s)}{\beta_1(s)}\right)\right]^{\frac{1}{\gamma_1(s)} - 1} \Upsilon'\left(\frac{\alpha_0(s) - y(s)}{\beta_1(s)}\right).\nonumber
\label{eq:Hprime}
\end{align}
 is the derivative of $\mathcal{H}_\theta(.)$ given in \eqref{inner-cdf} with respect to $y(s)$. Hence, the corresponding log likelihood for estimating parameters is defined as 
\[
\ell(\theta) = \sum_{j=1}^{n} \log f(y(s_j) | \theta),
\]
where $f(y(s_j) | \theta)$ is given in \eqref{dens-bat}.

\subsection{Return level estimates for dry and wet periods}\label{sec:return-level}

To estimate future return levels, we construct variables for dry and wet extremes from the calculated drought index at each location \( s \in S \) as
\begin{align}\label{dry-wet-var}
    Z_{(*)}(s) = 
    \begin{cases}
        Y(s) \mid Y(s) > 0, & \text{(wet events)} \\
        -Y(s) \mid Y(s) < 0, & \text{(dry events)},
    \end{cases}
\end{align}
where the asterisk \((*)\) represents either the wet \((w)\) or dry \((d)\) tails. The negative values in the dry tail are inverted for modeling flexibility. Both tails are modeled using the Extended Generalized Pareto Distribution (EGPD)~\citep{naveau2016modeling}.

For the wet tail, the model is:
\[
Z_{(w)}(s) \sim \text{EGPD}(\kappa_{(w)}(s), \sigma_{(w)}(s), \xi_{(w)}(s)),
\]
and for the dry tail:
\[
Z_{(d)}(s) \sim \text{EGPD}(\kappa_{(d)}(s), \sigma_{(d)}(s), \xi_{(d)}(s)).
\]

We adopt the first family of EGPD from \citet{naveau2016modeling}, where \( G(v) = v^{\kappa(s)} \). The cumulative distribution function (CDF) at each location \( s \in S \) is
\begin{equation}
    \mathcal{F}\big(z_{(*)}(s); \kappa(s), \sigma(s), \xi(s)\big) = \left[1 - \left(1 + \frac{z_{(*)}(s)\xi(s)}{\sigma(s)}\right)^{-1/\xi(s)}\right]^{\kappa(s)},
\end{equation}
and the corresponding quantile function is
\begin{equation}\label{egpd-qunatile}
    \mathcal{Q}p_{(*)}(p)(s)  = 
    \begin{cases}
        \dfrac{\sigma(s)}{\xi(s)} \left[ \left(1 - p^{1/\kappa(s)}\right)^{-\xi(s)} - 1 \right], & \text{if } \xi(s) > 0, \\
        -\sigma(s) \log\left(1 - p^{1/\kappa(s)}\right), & \text{if } \xi(s) = 0.
    \end{cases}
\end{equation}
For $s\in S$, the \( n(s) \) be the total number of events, and \( n_{(w)(s)} \), \( n_{(d)}(s) \) be the number of events in the wet and dry tails, respectively. The empirical exceedance rates are defined as $
p_{(w)}(s) = {n_{(w)}(s)}/{n(s)}, p_{(d)}(s) = {n_{(d)}(s)}/{n(s)}.$
For a return period \( T \) (in years), the corresponding exceedance probabilities are defined as $
p_{(w)}^T(s) = {1}/({T \cdot p_{(w)}(s)})$ and $ p_{(d)}^T(s) = {1}/({T \cdot p_{(d)}(s)}).
$

The return levels for the wet and dry tails at location \( s \) are then given by:
\begin{align}
    RL_{(w)}^T(s) &= \mathcal{Q}^Tp_{(w)}\left(1 - p^T_{(w)}(s)\right),\label{wet-rl} \\
    RL_{(d)}^T(s) &= -\mathcal{Q}^T p_{(d)}\left(1 - p^T_{(d)}(s)\right),\label{dry-rl}
\end{align}
where the negative sign in the dry return level reverses the earlier transformation applied to dry values. By replacing the parameters in \eqref{egpd-qunatile} by their estimates, one can estimate the future return levels.

\section{Results}\label{result-disc}
\subsection{Data description}
For the application of the proposed space-time modeling framework, we utilize data from the Puglia region, situated in southeastern Italy, which is a classic example of a Mediterranean climate zone characterized by relatively low rainfall compared to other parts of the country. These environmental and climatic variabilities make Puglia particularly relevant for drought research. For instance, understanding when droughts begin could contribute to the creation of early warning systems, enabling farmers to shift towards drought-resistant crops and prompting increased water transfers from adjacent areas. Given its susceptibility to drought conditions, it is essential to analyze the spatial and temporal patterns of drought across the region.

Puglia is bordered by the Adriatic sea to the east and the Ionian sea to the southeast. The region is characterized by a landscape dominated by plains and low hills and mountains located in the Daunian Sub-Apennines and the Gargano promontory subregions in Foggia. Around half of Apulia's land area is covered by the Tavoliere delle Puglie in Foggia, Italy's second-largest plain. There are also several smaller plains, such as Salento in Lecce, Terra di Bari, Valle d'Itria in Brindisi, and Arco Ionico-Tarantino in Taranto.
\begin{figure}[t]
	\centering
\includegraphics[width=15cm,height=10cm]{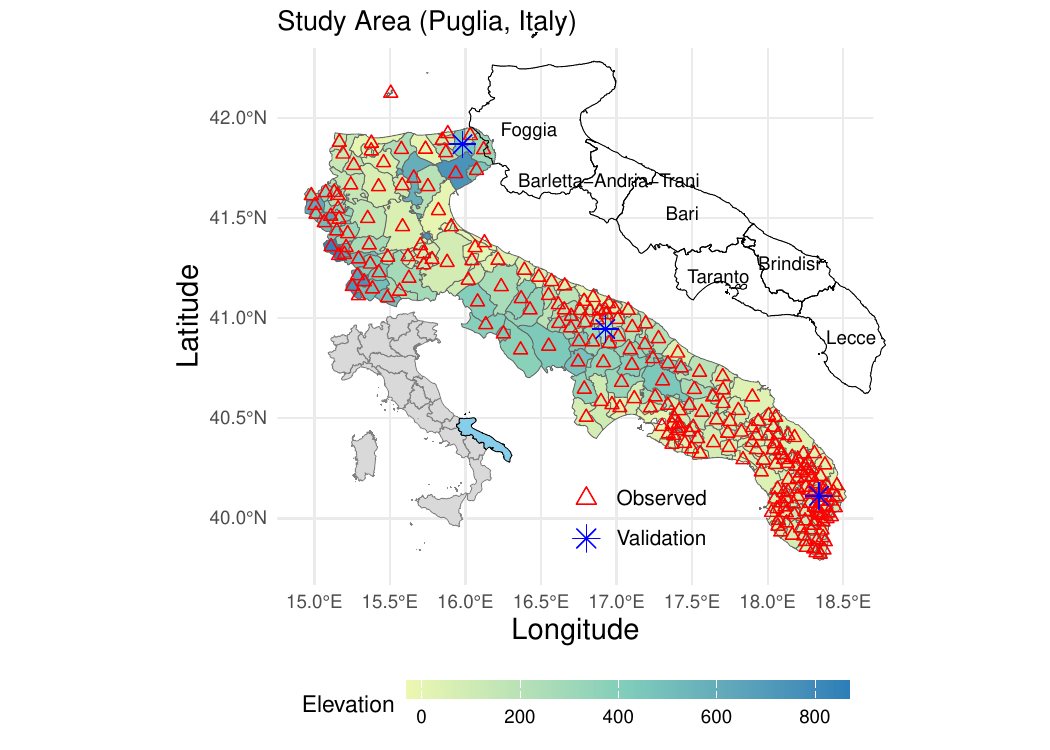}
	\caption{Study area map with spatial locations of the observed stations.  }
	\label{fig:study-map}
\end{figure}

Based on coordinates and elevation information, satellite data from 258 stations containing the variables precipitation and maximum temperature ranging (Jan-1983 to Dec-2023) are downloaded from the power NASA website\footnote{https://power.larc.nasa.gov/data-access-viewer/}. \citet{tayyeh2023analysis} compared NASA POWER data with ground-based observations and found coefficients of determination ranging from 0.74 to 0.91 for precipitation and 0.74 to 0.94 for maximum temperature, indicating that NASA POWER data can be reliably used in research where ground-based data are not readily available. Figure~\ref{fig:study-map} shows the spatial location of all 258 stations and subdivisions of Puglia.

\begin{figure}[H]
	\centering
\includegraphics[width=16.5cm,height=7cm]{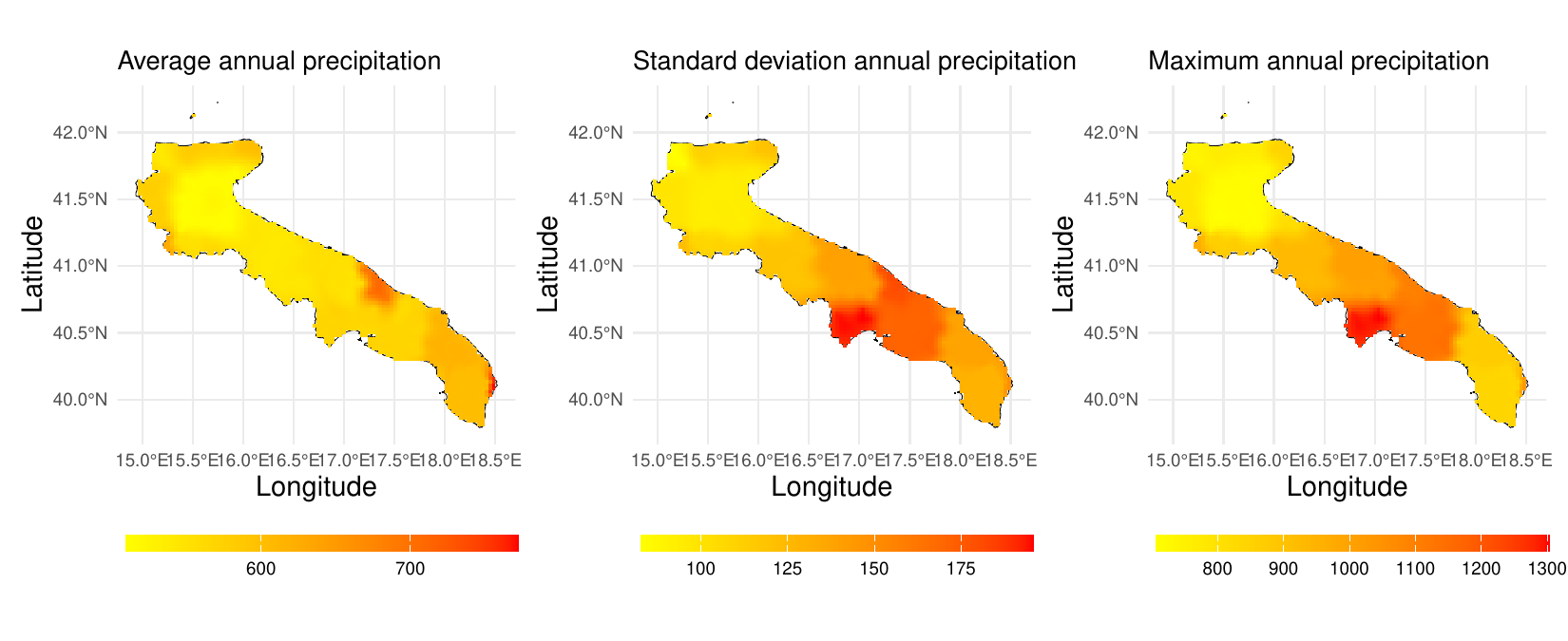}
	\caption{Descriptive statistics of the annual precipitation distribution for observed data.}
	\label{fig:descriptives}
\end{figure}

Figure~\ref{fig:descriptives} presents the descriptive statistics of precipitation across the study region, derived using ordinary Kriging interpolation. The estimated precipitation ranges from approximately 497mm to 782mm, with notable spatial variability. The lowest precipitation levels are observed in central Foggia, while the highest values occur in the Valle d'Itria area of Brindisi. Substantial spatial variation is evident across the gauges of Taranto and Brindisi, with certain southwestern stations in Brindisi recording annual precipitation totals of approximately 1300mm.

\subsection{Construction of drought index through space-time Gamma model}
At the first stage, we consider the space-time Gamma model, which allows variation over the year and between locations to model nonzero precipitation data accumulated over $m\in(1, 3, 6, 12)$ monthly temporal scale. This variation is primarily captured by expressing the Gamma distribution parameters $\alpha(s, t) > 0$ and $\psi(s, t) > 0$ using a tensor product of thin plate splines and cyclic cubic regression splines. The method enforces smooth spatiotemporal continuity across successive years.
To model the precipitation accumulated over $m$ months of temporal scale, the following basis functions are employed to construct $g_{*}$ in expression~\eqref{basis} 
\begin{eqnarray}
    \log (\psi(s,t))&=& \Tilde{w}\left(g_{\psi, tps, 40}(s), g_{\psi, ccs, 6}(\Tilde{ m}_{cyc}(t))\right) + g_{\psi, ncs, 15}(\Tilde{ m}_{temp}(s))\\
    \log (\alpha(s,t))&=& \Tilde{w}\left(g_{\alpha, tps, 40}(s), g_{\alpha, ccs, 6}(\Tilde{ m}_{cyc}(t))\right),
\end{eqnarray}
where $t$ is measured in years, $\Tilde{ m}_{cyc}(t)$ and  $\Tilde{ m}_{temp}(s)$ are the monthly mean of cyclical variations and temperature at locations $s\in S$ due to accumulation, $\Tilde{w}$ indicate a tensor product, $g_{*,tps,*}, g_{*,ncs,*},$ and $g_{*,ccs,*}$ denote thin plate, natural cubic, and
cyclic cubic regression splines, respectively, and $g{*,*,D}$ represent
a spline with basis dimension $ D$.

\begin{figure}[H]
	\begin{subfigure}{.5\textwidth}
		\centering
        \includegraphics[width=1\linewidth, height=0.3\textheight]{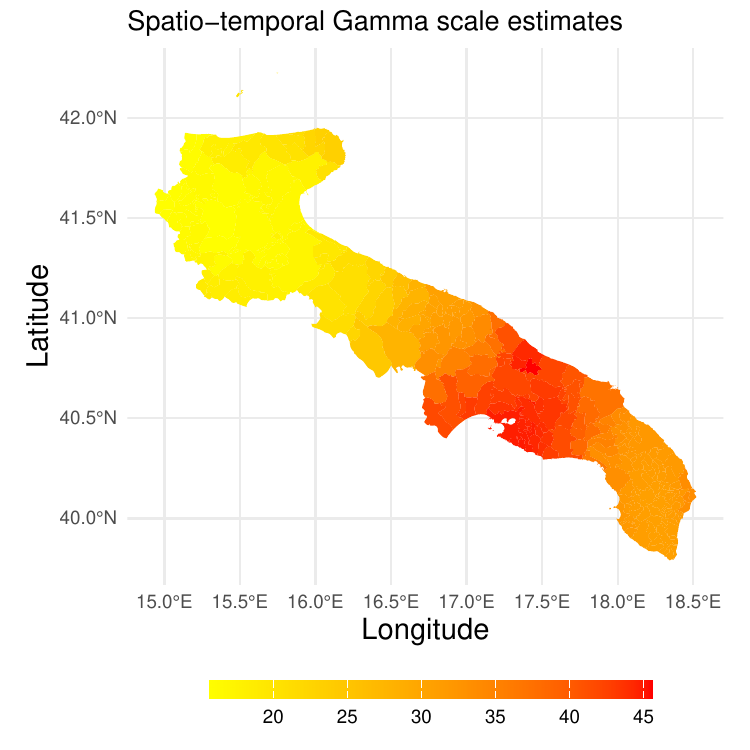}
	\end{subfigure}
	\begin{subfigure}{.5\textwidth}
		\centering
	\includegraphics[width=1\linewidth, height=0.3\textheight]{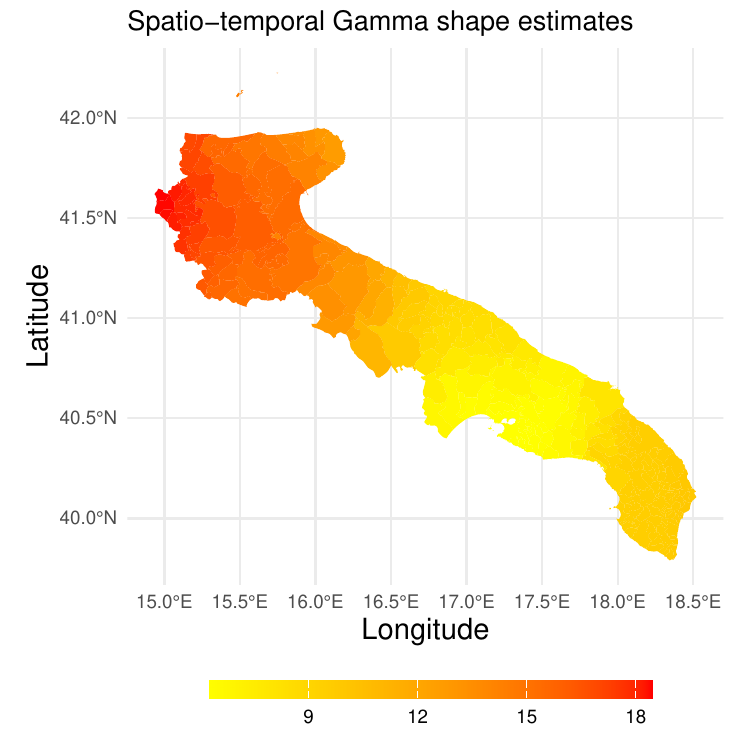}  
	\end{subfigure}
	\caption{Spatio-temporal estimates of fitted Gamma model for $m=6$ accumulation period. }
	\label{fig:fit-gamma6}
\end{figure}

\begin{figure}[H]
	\begin{subfigure}{.32\textwidth}
		\centering
		\includegraphics[width=1\linewidth, height=0.18\textheight]{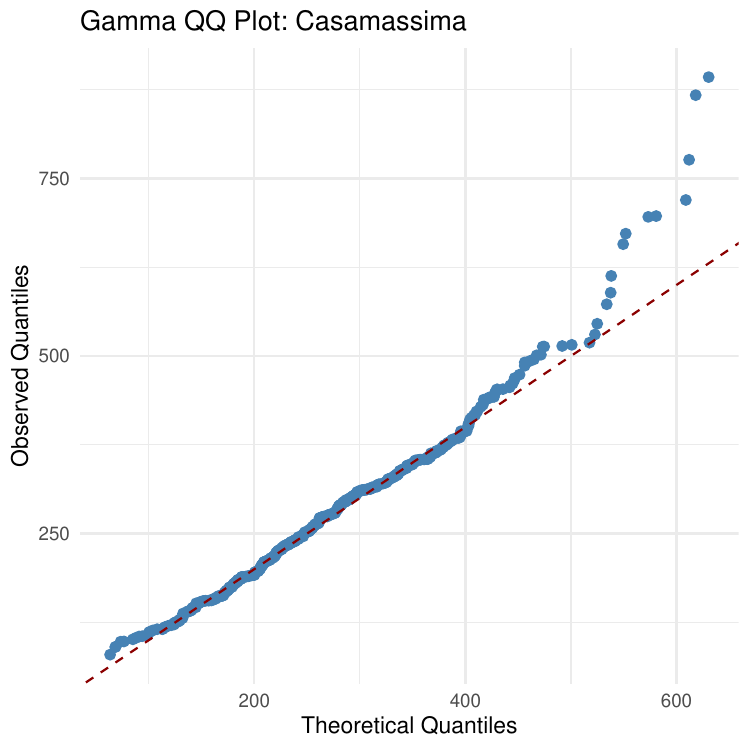}  
	\end{subfigure}
	\begin{subfigure}{.32\textwidth}
		\centering
		\includegraphics[width=1\linewidth, height=0.18\textheight]{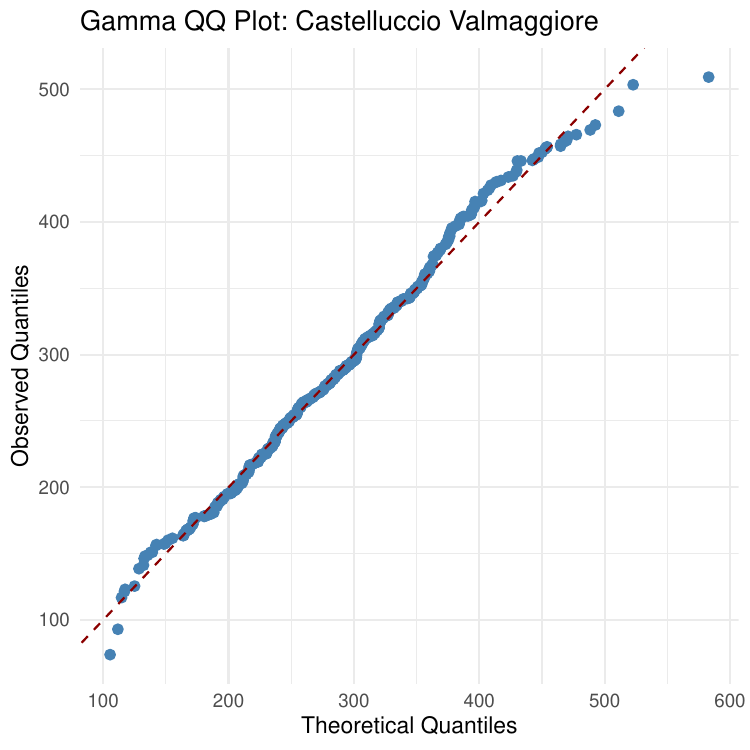}  
	\end{subfigure}
    \begin{subfigure}{.32\textwidth}
		\centering
		\includegraphics[width=1\linewidth, height=0.18\textheight]{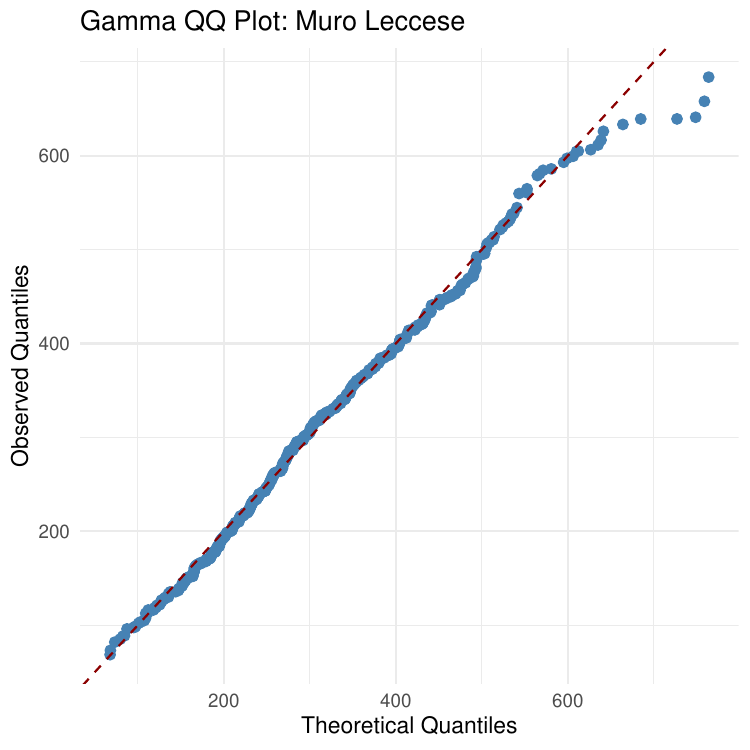}  
	\end{subfigure}
	\caption{QQ plots of fitted Gamma model for $m=6$ accumulation period at validation locations.  }
	\label{fig:qq-plot}
\end{figure}

The Gamma model proposed in Section~\ref{sec: space-time-gamma} is fitted by the likelihood defined in expression~\eqref{REML} to data obtained corresponding to accumulation scales $m \in \{1, 3, 6, 12\}$. Table~\ref{tab:edf} reports the effective degrees of freedom of fitted Gamma models, which remain below the maximum allowable values for each smoothing term. This suggests the selected basis dimensions are therefore high enough. Figure~\ref{fig:fit-gamma6} shows the mean estimates of the fitted space time Gamma model for $m=6$, while for other accumulation periods the parameters are presented in Figure~\ref{fig:pars-est-sup}. The scale parameter estimates of the fitted Gamma distribution indicate greater variability in precipitation for locations situated in Taranto and Brindisi, moderate variability in Lecce, and lower variability in Foggia, Barletta-Andria-Trani, and Bari. In contrast, the shape parameter estimates reveal that the precipitation distribution is less skewed in Foggia, Barletta-Andria-Trani, and Bari.
Conversely, locations in Taranto and Brindisi exhibit heavier tails, suggesting higher skewness, while Lecce shows an intermediate skewness with a moderately thick tail. We further randomly choose validation stations (see Figure~\ref{fig:study-map}) to assess the fitting of the Gamma model. For $m=6$, the QQ plots presented in Figure~\ref{fig:qq-plot} collectively provide strong evidence in support of the nonstationary Gamma model, indicating an adequate fit to the observed data under nonstationary conditions. Results for the accumulation periods are presented in Figure~\ref{fig:qq-plot-sup}. There is a possibility that the Gamma distribution may not adequately capture the upper tail of the precipitation distribution toward the rainiest sites, where GAM operates closer to interpolation limits than true extrapolation. It might be necessary to modify the model or refine data quality control to improve estimates for such locations.

\begin{figure}[H]
	\centering
\includegraphics[width=16.5cm,height=7cm]{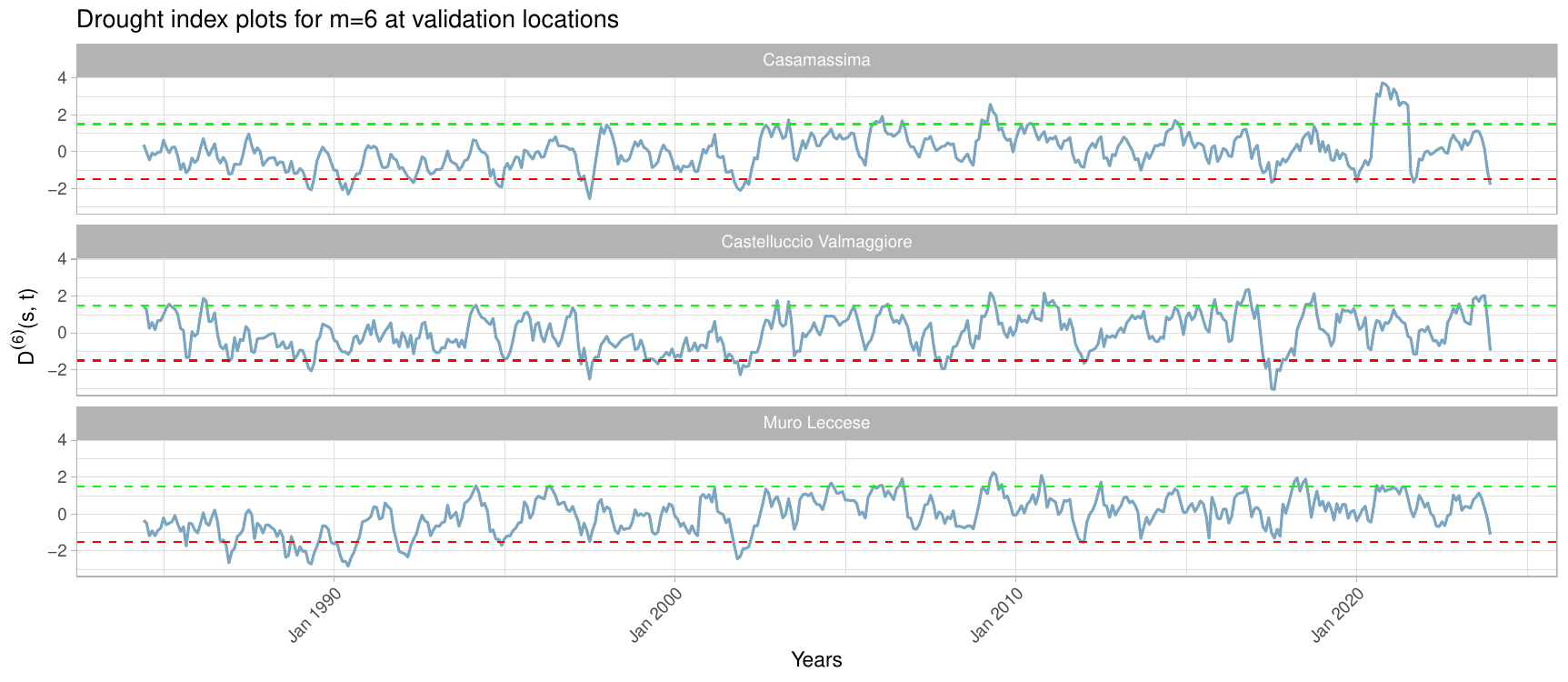}
	\caption{Calculated drought indices at validation locations for $m=6$ accumulation period.}
	\label{fig:SPI-validate}
\end{figure}

We use parameter estimates of the fitted space time Gamma model in Section~\ref{sec:drought-index} and construct the drought index $D^{(m)}(s, t)$ for scales $m \in \{1, 3, 6, 12\}$. Figure~\ref{fig:SPI-validate} illustrates the constructed $D^{(6)}(s, t)$ values at the validation locations. The other constructed $D^{(m)}(s, t)$, $m \in \{1, 3, 12\}$ on validation locations are illustrated in Figure~\ref{fig:SPI-validate-sup}. Drought events are clearly identifiable as instances where the SPI values fall below the red horizontal line representing the threshold of 
-1.5. According to the classification criteria presented in Table~\ref{tab:drought-catagories}, the values below the red line represent the severe or extreme drought conditions in the given period. Based on the identified drought events in observed stations, we can reveal a clear indication of drought risk through spatial and temporal patterns across the Puglia region. 
\begin{table}[t]
    \centering
    \caption{Effective degrees of freedom for smooth terms in the Gamma model for $m \in \{1, 3, 6, 12\}$. The first row is corresponding to $\Tilde{w}\left(g_{\psi, tps, *}(s), g_{\psi, ccs, *}(\Tilde{ m}_{cyc}(t))\right)$ smooth term and the second row is associate with $g_{\psi, ncs, *}(\Tilde{ m}_{temp}(s))$ smooth term. Numbers in parentheses indicate maximum possible values. }
    \begin{tabular}{p{1.7cm} p{1.7cm}  p{1.7cm} p{1.7cm}  p{1.7cm} p{1.7cm}  p{1.7cm} p{1.7cm}}
        \toprule
        \multicolumn{2}{c}{$m=1$} 
        & \multicolumn{2}{c}{$m=3$} 
        & \multicolumn{2}{c}{$m=6$} 
        & \multicolumn{2}{c}{$m=12$} \\
        \cmidrule(l){1-2} \cmidrule(l){3-4} \cmidrule(l){5-6} \cmidrule(l){7-8}
        $\psi(s,t)$ & $\alpha(s,t)$
        & $\psi(s,t)$ & $\alpha(s,t)$
        & $\psi(s,t)$ & $\alpha(s,t)$
        & $\psi(s,t)$ & $\alpha(s,t)$ \\
        \midrule
        173.95(199)&88.83(199)&148.50(199)&106.24(199) &121.59(199)&111.19(199)&120.38(199)&91.7(199)\\
        13.66(14)&--&13.79(14)&--&13.90(14)&--&13.88(14)&--\\
        \bottomrule
    \end{tabular}\label{tab:edf}
\end{table}
\paragraph{Drought risk:} To estimate drought risk (Section~\ref{sec:droughtrisk}) at each location, we initially employed the two-tailed GP-N-GP mixture model (Section~\ref{sec:gng}), which separately models the lower and upper tails of the distribution of $Y^{(m)}(s)=D^{(m)}(s,t)$ using GPDs, while a Gaussian distribution captures the central bulk of the data. While fitting the GP-N-GP model, the lower and upper thresholds were fixed at the empirical 0.10 and 0.90 quantiles of $Y^{(m)}(s) $, $ s\in S$, respectively.  This structure allows for flexible modeling of extremes—such as severe droughts or unusually wet periods—while maintaining interpretability in the central regime. Using parameter estimates of GP-N-GP at each location in expression~\eqref {cdfrisk} and setting the threshold $u_c=-2$, we derive the corresponding probabilistic extreme drought risk values. However, as discussed in Remark~\ref{remark1}, a key limitation of this approach lies in the subjective choice of thresholds at the time of fitting the GP-N-GP model, which strongly influence the stability and accuracy of the estimated tail behavior. Improper threshold selection may lead to either overfitting in non-extreme regions or poor tail estimation due to insufficient data.

To mitigate the sensitivity of tail estimates to subjective threshold choices inherent in the GP-N-GP framework, we adopt the BATs model (described in Section~\ref{sec:bats}), as an alternative, threshold-free approach. The BATs model introduces a smooth, continuous, and unified distributional representation that captures both central and tail behavior without pre-specifying truncation points. The parametric formulation enables flexible control of the tail heaviness through dedicated shape parameters, while maintaining a well-defined and interpretable central region. This makes BATs particularly well-suited for extremes modeling in environmental contexts, where extremes may not always exhibit sharp transitions from the bulk of the data.

\begin{figure}[t]
	\begin{subfigure}{.32\textwidth}
		\centering
		\includegraphics[width=1\linewidth, height=0.15\textheight]{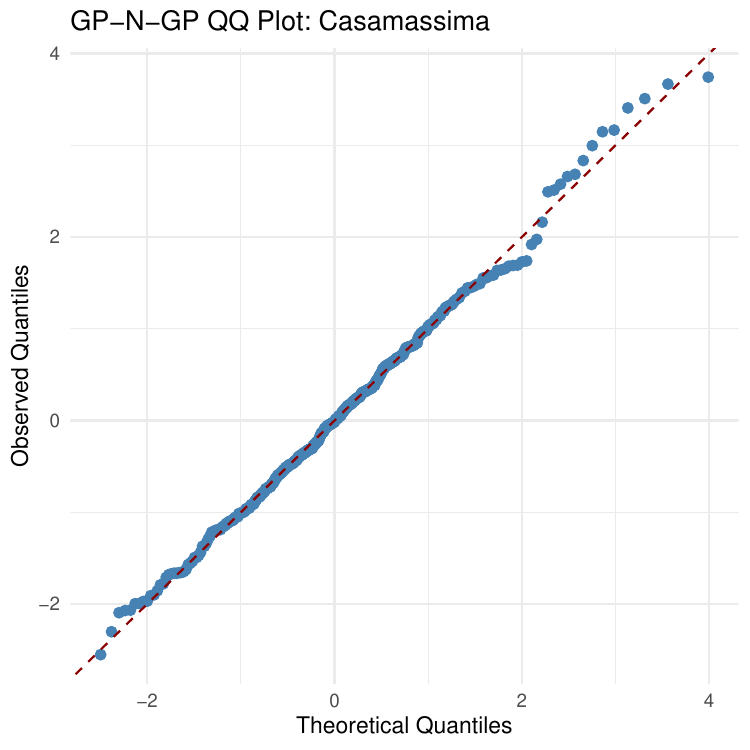}  
	\end{subfigure}
	\begin{subfigure}{.32\textwidth}
		\centering
		\includegraphics[width=1\linewidth, height=0.15\textheight]{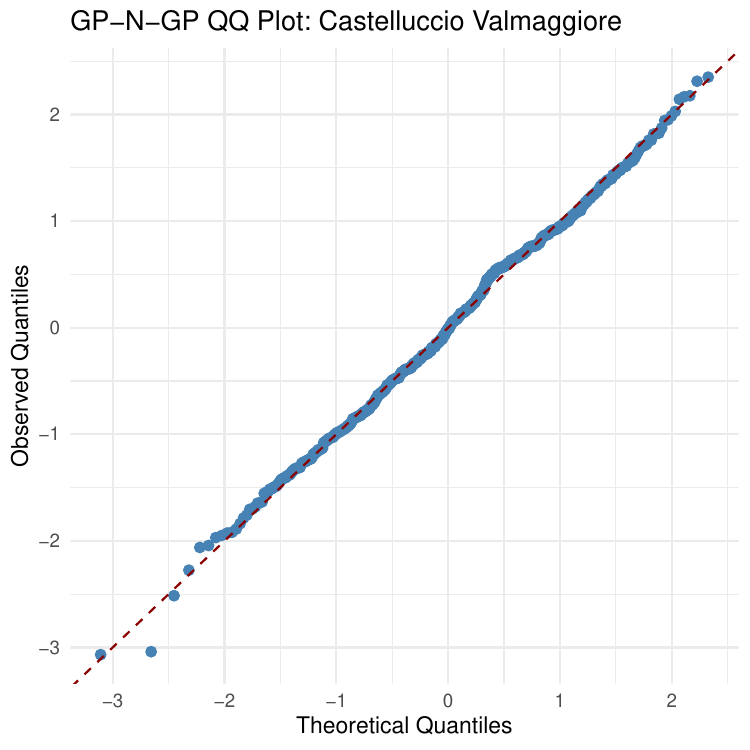}  
	\end{subfigure}
    \begin{subfigure}{.32\textwidth}
		\centering
		\includegraphics[width=1\linewidth, height=0.15\textheight]{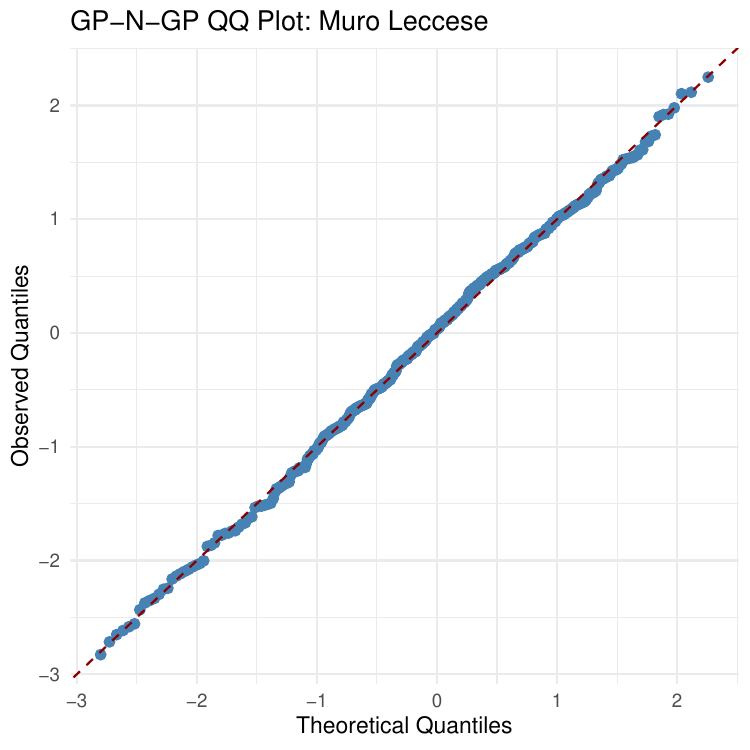}  
	\end{subfigure}
	\newline
    \begin{subfigure}{.32\textwidth}
		\centering
		\includegraphics[width=1\linewidth, height=0.15\textheight]{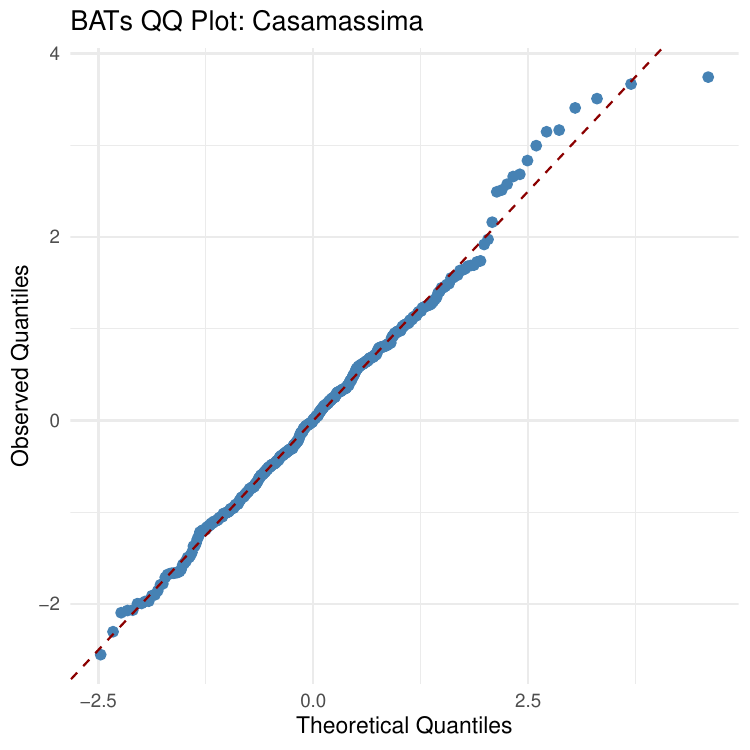}  
	\end{subfigure}
	\begin{subfigure}{.32\textwidth}
		\centering
		\includegraphics[width=1\linewidth, height=0.15\textheight]{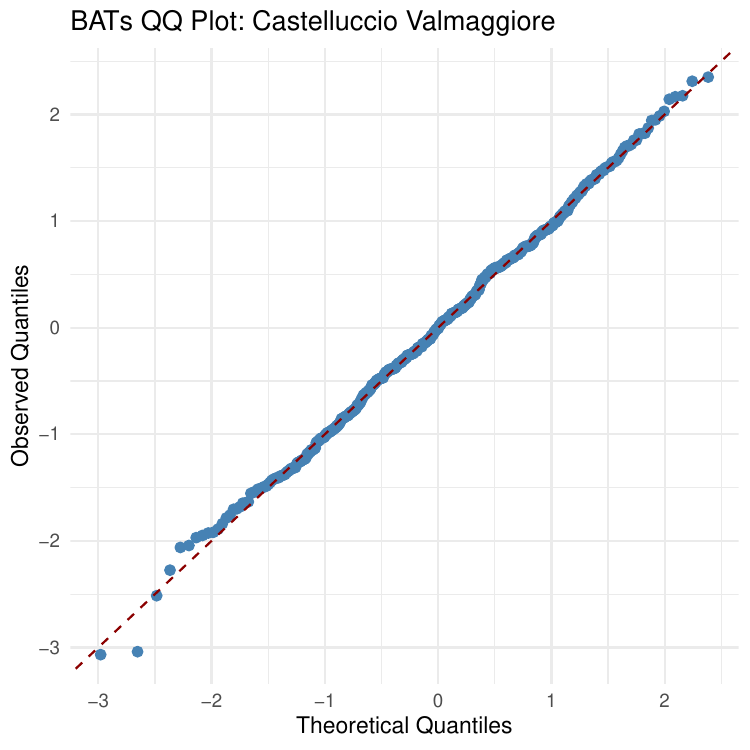}  
	\end{subfigure}
    \begin{subfigure}{.32\textwidth}
		\centering
		\includegraphics[width=1\linewidth, height=0.15\textheight]{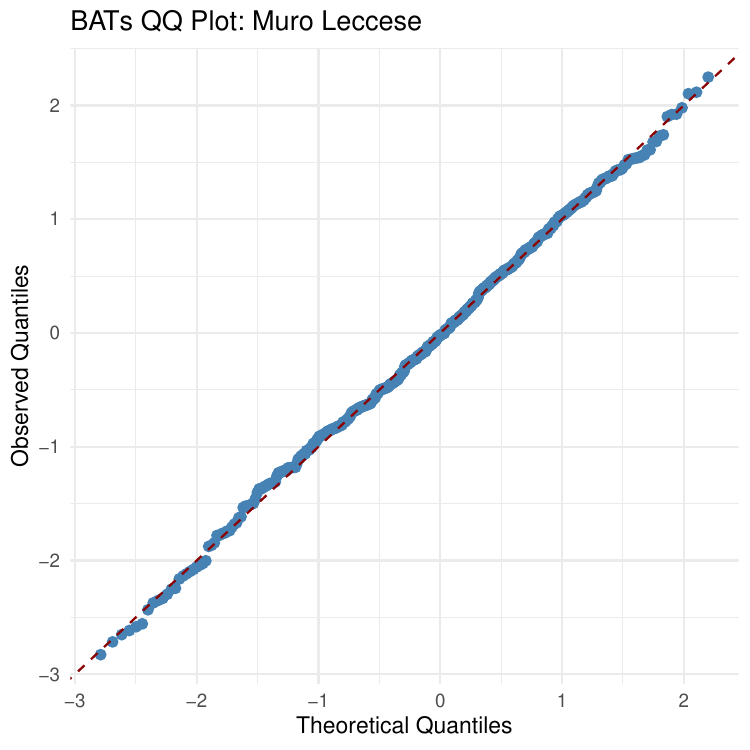}  
	\end{subfigure}

    \caption{QQ plots of the fitted GP-N-GP model for $m = 6$ at validation locations (top-row), and QQ plots of the fitted BATs model for $m = 6$ at the same validation locations (bottom-row). }
	\label{fig:qq-plot-gng-bat}
\end{figure}

\begin{figure}[t]
	\begin{subfigure}{.5\textwidth}
		\centering
\includegraphics[width=1\linewidth, height=0.3\textheight]{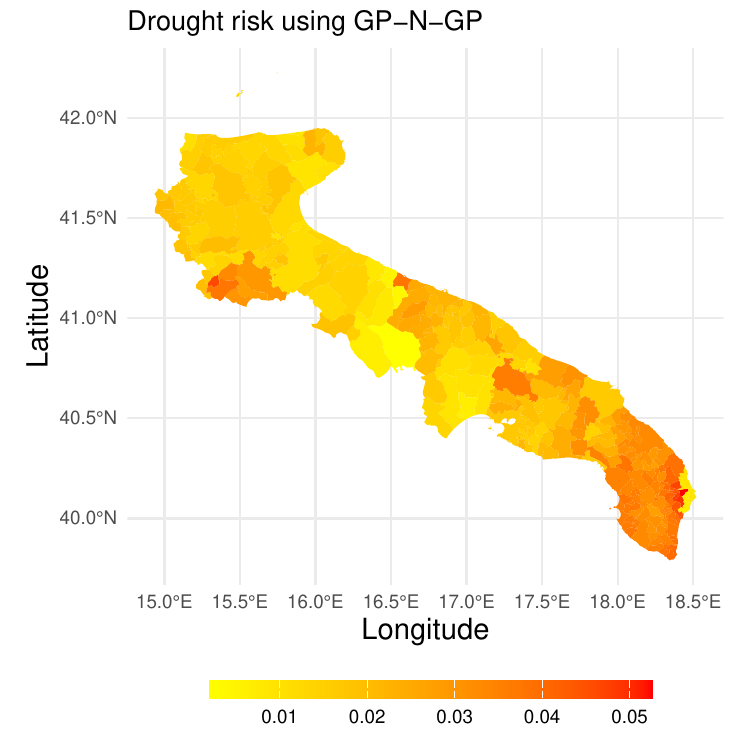}  
	\end{subfigure}
	\begin{subfigure}{.5\textwidth}
		\centering
\includegraphics[width=1\linewidth, height=0.3\textheight]{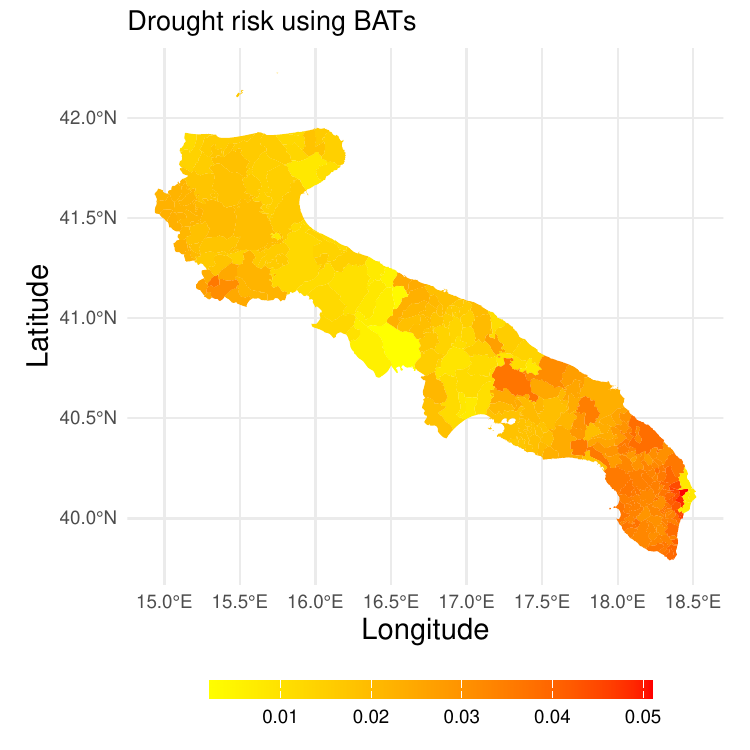}  
	\end{subfigure}
	\caption{Estimated extreme drought risk through GP-N-GP (left), and BATs (right).}
	\label{fig:drought-risk}
\end{figure}
We fitted the BATs model to the $Y^{(m)}(s)$, and computed probabilistic drought risk estimates in the same manner as with the GP-N-GP model, again using 
$u_c=-2$ in Equation~\eqref{cdfrisk}.  Figure~\ref{fig:qq-plot-gng-bat} presents diagnostic plots at selected validation locations, highlighting that the GP-N-GP and BATs models exhibit a good fit in capturing extreme tail behavior. The GP-N-GP model may exhibit poor fit when higher thresholds are used, due to limited data in the tail regions, leading to unstable parameter estimates~\citep{zhao2014extreme}. Additional plots regarding diagnostics are given in the Supplement file (see Figures~\ref{fig:qq-plot-gng-sup},\ref{fig:qq-plot-bat-sup}).
For the fixed drought temporal scale 
$m=6$, the spatial distribution of drought risk estimates from both models is visualized in Figure~\ref{fig:drought-risk}.  Both the GP-N-GP and BATs models exhibit a high degree of agreement across locations, with the GP-N-GP model showing a marginally higher sensitivity to extreme events in certain areas.
Notably, regions such as Taranto, Brindisi, Lecce, and parts of northwestern to southwestern Foggia exhibit moderate to high drought risk. In contrast, the Barletta–Andria–Trani region and the central, northeastern, and southwestern zones of Bari exhibit relatively low drought risk. The southeastern and southern regions of Bari are characterized by intermediate (moderate) drought risk levels.

Drought risk estimates for additional temporal aggregation scales $m \in \{1, 3, 12\}$  are provided in the supplementary Figure~\ref{fig:drought-risk-sup}, highlighting the consistency and temporal robustness of the modeled drought risk patterns across varying timescales.
\paragraph{Return level estimates:} One of the important objectives of this study is to predict future events during both dry and wet periods. To estimate future return levels, we followed the procedure detailed in Section~\ref{sec:return-level}. First, we separated the $Y^{(m)}(s)$ into dry and wet events using the classification rule in \eqref{dry-wet-var}, and constructed new variables $Z_{(w)}(s)$ and $Z_{(d)}(s)$ corresponding to wet and dry periods, respectively. We then fitted the EGPD model, as described in Section~\ref{sec:return-level}, separately to the upper and lower tails using the maximum likelihood estimation method.

\begin{figure}[t]
	\begin{subfigure}{.32\textwidth}
		\centering
		\includegraphics[width=1\linewidth, height=0.15\textheight]{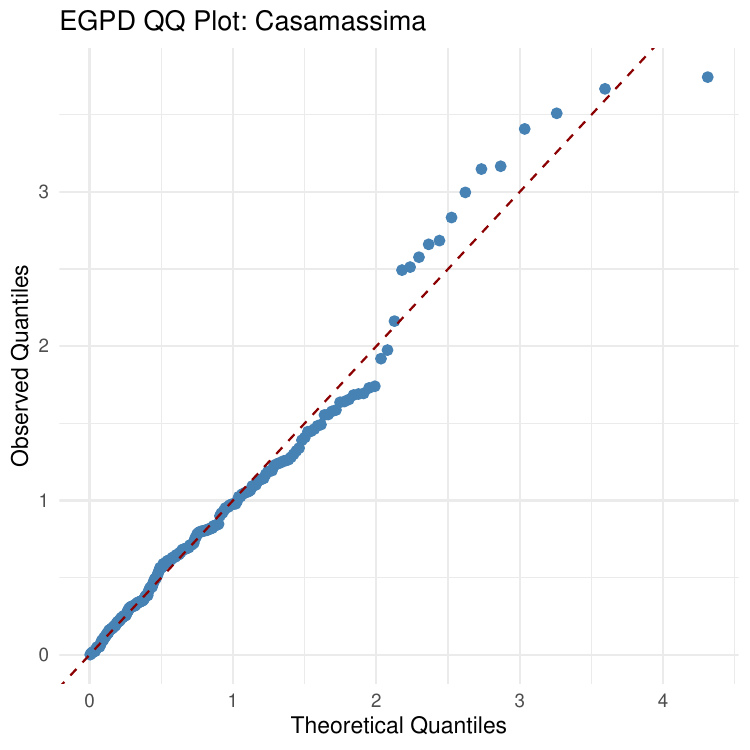}  
	\end{subfigure}
	\begin{subfigure}{.32\textwidth}
		\centering
		\includegraphics[width=1\linewidth, height=0.15\textheight]{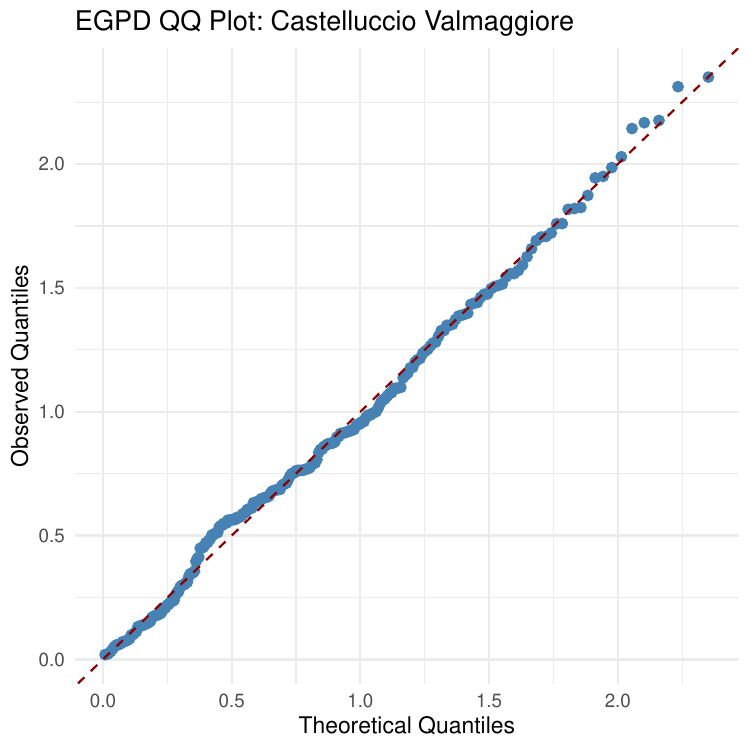}  
	\end{subfigure}
    \begin{subfigure}{.32\textwidth}
		\centering
		\includegraphics[width=1\linewidth, height=0.15\textheight]{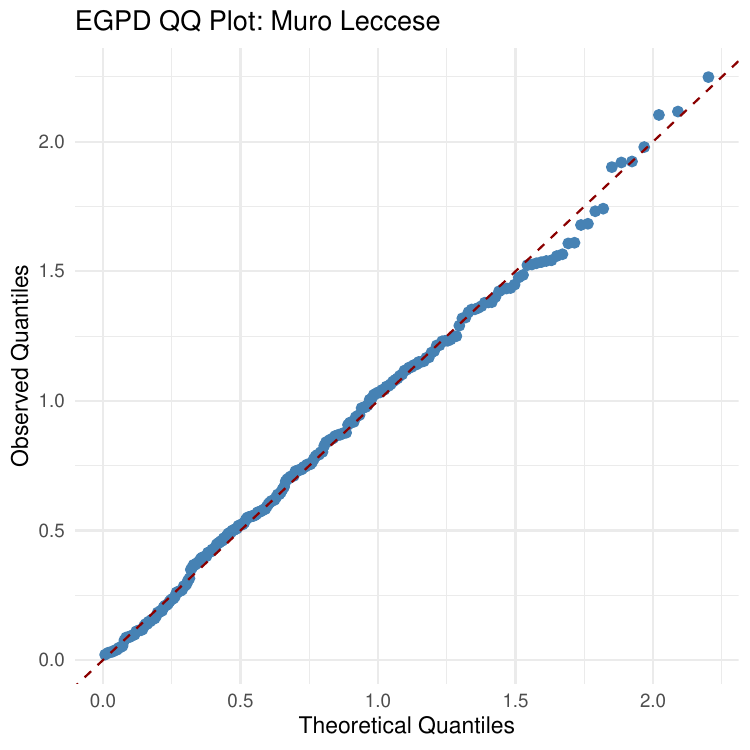}  
	\end{subfigure}
	\newline
    \begin{subfigure}{.32\textwidth}
		\centering
		\includegraphics[width=1\linewidth, height=0.15\textheight]{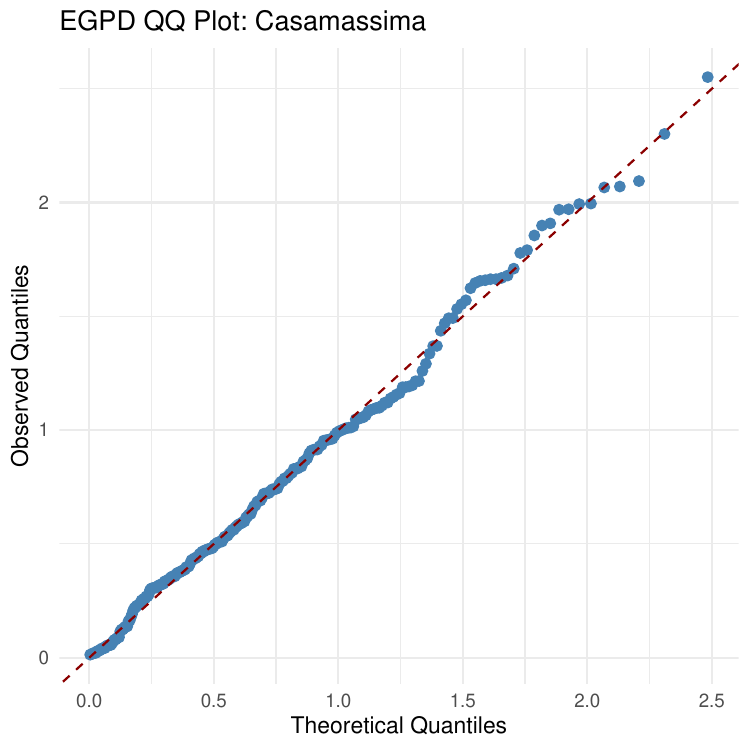}  
	\end{subfigure}
	\begin{subfigure}{.32\textwidth}
		\centering
		\includegraphics[width=1\linewidth, height=0.15\textheight]{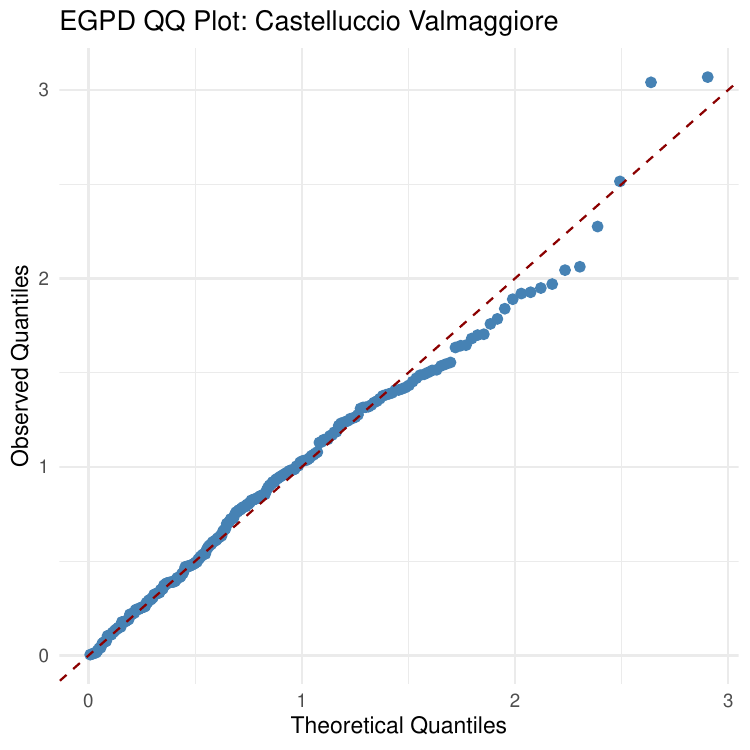}  
	\end{subfigure}
    \begin{subfigure}{.32\textwidth}
		\centering
		\includegraphics[width=1\linewidth, height=0.15\textheight]{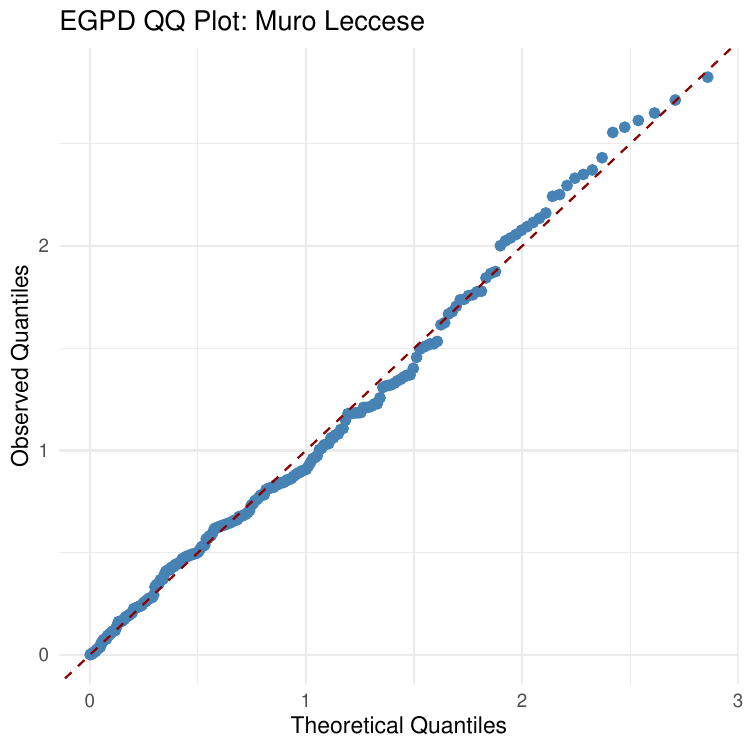}  
	\end{subfigure}

    \caption{QQ plots of fitted EGPD model to wet periods at validation locations for $m=6$ (top row) and QQ plots of fitted EGPD model to inverted dry periods at validation locations for $m=6$ (bottom row). }
	\label{fig:qqplot-egpd}
\end{figure}

Since the EGPD is defined over the positive real line, we transformed the dry-event data by inverting the sign of the negative values prior to model fitting. In contrast to \citet{naveau2016modeling}, who employed a censored likelihood approach to address discretization issues near zero in hourly precipitation data, our dataset does not exhibit such discretization effects. Therefore, we used the full (uncensored) likelihood to estimate the model parameters. Figure~\ref{fig:qqplot-egpd} presents QQ plots at validation sites to evaluate the fit of the EGPD model to both $Z_{(w)}(s)$ and $Z_{(d)}(s)$, based on a time scale of $m = 6$. The top row corresponds to the QQ plots for the wet spell data, while the bottom row shows the QQ plots for the inverted dry spell data. Additional QQ plots for $m = 1, 3, 12$ are given in~\ref{fig:qqplot-egpd-pos}, \ref{fig:qqplot-egpd-neg} in the Supplement file.   Using the estimated parameters and the quantile function of the EGPD, we computed return levels for return periods $T \in \{5, 10, 20, 50 \}$. As the dry and wet periods were modeled separately, we adjusted the return periods by multiplying $T$ by the empirical rate of exceedances, as explained in Section~\ref{sec:return-level}.

\begin{figure}[H]
	\begin{subfigure}{.5\textwidth}
		\centering
\includegraphics[width=1\linewidth, height=0.25\textheight]{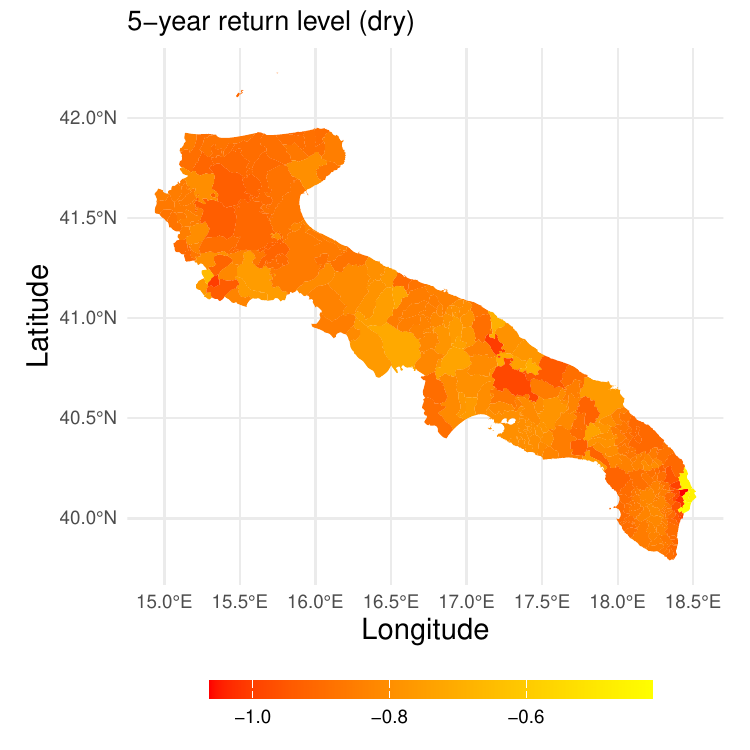}  
	\end{subfigure}
	\begin{subfigure}{.5\textwidth}
		\centering
\includegraphics[width=1\linewidth, height=0.25\textheight]{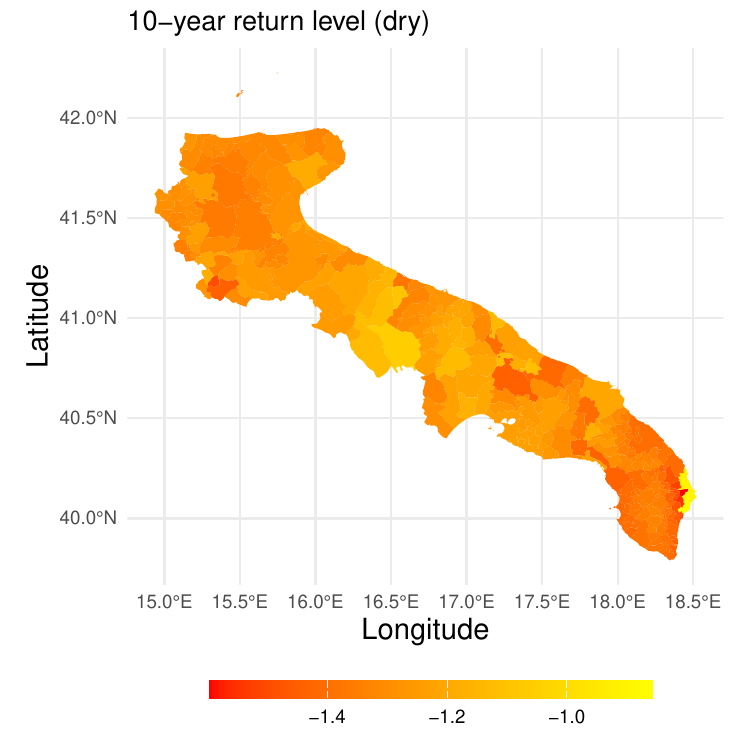}  
	\end{subfigure}
    \newline
    \begin{subfigure}{.5\textwidth}
		\centering
\includegraphics[width=1\linewidth, height=0.25\textheight]{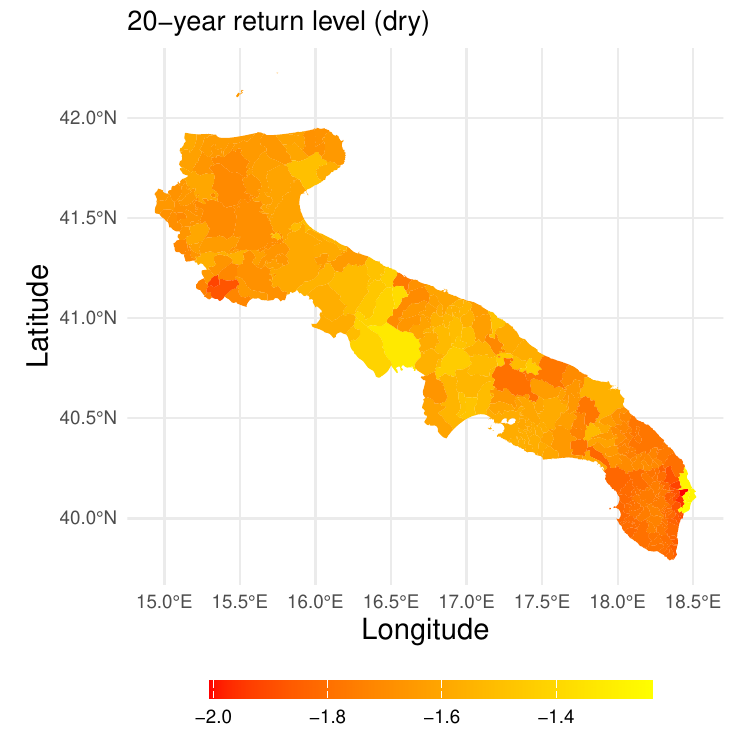}  
	\end{subfigure}
	\begin{subfigure}{.5\textwidth}
		\centering
\includegraphics[width=1\linewidth, height=0.25\textheight]{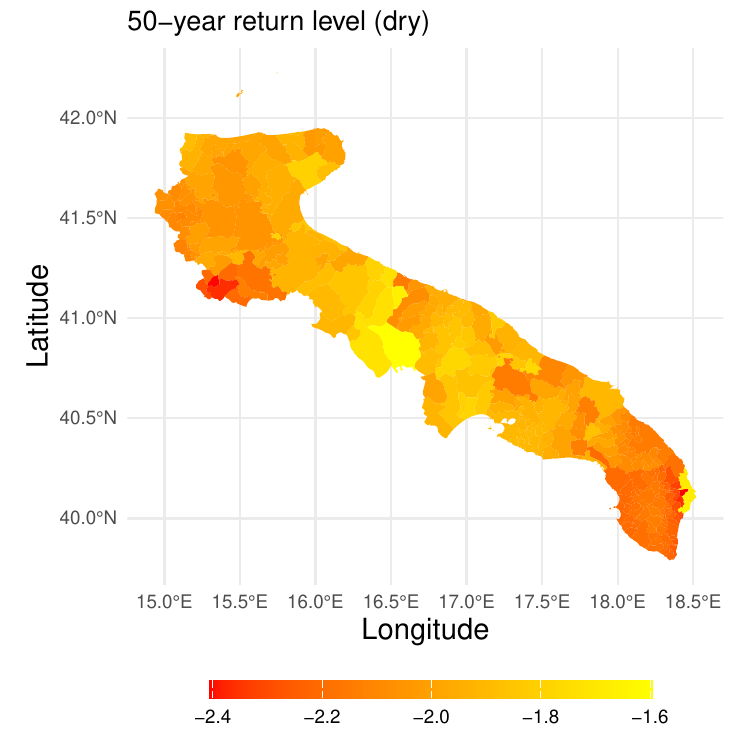}  
	\end{subfigure}
	\caption{Spatial maps of the estimated return levels for 5-, 10-, 20-, and 50-year return periods, based on an accumulation period of $m = 6$. }
	\label{fig:rl-dry-m6}
\end{figure}

The return level plots shown in Figure~\ref{fig:rl-dry-m6} based on a ($m = 6$) month accumulation period, illustrate the expected severity of future drought events corresponding to different return periods $T = 5, 10, 20,$ and $50$ years. These return levels represent the rainfall deficit that are expected to be exceeded, on average, once every $T$ years. To better understand these results, drought intestines obtained through return levels are interpreted using the classification thresholds provided in Table~\ref{tab:drought-catagories}, which identify drought categories such as moderate, severe, and extreme drought. For the 5-year return period ($T = 5$), some provincesof Puglia exhibit noticeable rainfall deficits. In particular, the Lecce province, along with the southeastern part of Bari and central areas of Foggia, are highlighted as experiencing rainfall deficits consistent with at least low drought conditions. This indicates that drought events of this frequency already pose a significant hydrological risk to these subregions. At the 10-year return period ($T = 10$), the spatial extent and intensity of the drought conditions become more pronounced. The Lecce region, in particular, emerges as a vulnerable area, with return level values reaching $-1.5$, indicating moderate drought conditions. Thus, droughts occurring on a decadal scale are likely to have a significant impact on local water resources and agricultural systems.

For the 20- and 50-year return periods ($T = 20$ and $T = 50$), the analysis reveals the most severe rainfall deficits, pointing to extreme drought events. The southwestern part of Lecce, large areas within Foggia and southeastern location of Bari exhibit return levels that exceed the $-1.8$ and $-2$ respectively for very dry to extreme drought as defined in Table~\ref{tab:drought-catagories}. In light of these high-magnitude, low-frequency events, these subregions of Puglia may be at risk from long-term climatic extremes, which could impact ecosystem sustainability, groundwater recharge, and regional drought management policies. Additional results of return levels based on other accumulation periods $m=1, 3, 12$ and wet events are provided in the Figures~\ref{fig:rl-dry-m1}--\ref{fig:rl-wet-m12} in the Supplementary File.

\section{Conclusion }\label{sec:concl}
To construct the drought index, GAM forms are employed to capture variations in the marginal distribution of monthly accumulated precipitation. Particular attention is paid to spatiotemporal variation, which is often evident in drought phenomena. Thin plate regression splines with relatively high smoothness are used to model continuous spatial variation. These splines are a particular generalization of Duchon splines \citep{duchon1977splines}, offering greater flexibility, such as handling anisotropy, while maintaining differentiability. This differentiability constraint, however, is not required when using Gaussian processes with suitable covariance functions. For modeling the temporal component of accumulated precipitation, cubic regression splines are applied to represent continuous one-dimensional variation, with cyclic versions ensuring continuity at temporal endpoints. Tensor products of cyclic cubic splines and thin plate splines enable the modeling of smooth spatio-temporal variation, allowing the integration of year-round data and helping to mitigate data scarcity in extreme precipitation events. The resulting spatio-temporal Gamma distribution estimates and derived drought index align well with the established understanding of rainfall deficits, supporting the validity of this modeling approach.

The proposed space time Gamma model offers a flexible framework for capturing climate-driven nonstationarity in drought indices. By leveraging the full dataset through a tensor product spline, the model effectively incorporates spatial and temporal variation, in contrast to existing models that rely solely on smooth functions and overlook spatial structure. This approach captures the true underlying nonstationarity of the climate system.

This work also introduces a flexible dual-tail model to improve drought risk estimation across the region. In the GP-N-GP framework, a key challenge lies in selecting prespecified thresholds to distinguish between upper and lower extremes. While the model performs well when optimal thresholds are chosen, this selection process can be problematic when applied to different regions. To address this limitation, we use the BATs model, which estimates extreme tail behavior using the tail indices of the GPD, eliminating the need for threshold selection. Drought risk estimates from the BATs model closely align with those from the GP-N-GP model. However, BATs offer several advantages: it has less number parameters, avoid threshold specification, and are simpler to implement.

To estimate future drought and wet events, we introduce the EGPD model (see Section~\ref{sec:return-level}), which models the full distribution of dry and wet periods without requiring the selection of thresholds. Since EGPD is defined for positive data, dry events were inverted prior to model fitting, and the resulting return levels were subsequently transformed back to represent the original scale. The return level analysis highlights significant spatial variability in drought risk across the study area, with Lecce consistently identified as a drought-prone hotspot, particularly at higher return periods. These findings underscore the importance of integrating return level-based drought assessments into regional planning and climate adaptation strategies.

\section*{Data availability}
The data and codes prepared for this paper are available for reproducing the results and can be obtained from the corresponding author upon reasonable request. 
\\
\\
\bibliographystyle{apalike}
\begin{center}
    \bibliography{References}
\end{center}

\newpage

\renewcommand{\thetable}{S.\arabic{table}}
\setcounter{table}{0}
\renewcommand{\thefigure}{S.\arabic{figure}}
\setcounter{figure}{0}

\renewcommand{\theequation}{S.\arabic{equation}}
\setcounter{equation}{0}
\renewcommand{\thesection}{S.\arabic{section}}
\setcounter{section}{0}
\setcounter{page}{1}

\begin{center}
	\section*{Supplementary material for \\ ``Space time modeling for drought classification and
prediction''}
\end{center}

\begin{figure}[H]
	\begin{subfigure}{.5\textwidth}
		\centering
        \includegraphics[width=1\linewidth, height=0.25\textheight]{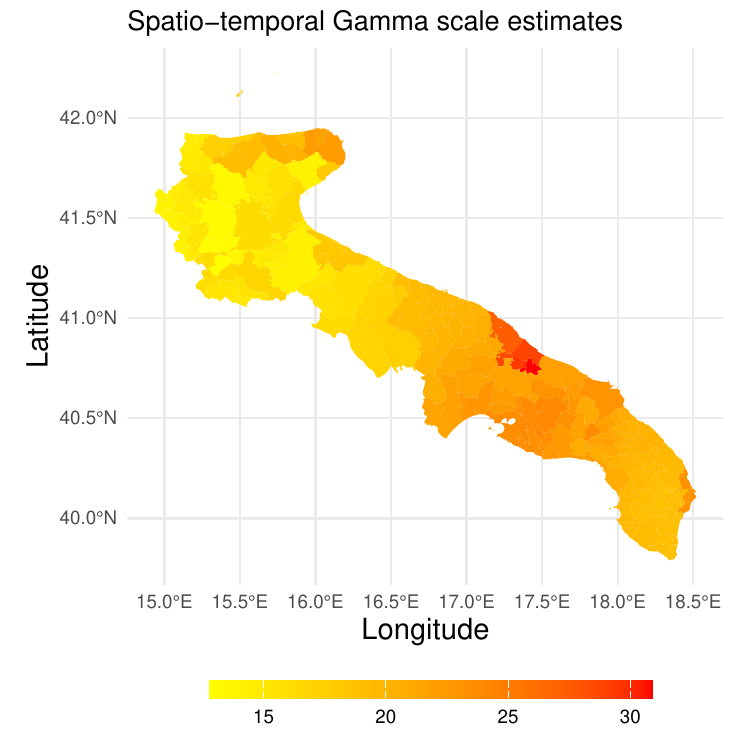}  
	\end{subfigure}
	\begin{subfigure}{.5\textwidth}
		\centering
		\includegraphics[width=1\linewidth, height=0.25\textheight]{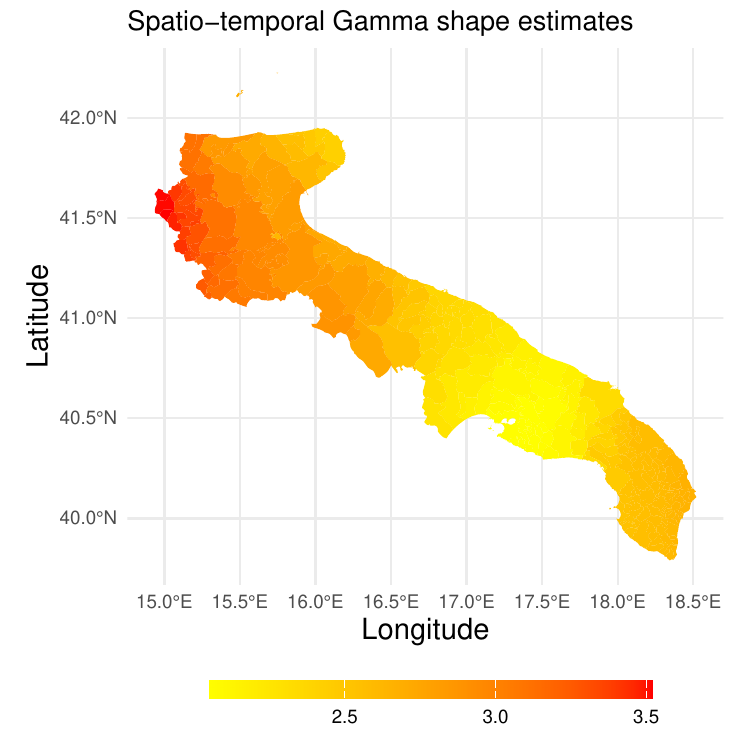}  
	\end{subfigure}
        \newline
       \begin{subfigure}{.5\textwidth}
		\centering
		\includegraphics[width=1\linewidth, height=0.25\textheight]{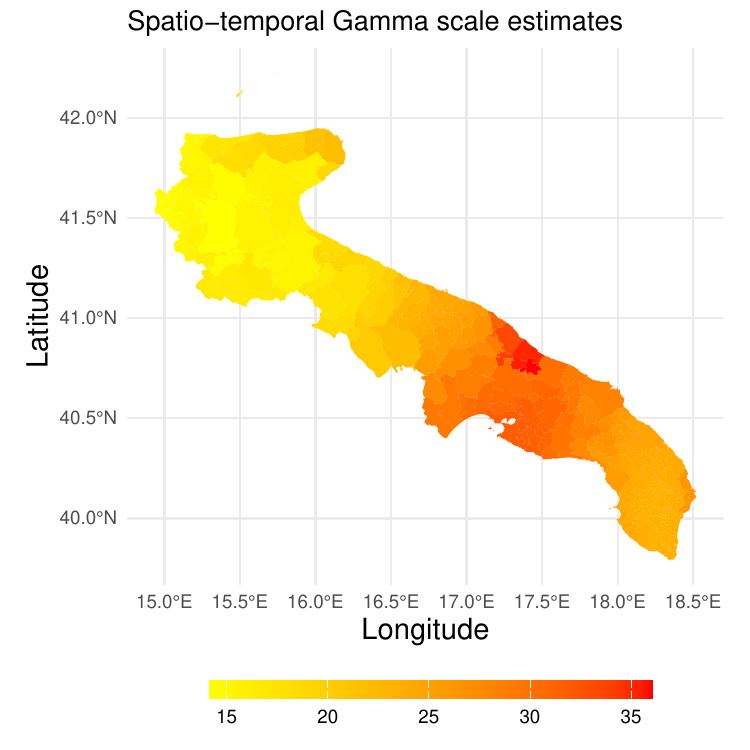}  
	\end{subfigure}
	\begin{subfigure}{.5\textwidth}
		\centering
		\includegraphics[width=1\linewidth, height=0.25\textheight]{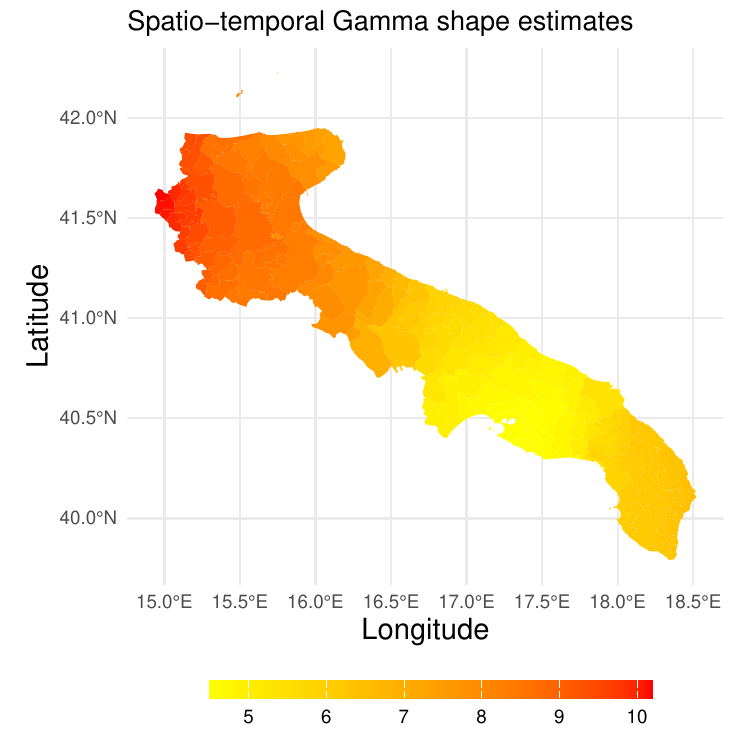}  
	\end{subfigure}
    \newline
       \begin{subfigure}{.5\textwidth}
		\centering
		\includegraphics[width=1\linewidth, height=0.25\textheight]{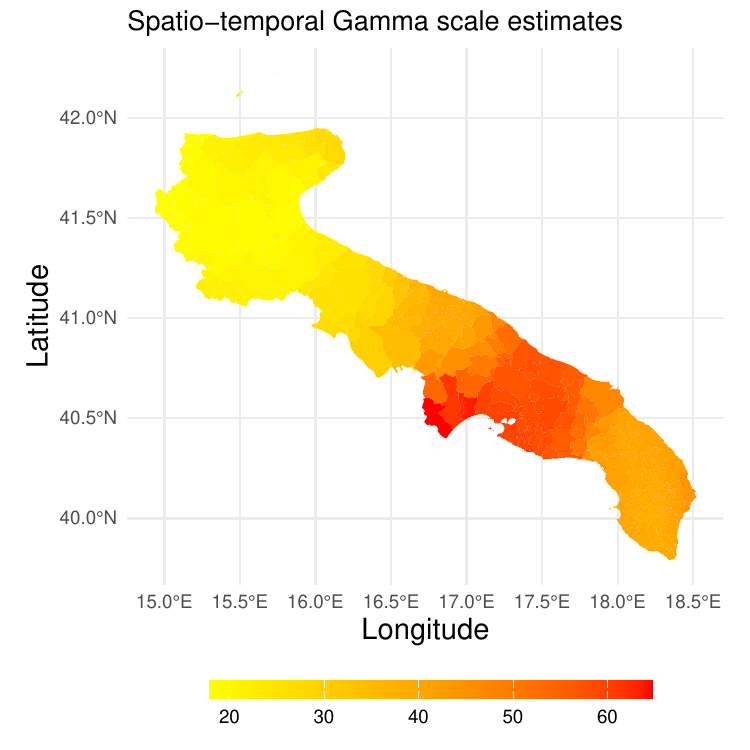}  
	\end{subfigure}
	\begin{subfigure}{.5\textwidth}
		\centering
		\includegraphics[width=1\linewidth, height=0.25\textheight]{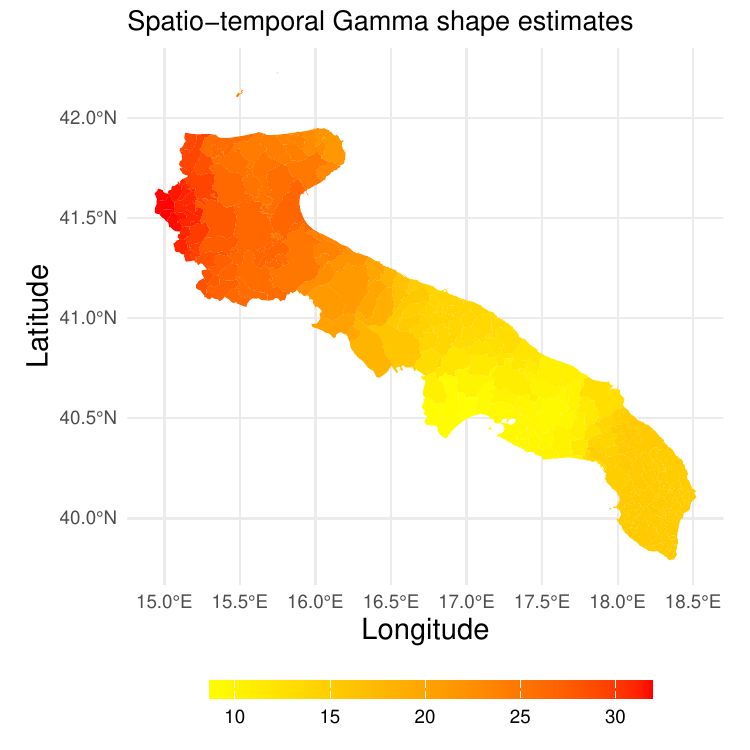}  
	\end{subfigure}
	\caption{Spatio-temporal estimates of fitted Gamma model for $m=1, 3, 12$ accumulation period (top to bottom rows), respectively.}
	\label{fig:pars-est-sup}
\end{figure}

\begin{figure}[H]
	\begin{subfigure}{.32\textwidth}
		\centering
		\includegraphics[width=1\linewidth, height=0.18\textheight]{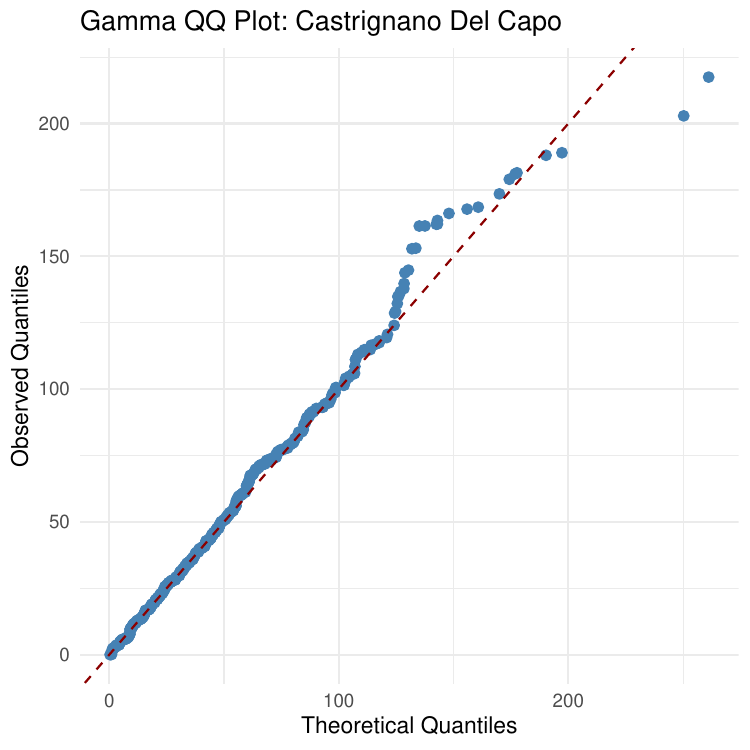}  
	\end{subfigure}
	\begin{subfigure}{.32\textwidth}
		\centering
		\includegraphics[width=1\linewidth, height=0.18\textheight]{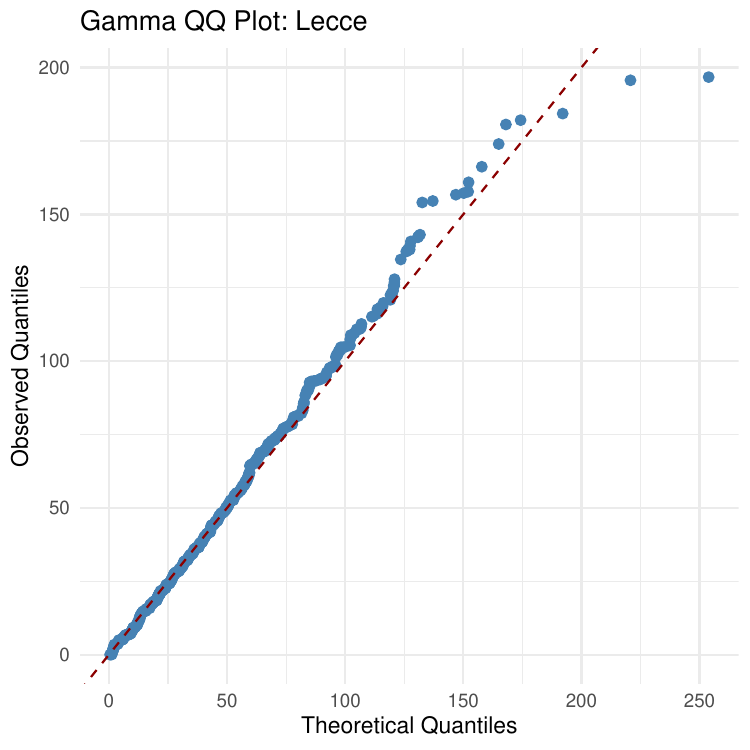}  
	\end{subfigure}
    \begin{subfigure}{.32\textwidth}
		\centering
		\includegraphics[width=1\linewidth, height=0.18\textheight]{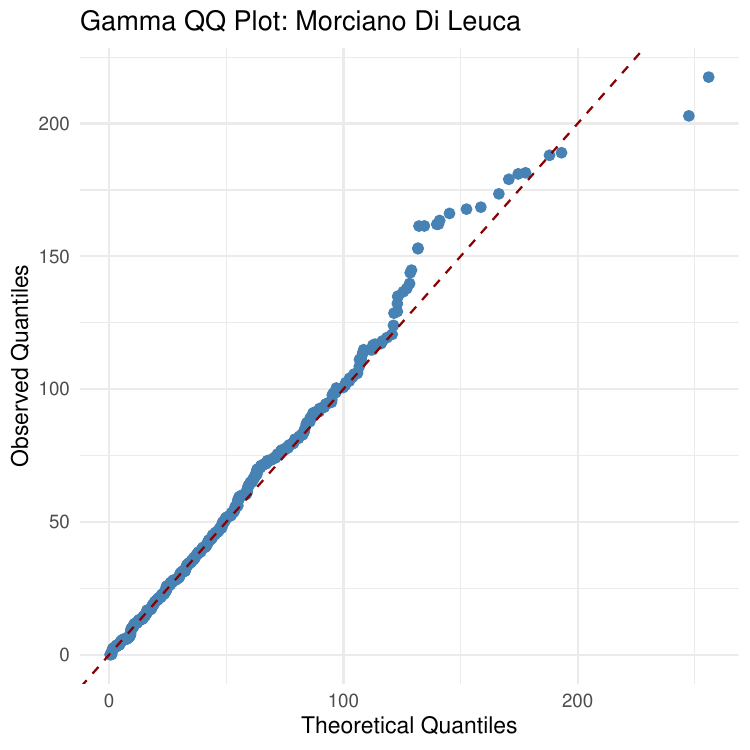}  
	\end{subfigure}
    \begin{subfigure}{.32\textwidth}
		\centering
		\includegraphics[width=1\linewidth, height=0.18\textheight]{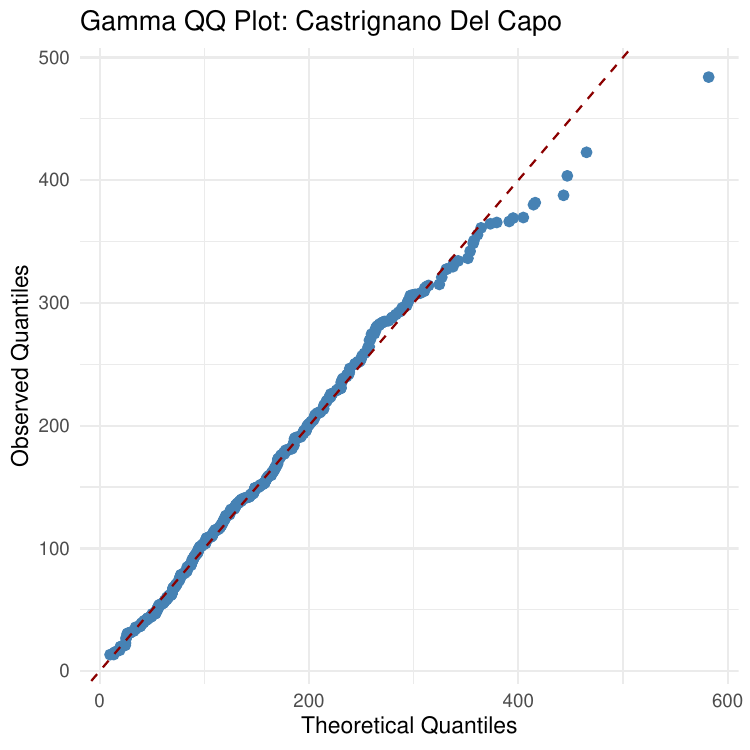}  
	\end{subfigure}
	\begin{subfigure}{.32\textwidth}
		\centering
		\includegraphics[width=1\linewidth, height=0.18\textheight]{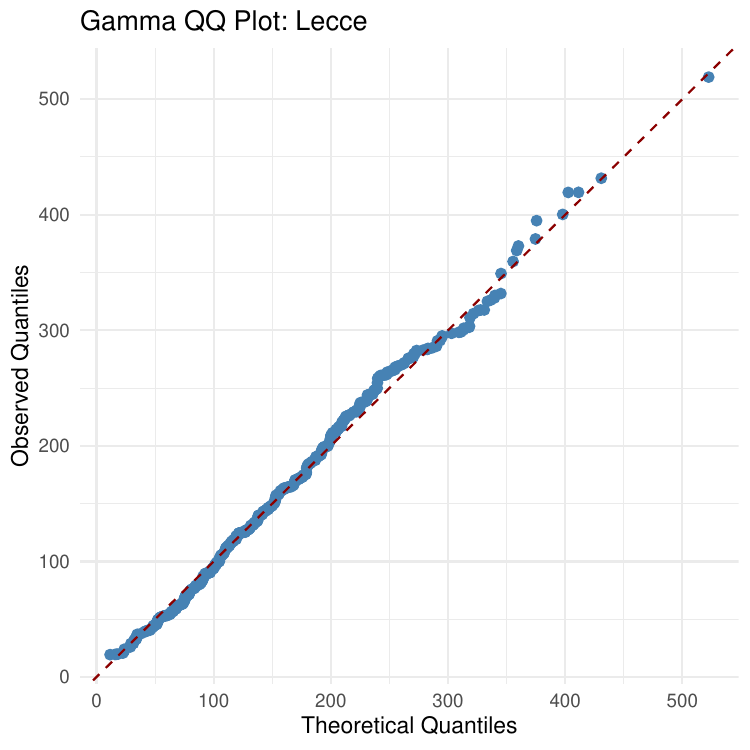}  
	\end{subfigure}
    \begin{subfigure}{.32\textwidth}
		\centering
		\includegraphics[width=1\linewidth, height=0.18\textheight]{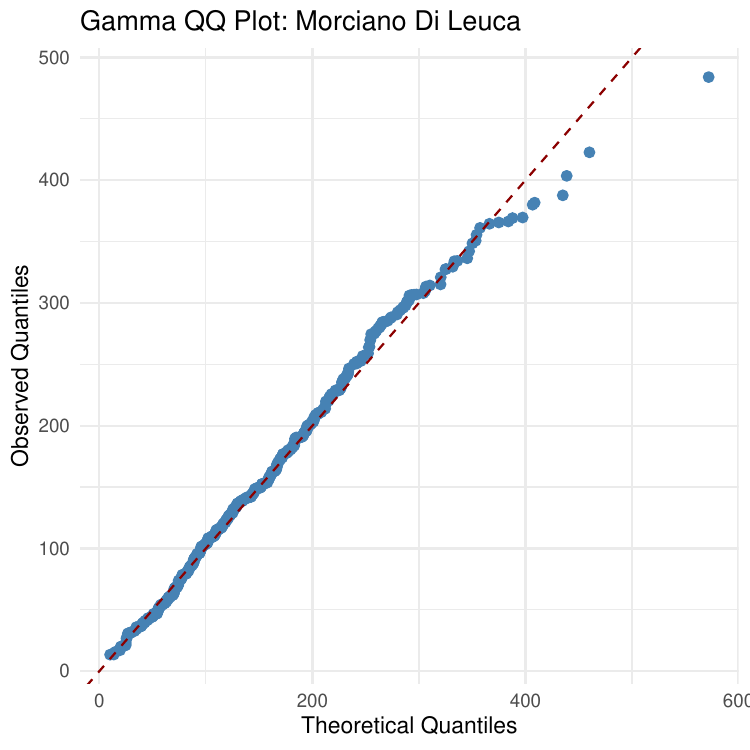}  
	\end{subfigure}
    \begin{subfigure}{.32\textwidth}
		\centering
		\includegraphics[width=1\linewidth, height=0.18\textheight]{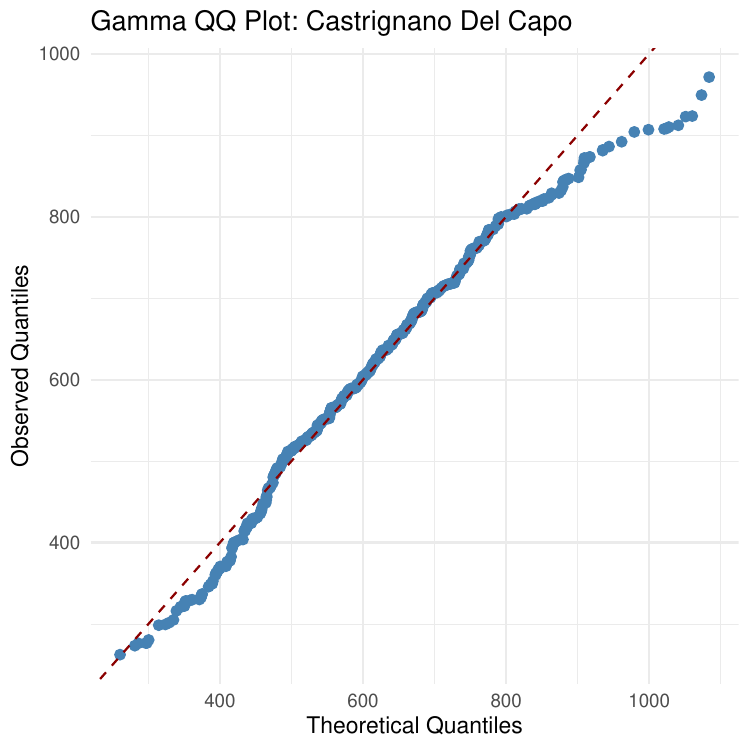}  
	\end{subfigure}
	\begin{subfigure}{.32\textwidth}
		\centering
		\includegraphics[width=1\linewidth, height=0.18\textheight]{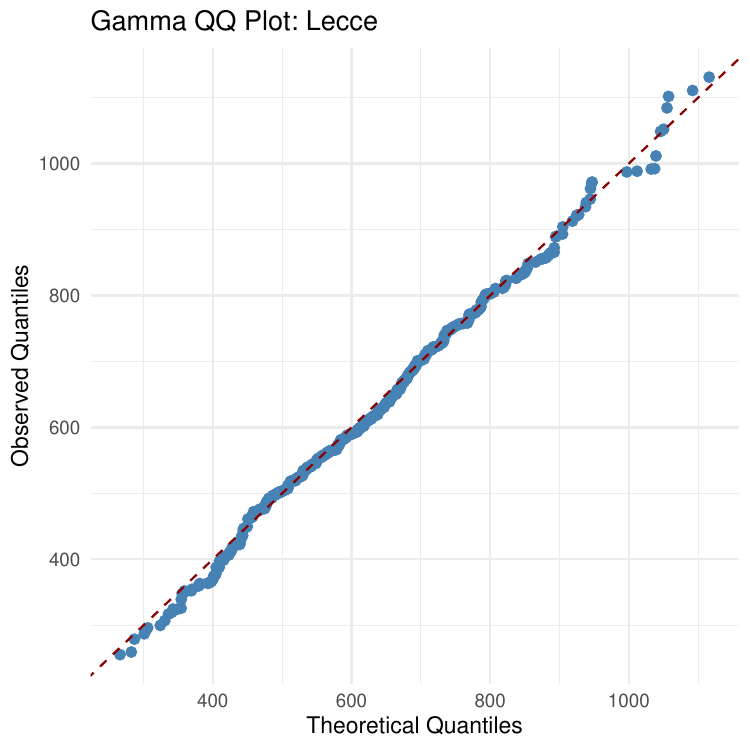}  
	\end{subfigure}
    \begin{subfigure}{.32\textwidth}
		\centering
		\includegraphics[width=1\linewidth, height=0.18\textheight]{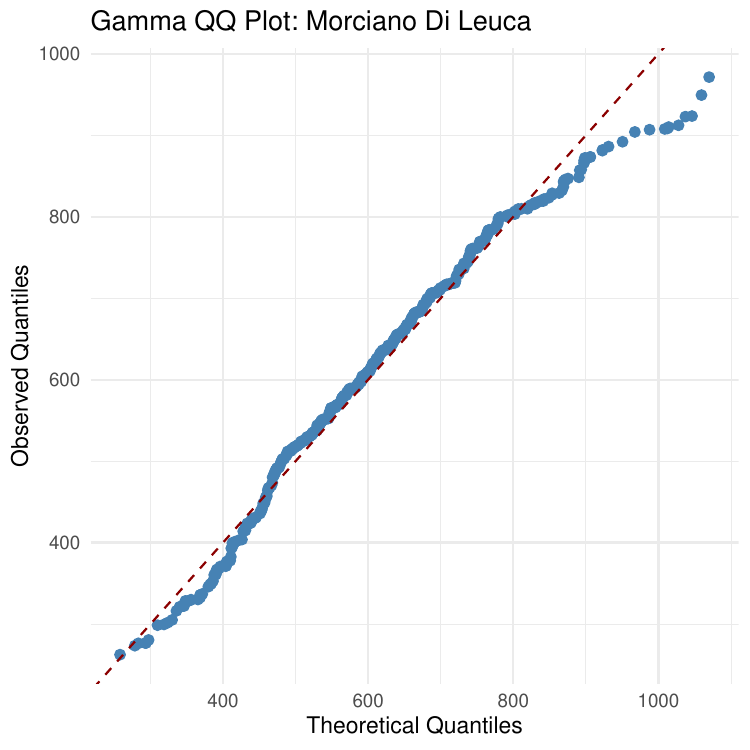}  
	\end{subfigure}
	\caption{QQ plots of the fitted Gamma model for accumulation periods \( m = 1 \) (top), \( m = 3 \) (middle), and \( m = 12 \) (bottom) at validation locations.}
	\label{fig:qq-plot-sup}
\end{figure}

\begin{figure}[H]
	\centering
\includegraphics[width=16.5cm,height=7cm]{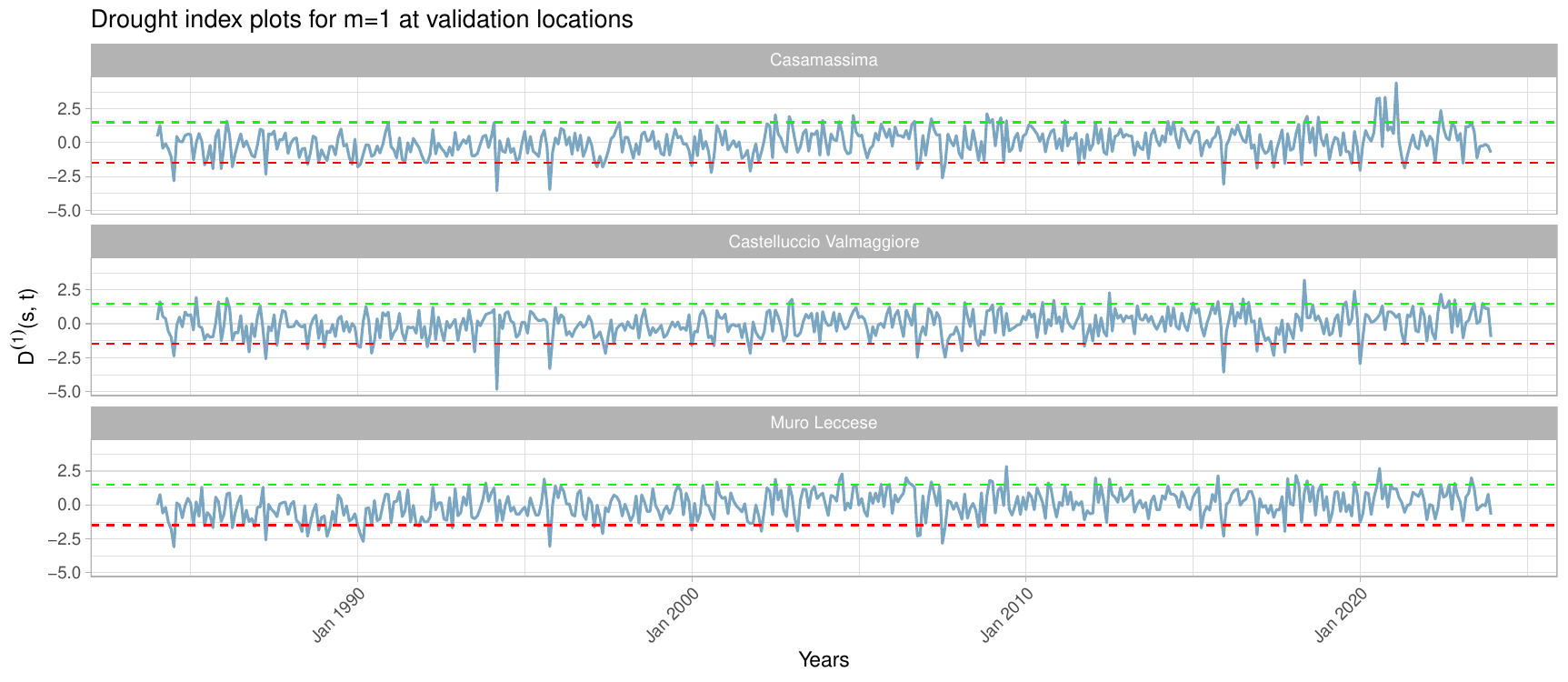}
\includegraphics[width=16.5cm,height=7cm]{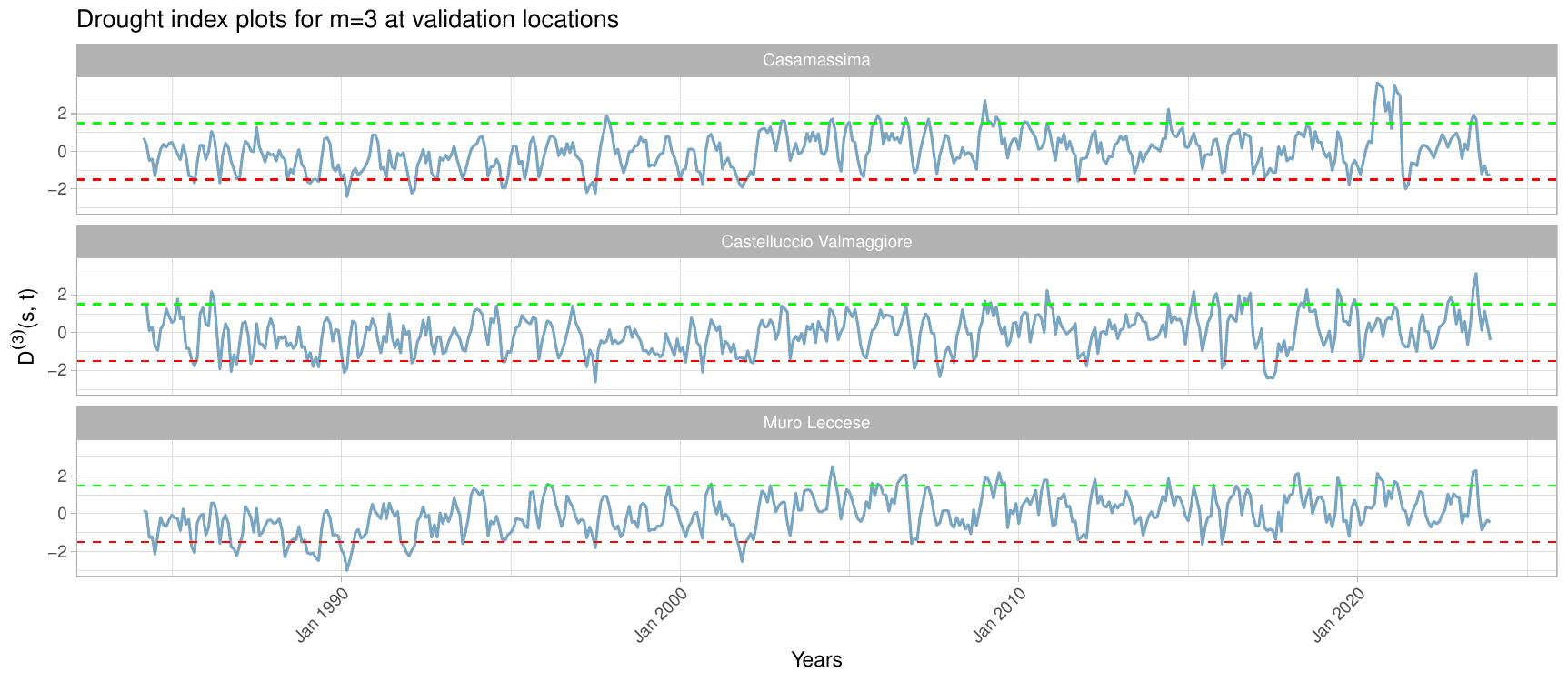}
\includegraphics[width=16.5cm,height=7cm]{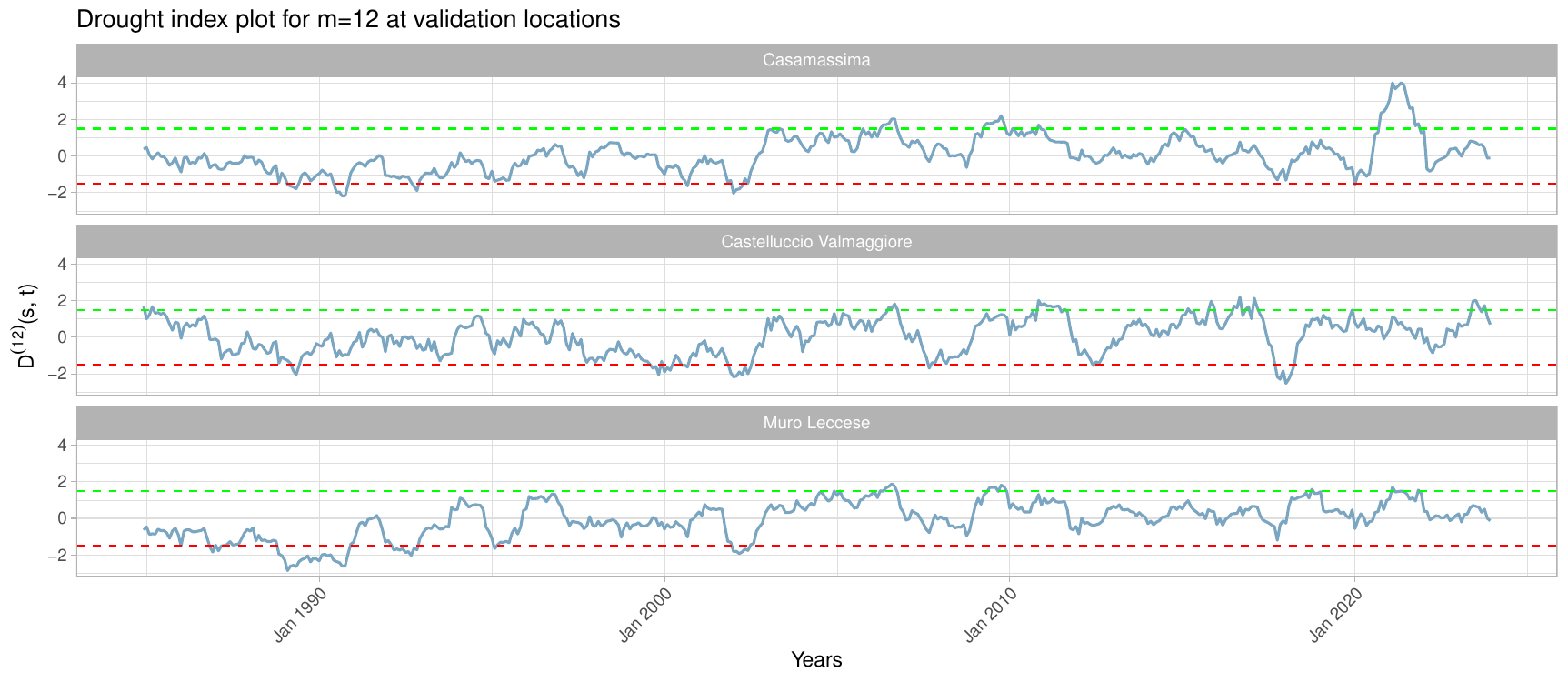}
	\caption{Calculated drought indices at validation locations for $m=1, 3, 12$ accumulation periods.}
	\label{fig:SPI-validate-sup}
\end{figure}

\begin{figure}[t]
	\begin{subfigure}{.32\textwidth}
		\centering
		\includegraphics[width=1\linewidth, height=0.15\textheight]{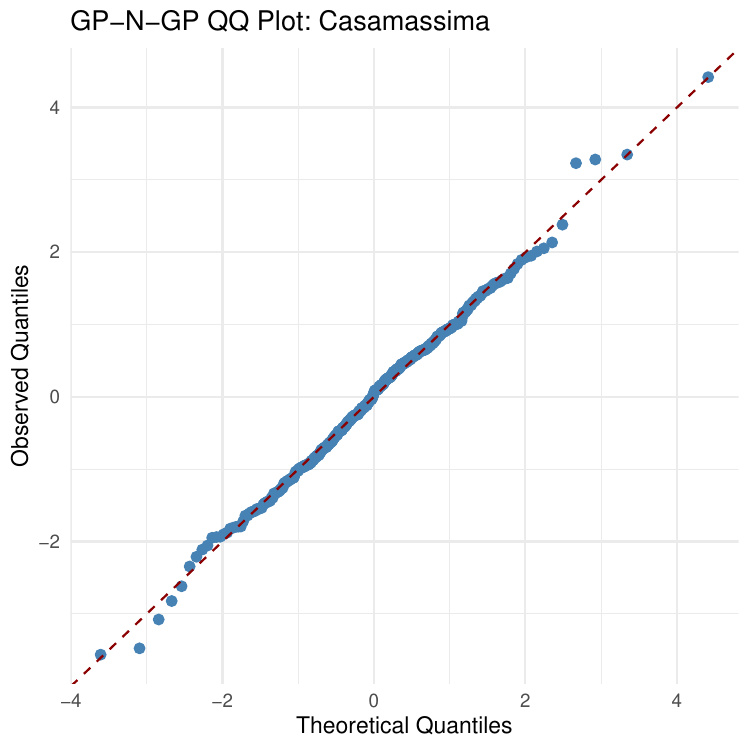}  
	\end{subfigure}
	\begin{subfigure}{.32\textwidth}
		\centering
		\includegraphics[width=1\linewidth, height=0.15\textheight]{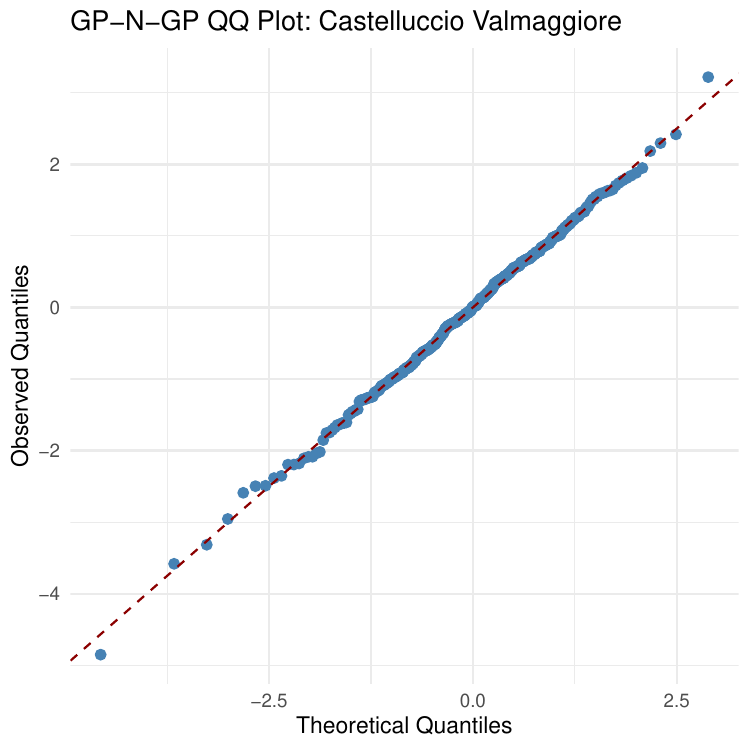}  
	\end{subfigure}
    \begin{subfigure}{.32\textwidth}
		\centering
		\includegraphics[width=1\linewidth, height=0.15\textheight]{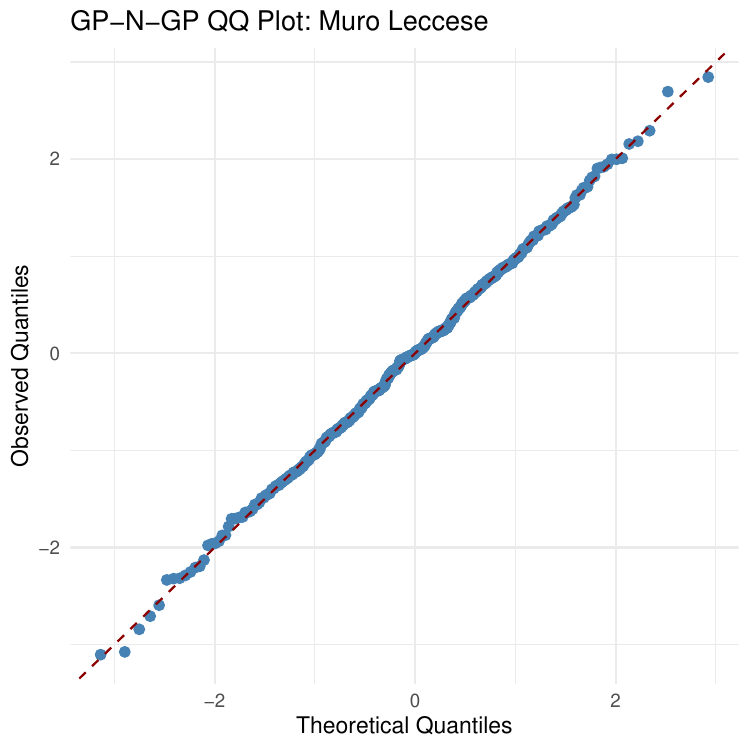}  
	\end{subfigure}
	\newline
    \begin{subfigure}{.32\textwidth}
		\centering
		\includegraphics[width=1\linewidth, height=0.15\textheight]{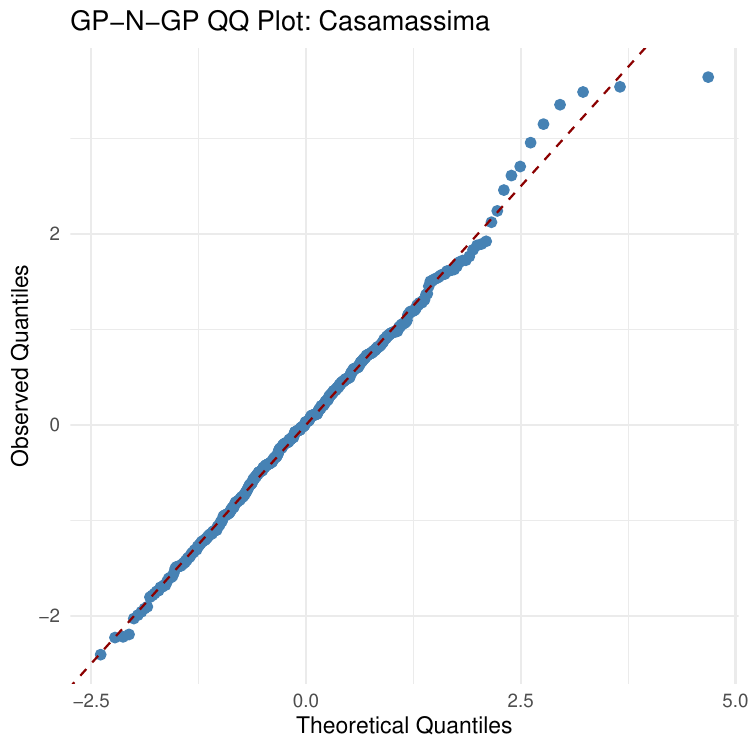}  
	\end{subfigure}
	\begin{subfigure}{.32\textwidth}
		\centering
		\includegraphics[width=1\linewidth, height=0.15\textheight]{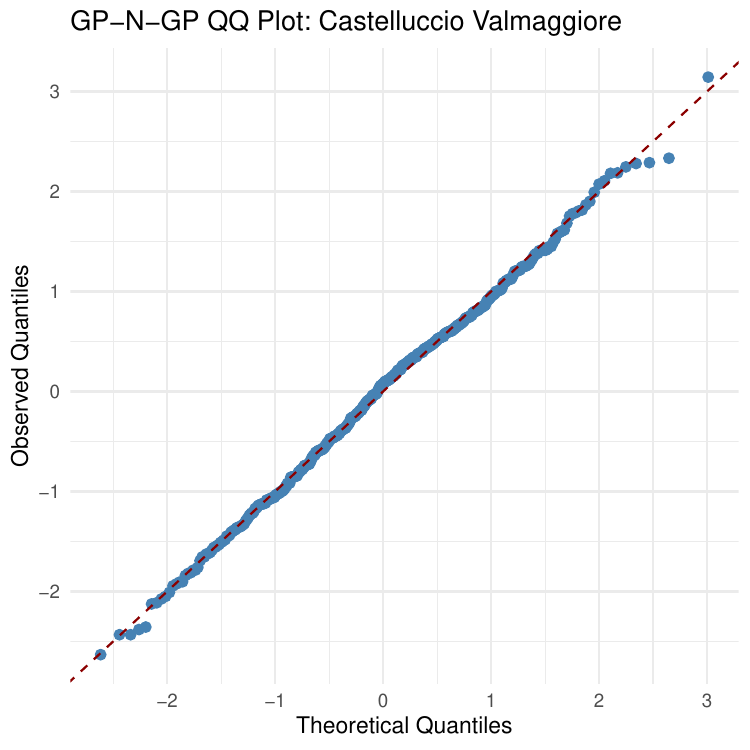}  
	\end{subfigure}
    \begin{subfigure}{.32\textwidth}
		\centering
		\includegraphics[width=1\linewidth, height=0.15\textheight]{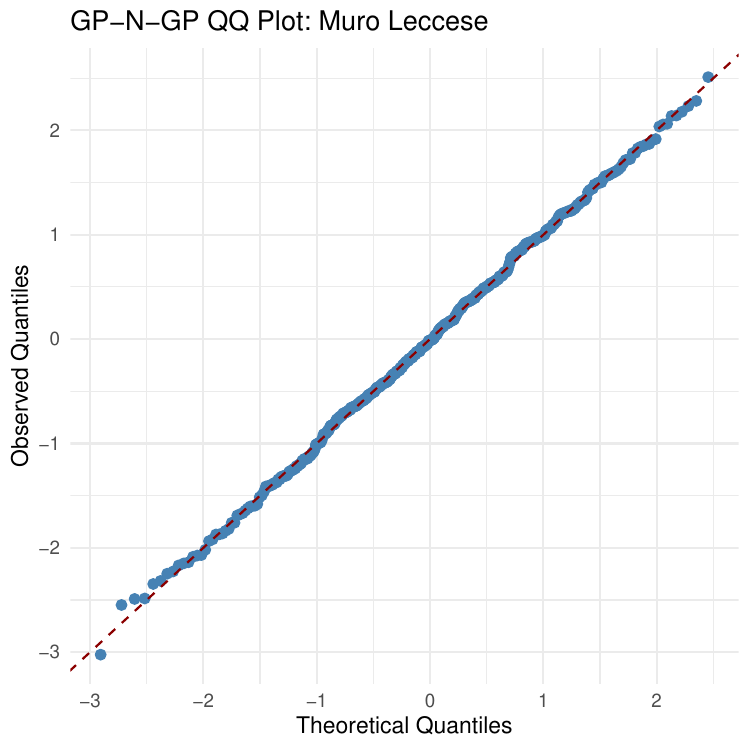}  
	\end{subfigure}
\newline
    \begin{subfigure}{.32\textwidth}
		\centering
		\includegraphics[width=1\linewidth, height=0.15\textheight]{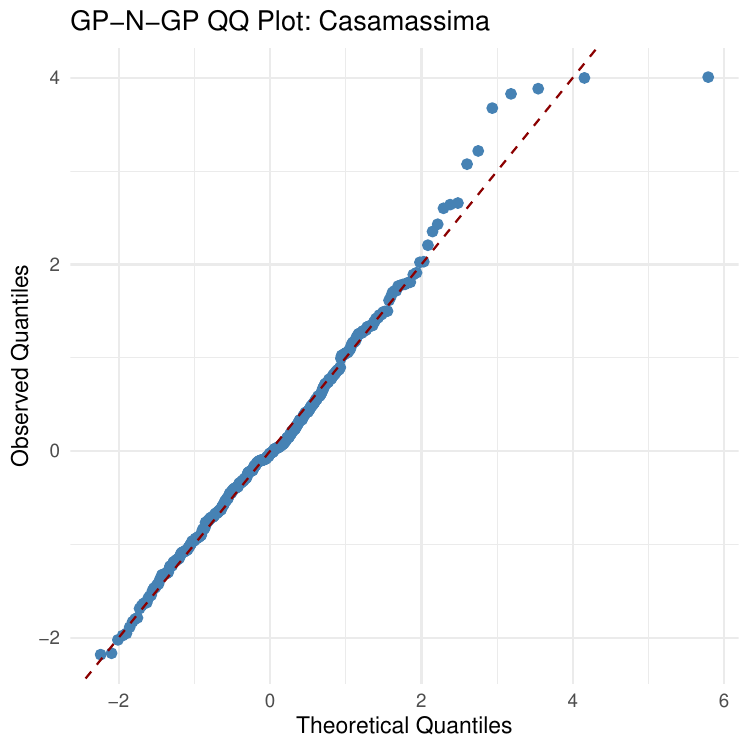}  
	\end{subfigure}
	\begin{subfigure}{.32\textwidth}
		\centering
		\includegraphics[width=1\linewidth, height=0.15\textheight]{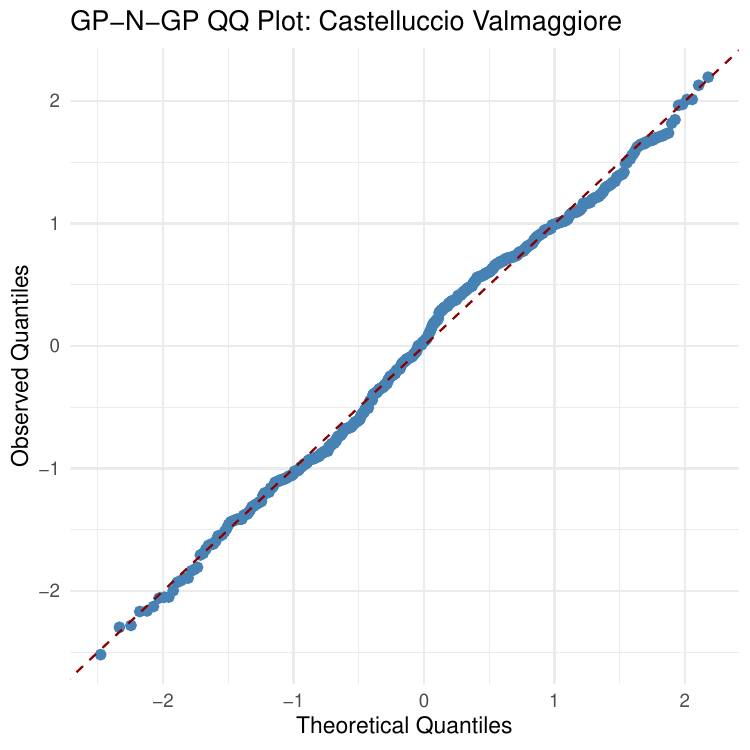}  
	\end{subfigure}
    \begin{subfigure}{.32\textwidth}
		\centering
		\includegraphics[width=1\linewidth, height=0.15\textheight]{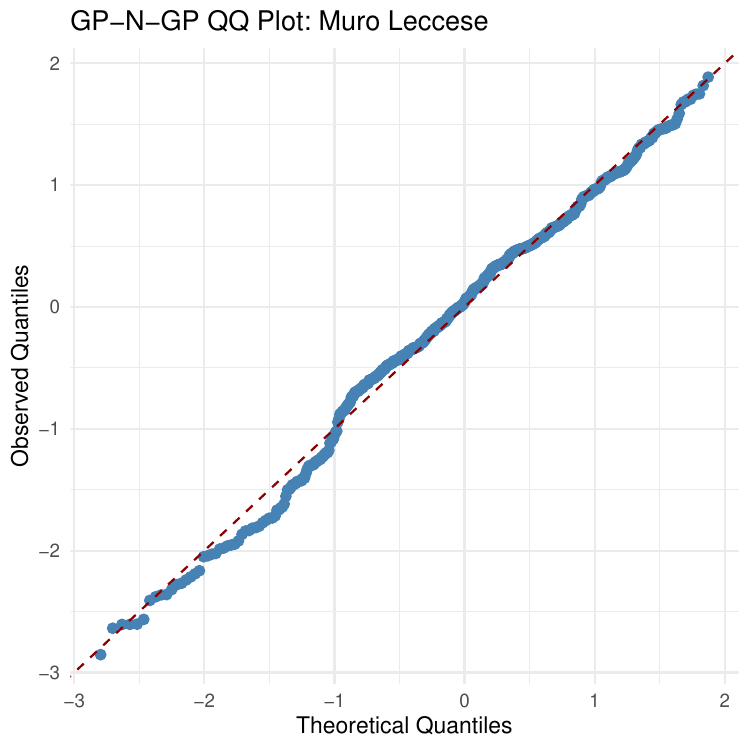}  
	\end{subfigure}
    
    \caption{QQ plots of the fitted GP-N-GP model for $m = 1, 3, 12$ (top to bottom rows) at validation locations. }
	\label{fig:qq-plot-gng-sup}
\end{figure}

\begin{figure}[t]
	\begin{subfigure}{.32\textwidth}
		\centering
		\includegraphics[width=1\linewidth, height=0.15\textheight]{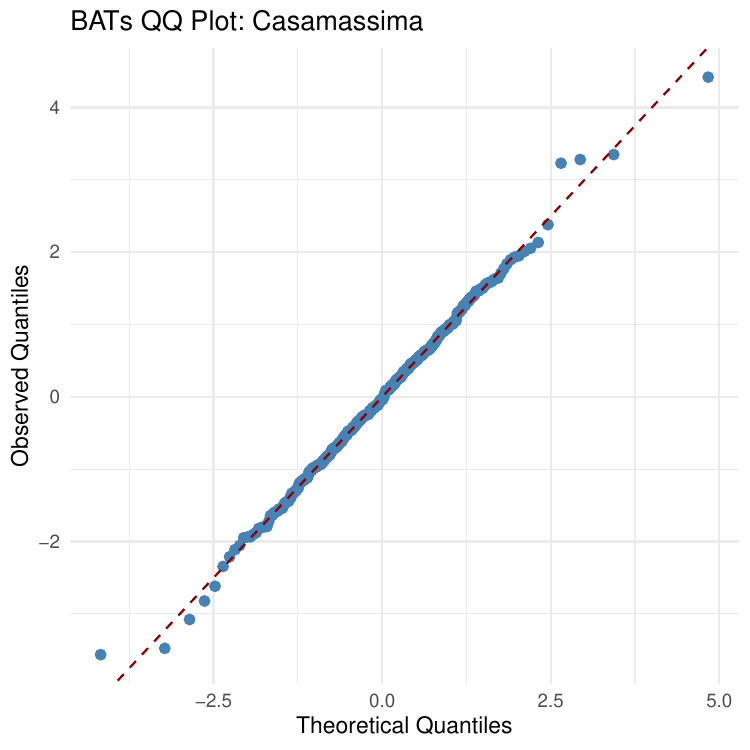}  
	\end{subfigure}
	\begin{subfigure}{.32\textwidth}
		\centering
		\includegraphics[width=1\linewidth, height=0.15\textheight]{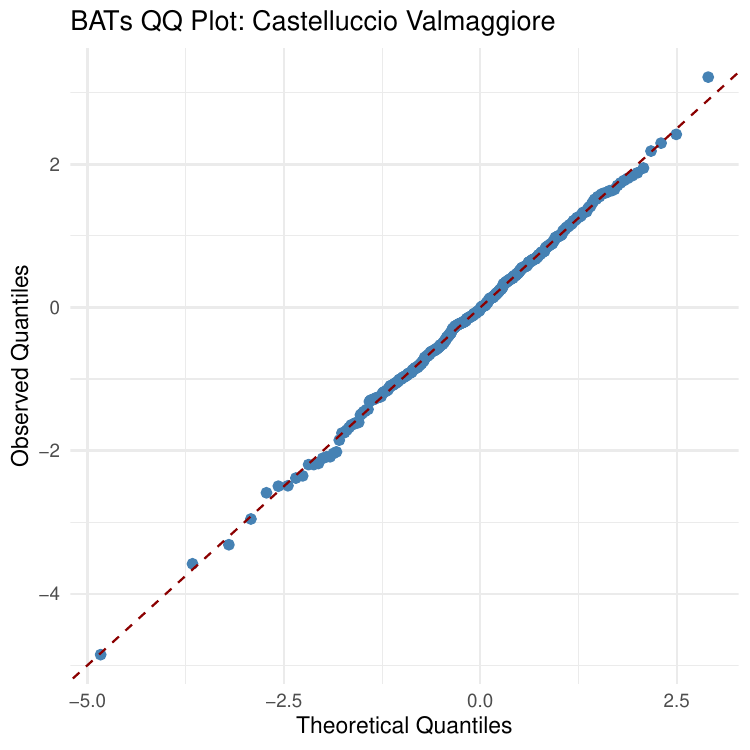}  
	\end{subfigure}
    \begin{subfigure}{.32\textwidth}
		\centering
		\includegraphics[width=1\linewidth, height=0.15\textheight]{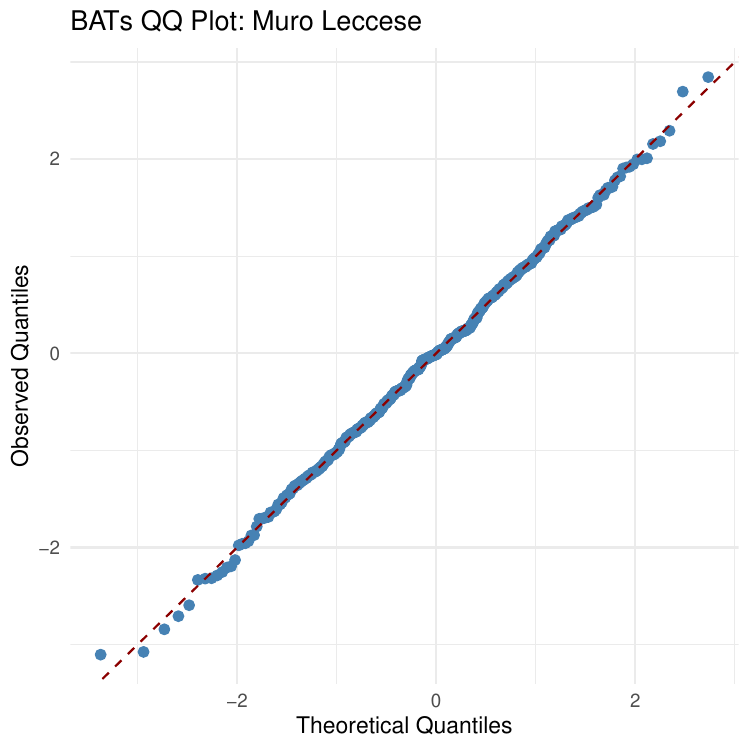}  
	\end{subfigure}
	\newline
    \begin{subfigure}{.32\textwidth}
		\centering
		\includegraphics[width=1\linewidth, height=0.15\textheight]{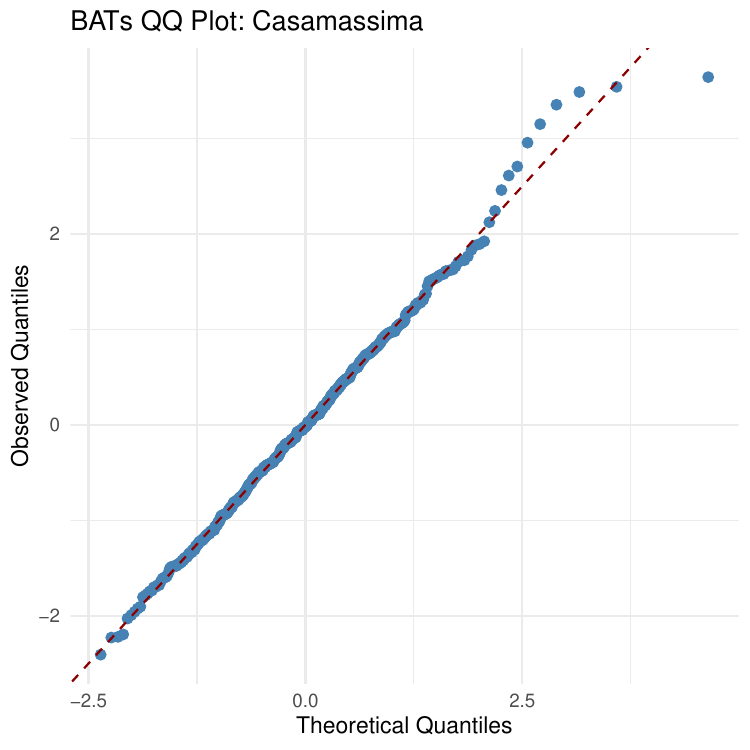}  
	\end{subfigure}
	\begin{subfigure}{.32\textwidth}
		\centering
		\includegraphics[width=1\linewidth, height=0.15\textheight]{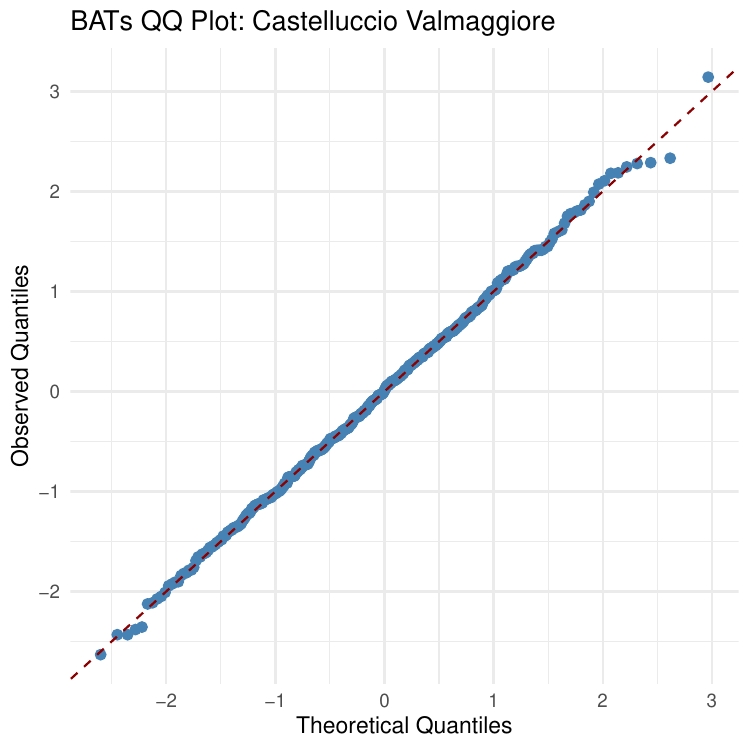}  
	\end{subfigure}
    \begin{subfigure}{.32\textwidth}
		\centering
		\includegraphics[width=1\linewidth, height=0.15\textheight]{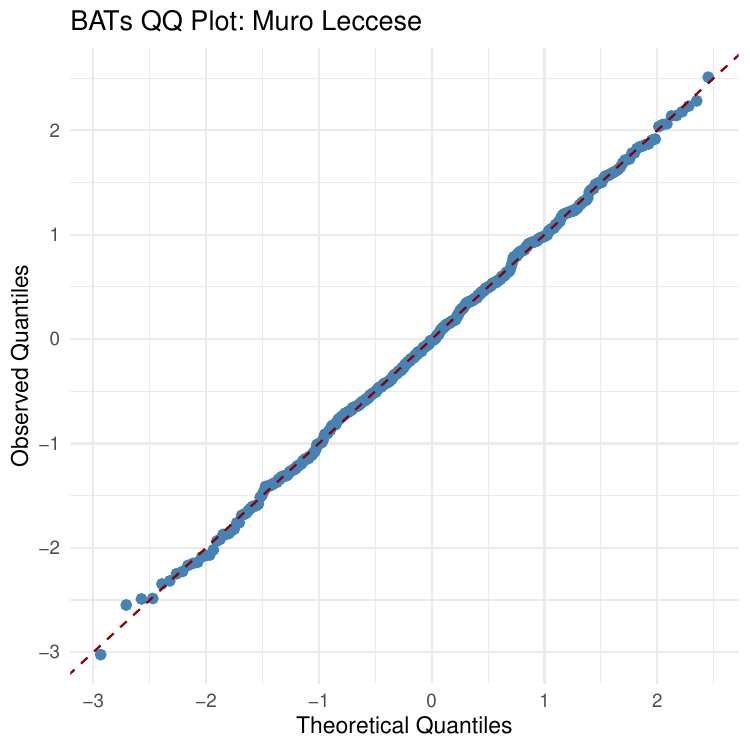}  
	\end{subfigure}
\newline
    \begin{subfigure}{.32\textwidth}
		\centering
		\includegraphics[width=1\linewidth, height=0.15\textheight]{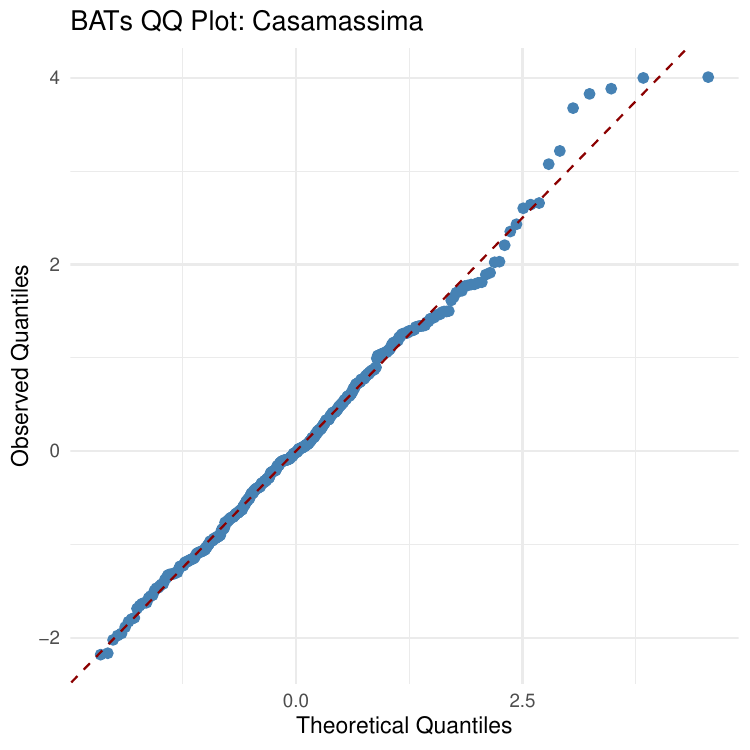}  
	\end{subfigure}
	\begin{subfigure}{.32\textwidth}
		\centering
		\includegraphics[width=1\linewidth, height=0.15\textheight]{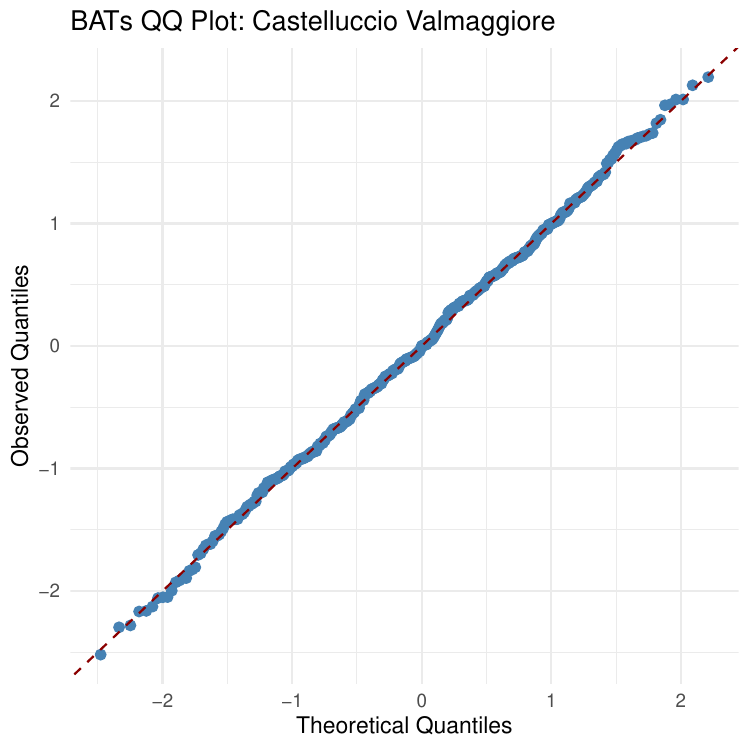}  
	\end{subfigure}
    \begin{subfigure}{.32\textwidth}
		\centering
		\includegraphics[width=1\linewidth, height=0.15\textheight]{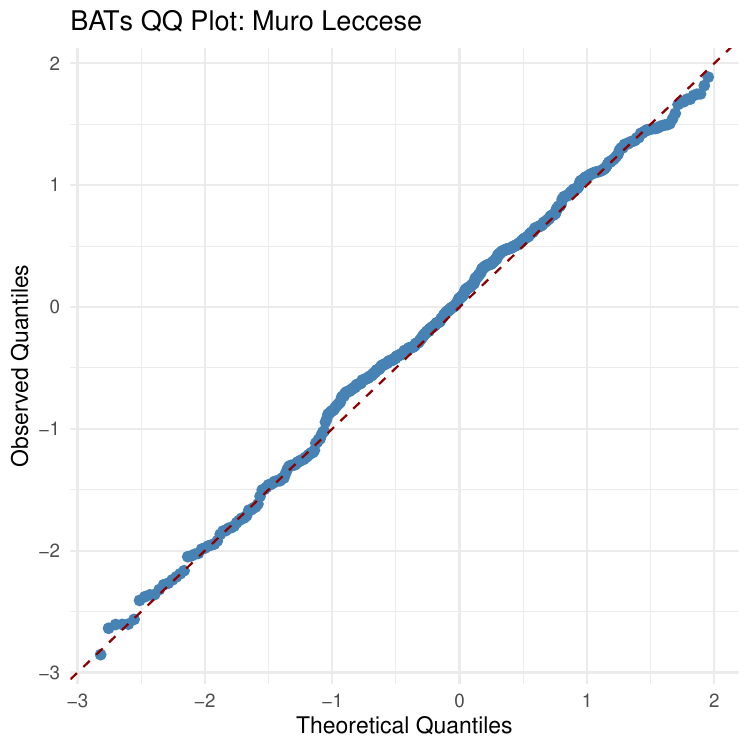}  
	\end{subfigure}
    
    \caption{QQ plots of the fitted BATs model for $m = 1, 3, 12$ (top to bottom rows) at validation locations.  }
	\label{fig:qq-plot-bat-sup}
\end{figure}

\begin{figure}[t]
	\begin{subfigure}{.5\textwidth}
		\centering
        \includegraphics[width=1\linewidth, height=0.3\textheight]{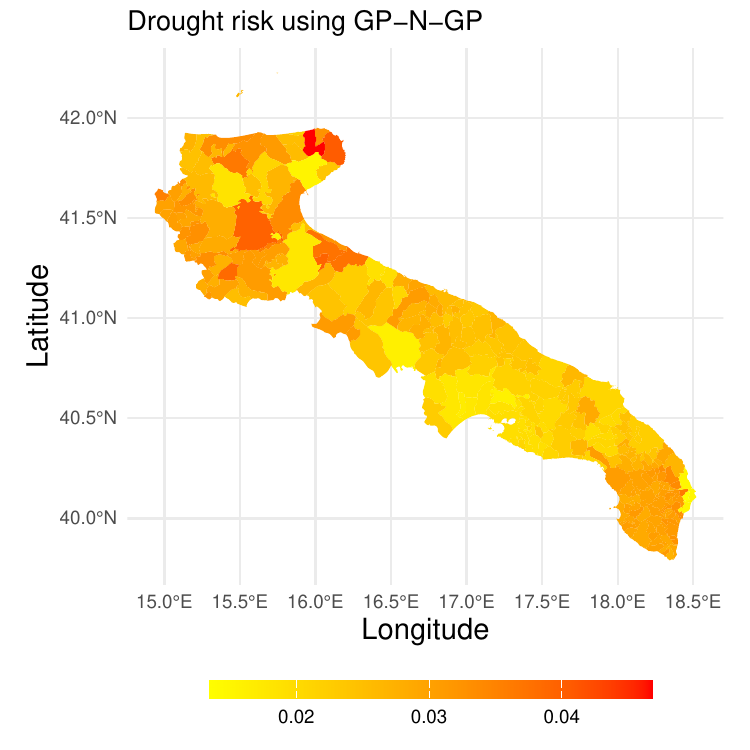}  
	\end{subfigure}
	\begin{subfigure}{.5\textwidth}
		\centering
		\includegraphics[width=1\linewidth, height=0.3\textheight]{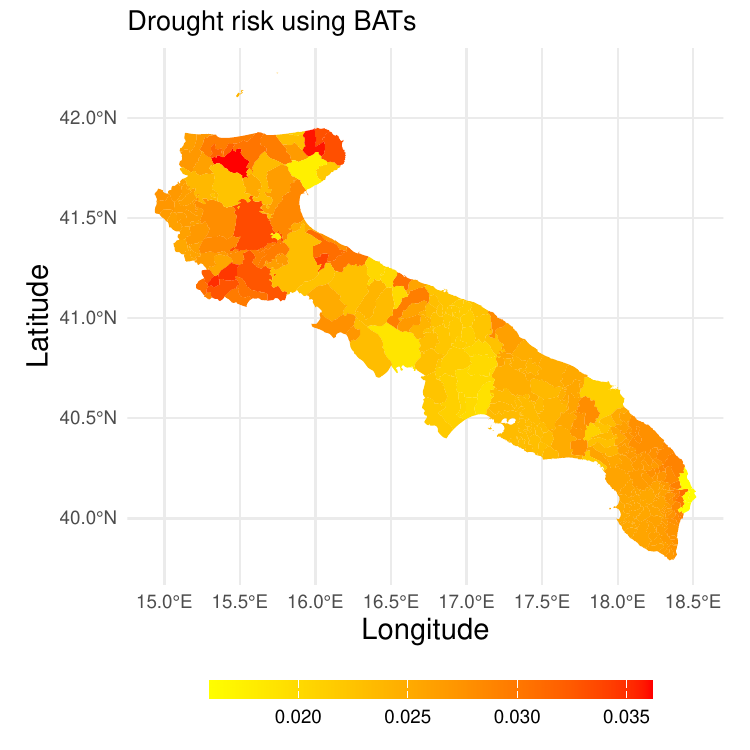}  
	\end{subfigure}
        \newline
       \begin{subfigure}{.5\textwidth}
		\centering
		\includegraphics[width=1\linewidth, height=0.3\textheight]{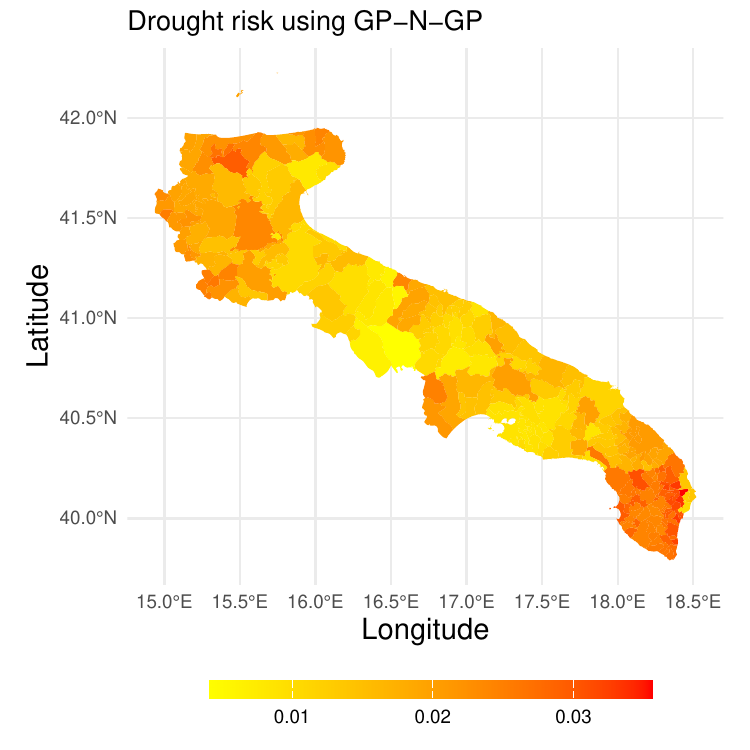}  
	\end{subfigure}
	\begin{subfigure}{.5\textwidth}
		\centering
		\includegraphics[width=1\linewidth, height=0.3\textheight]{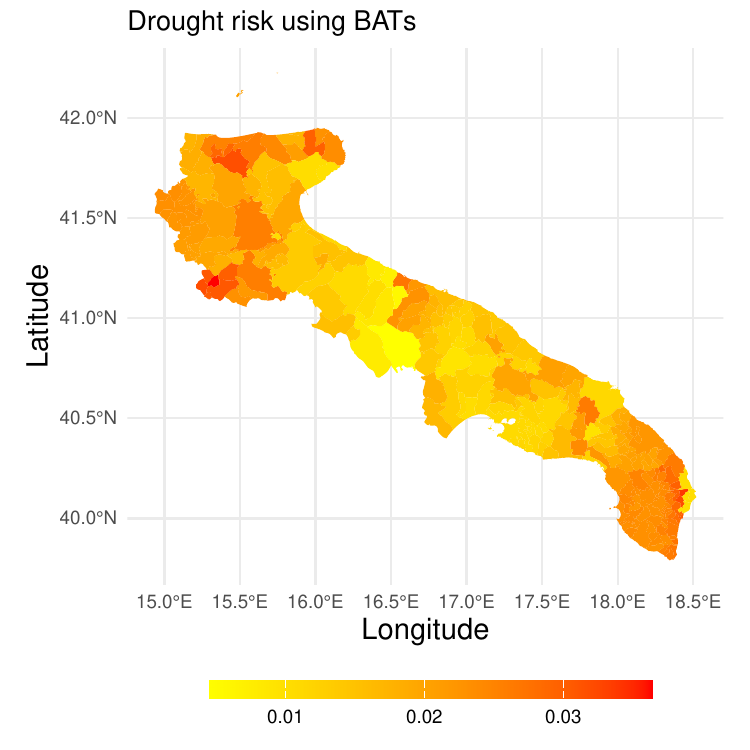}  
	\end{subfigure}
    \newline
       \begin{subfigure}{.5\textwidth}
		\centering
		\includegraphics[width=1\linewidth, height=0.3\textheight]{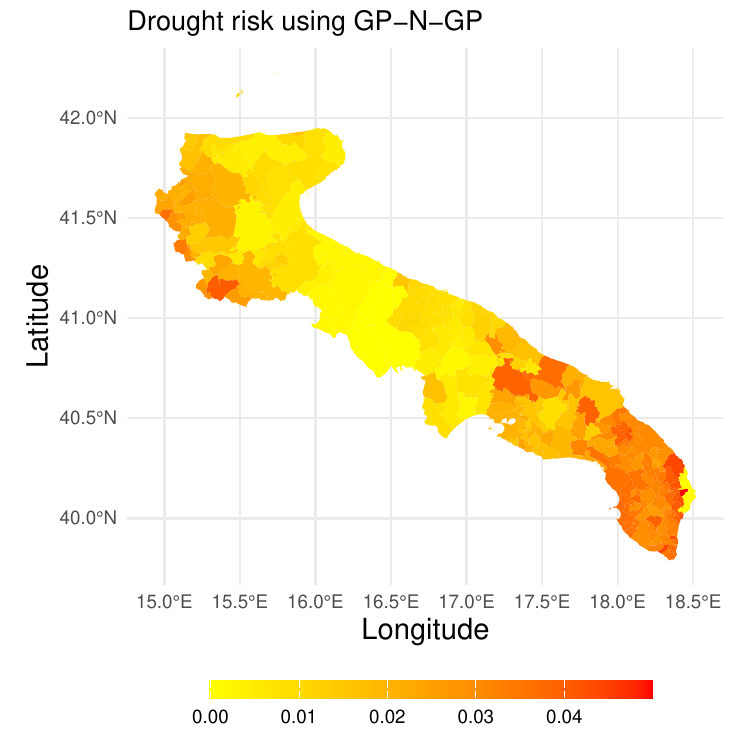}  
	\end{subfigure}
	\begin{subfigure}{.5\textwidth}
		\centering
		\includegraphics[width=1\linewidth, height=0.3\textheight]{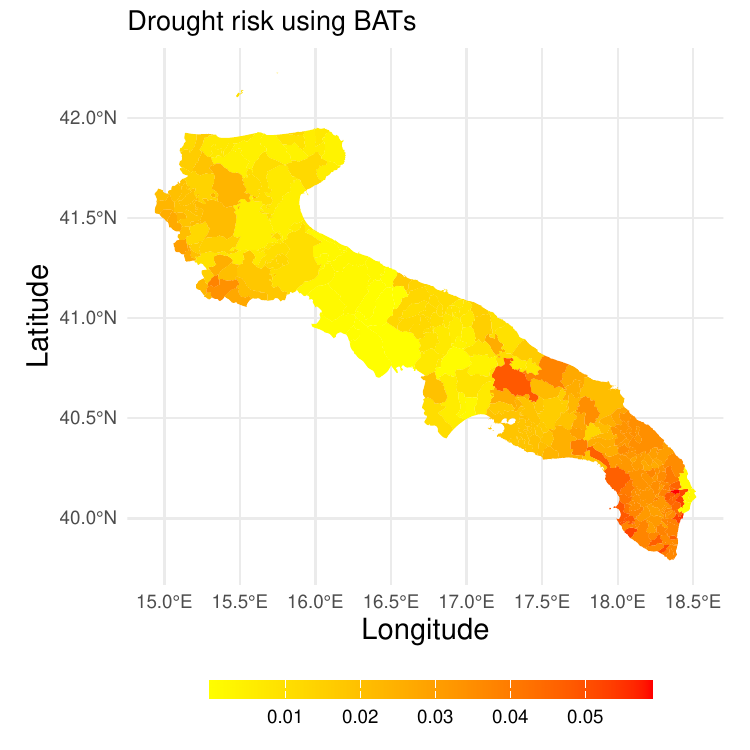}  
	\end{subfigure}
	\caption{Estimated drought risk for $m=1, 3, 12$ for GP-N-GP (left column) and BATs (right column), respectively. }
	\label{fig:drought-risk-sup}
\end{figure}

\begin{figure}[t]
	\begin{subfigure}{.32\textwidth}
		\centering
		\includegraphics[width=1\linewidth, height=0.15\textheight]{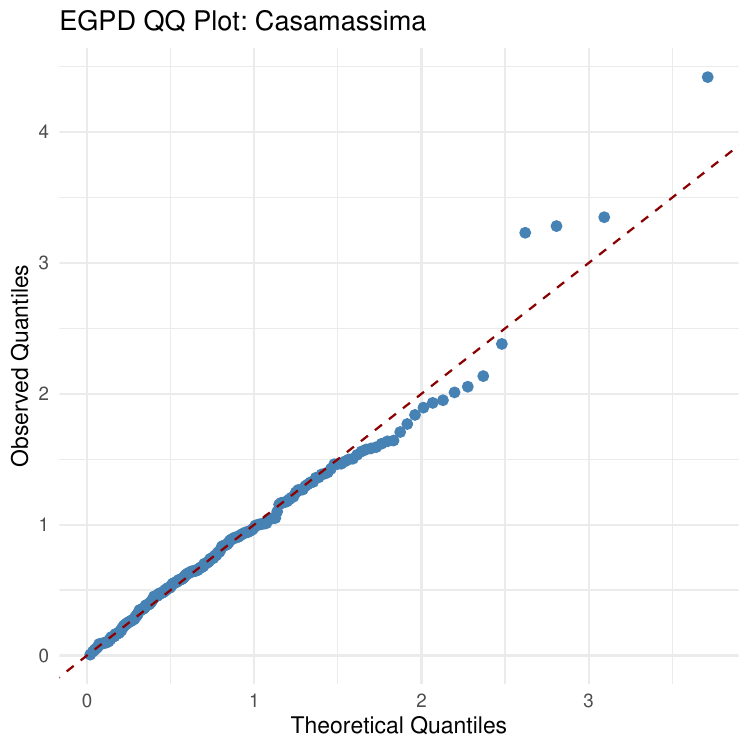}  
	\end{subfigure}
	\begin{subfigure}{.32\textwidth}
		\centering
		\includegraphics[width=1\linewidth, height=0.15\textheight]{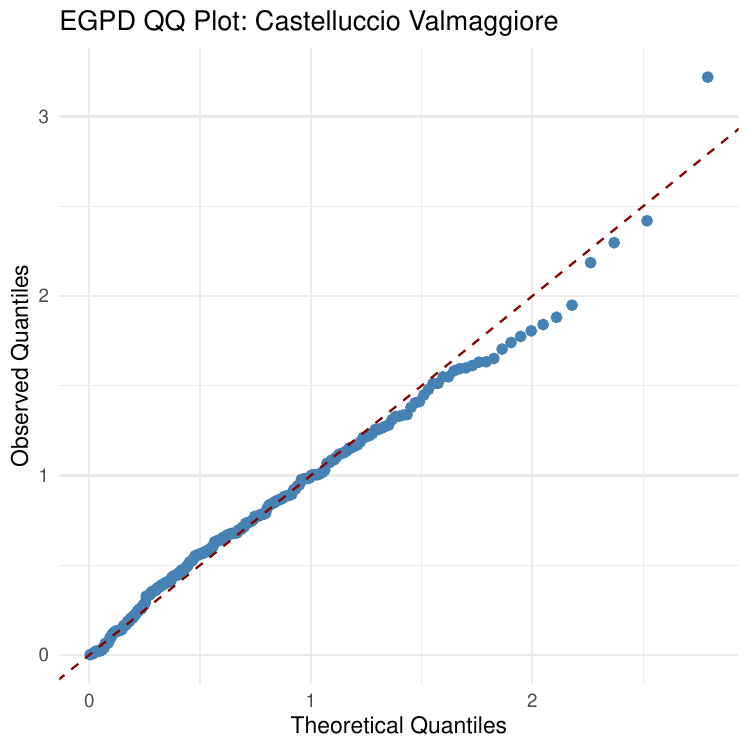}  
	\end{subfigure}
    \begin{subfigure}{.32\textwidth}
		\centering
		\includegraphics[width=1\linewidth, height=0.15\textheight]{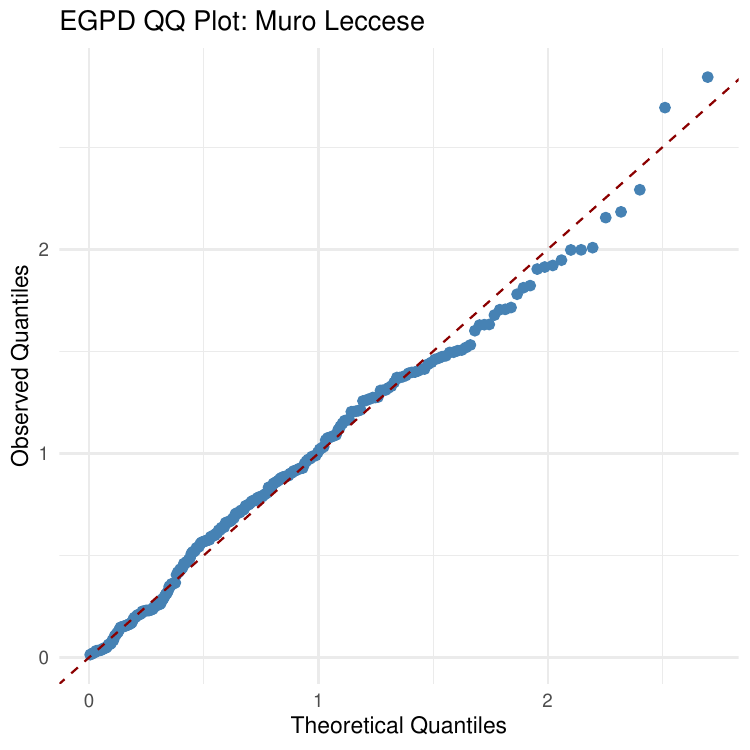}  
	\end{subfigure}
	\newline
    \begin{subfigure}{.32\textwidth}
		\centering
		\includegraphics[width=1\linewidth, height=0.15\textheight]{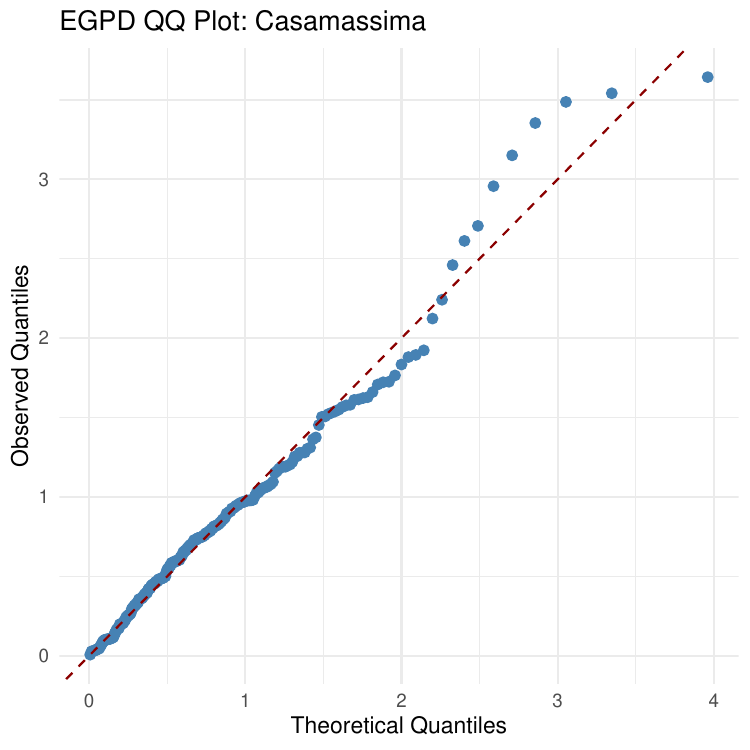}  
	\end{subfigure}
	\begin{subfigure}{.32\textwidth}
		\centering
		\includegraphics[width=1\linewidth, height=0.15\textheight]{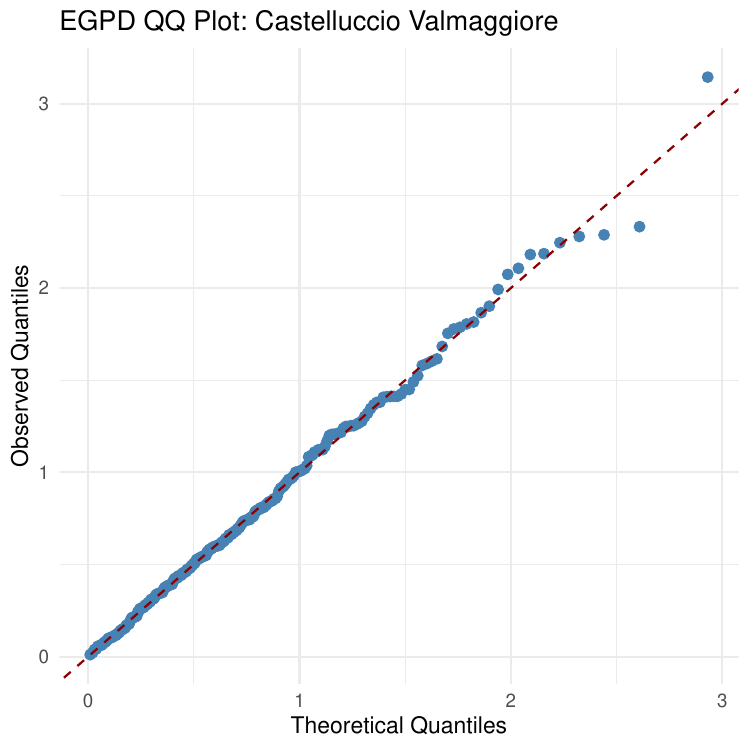}  
	\end{subfigure}
    \begin{subfigure}{.32\textwidth}
		\centering
		\includegraphics[width=1\linewidth, height=0.15\textheight]{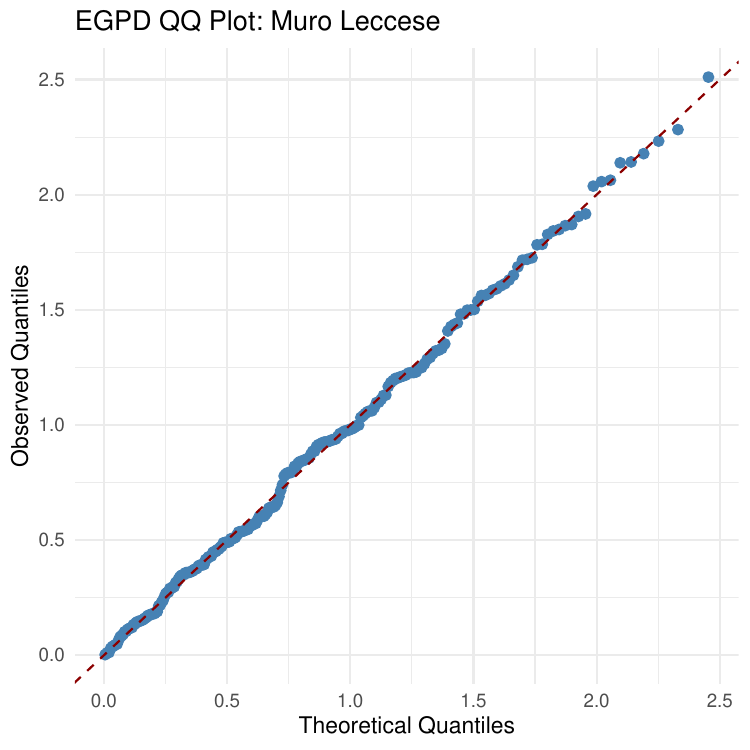}  
	\end{subfigure}
    \newline
    \begin{subfigure}{.32\textwidth}
		\centering
		\includegraphics[width=1\linewidth, height=0.15\textheight]{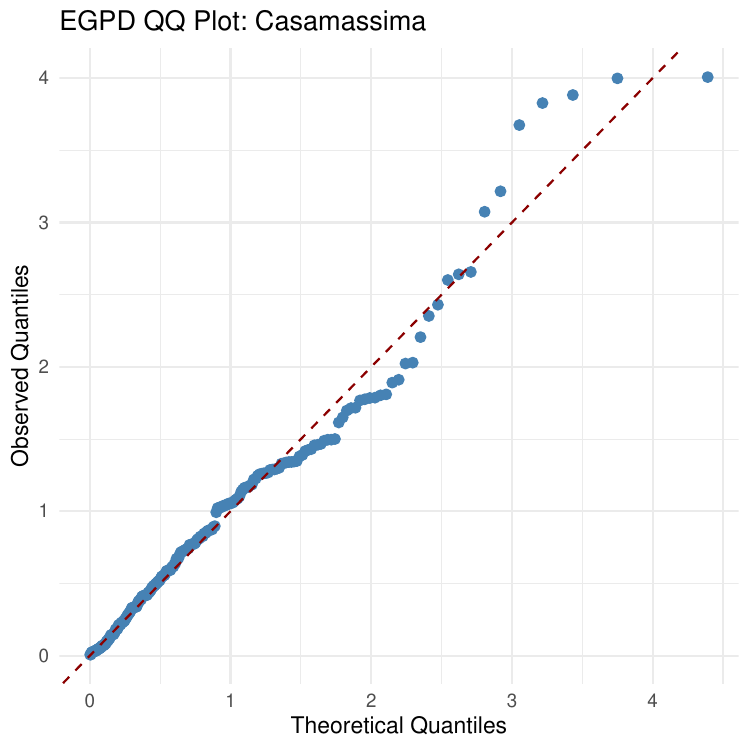}  
	\end{subfigure}
	\begin{subfigure}{.32\textwidth}
		\centering
		\includegraphics[width=1\linewidth, height=0.15\textheight]{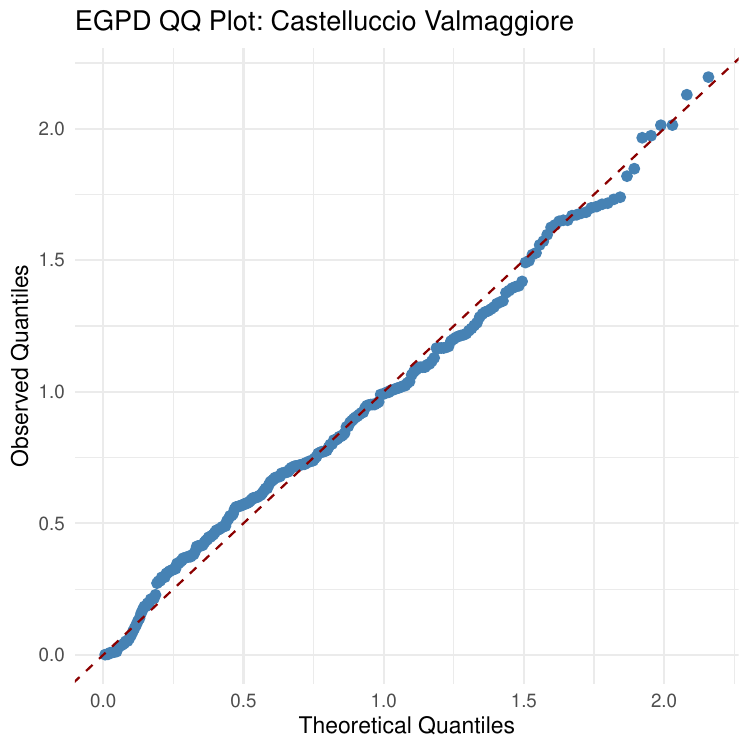}  
	\end{subfigure}
    \begin{subfigure}{.32\textwidth}
		\centering
		\includegraphics[width=1\linewidth, height=0.15\textheight]{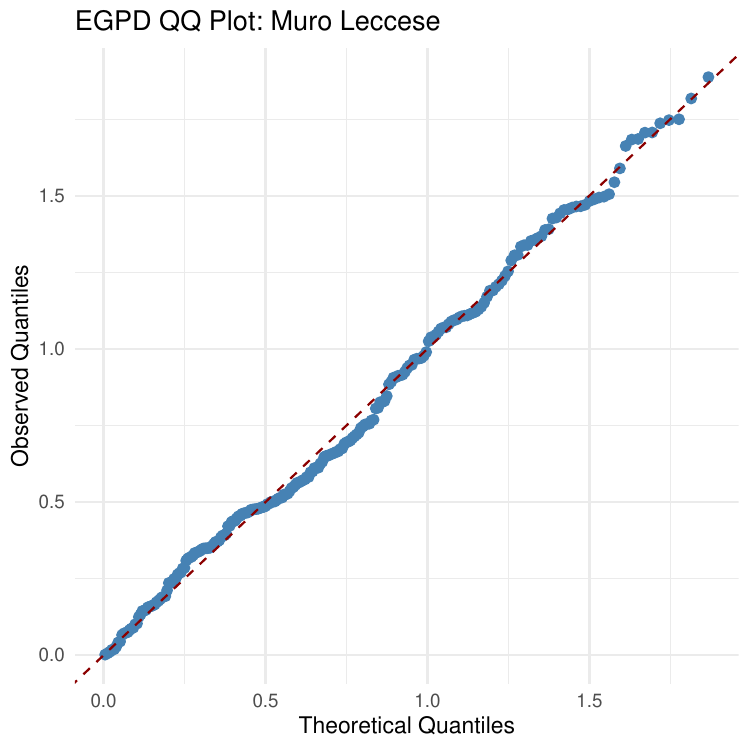}  
	\end{subfigure}
    
    \caption{QQ plots of fitted EGPD model to wet periods at validation locations for $m=1, 3, 12$ (top to bottom rows), respectively. }
	\label{fig:qqplot-egpd-pos}
\end{figure}

\begin{figure}[t]
	\begin{subfigure}{.32\textwidth}
		\centering
		\includegraphics[width=1\linewidth, height=0.15\textheight]{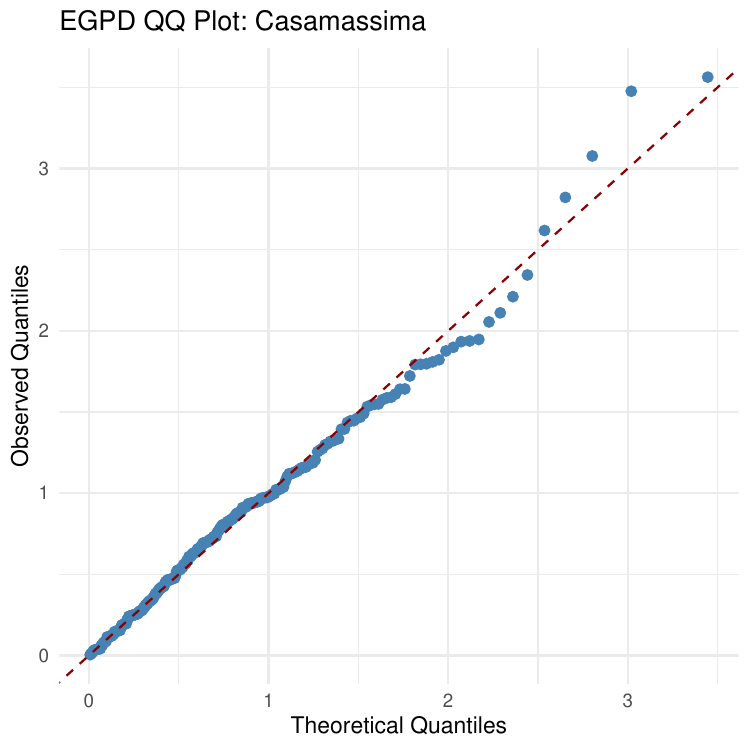}  
	\end{subfigure}
	\begin{subfigure}{.32\textwidth}
		\centering
		\includegraphics[width=1\linewidth, height=0.15\textheight]{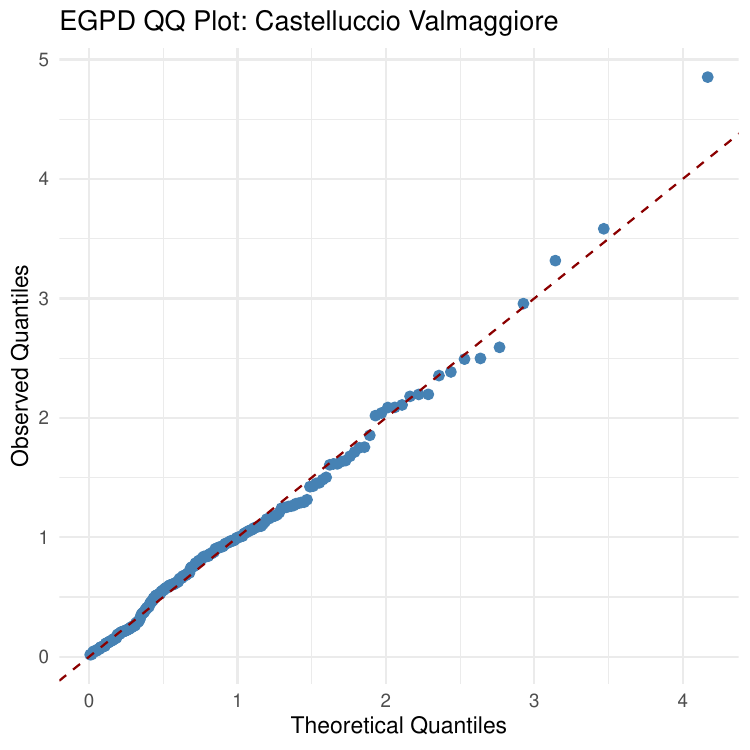}  
	\end{subfigure}
    \begin{subfigure}{.32\textwidth}
		\centering
		\includegraphics[width=1\linewidth, height=0.15\textheight]{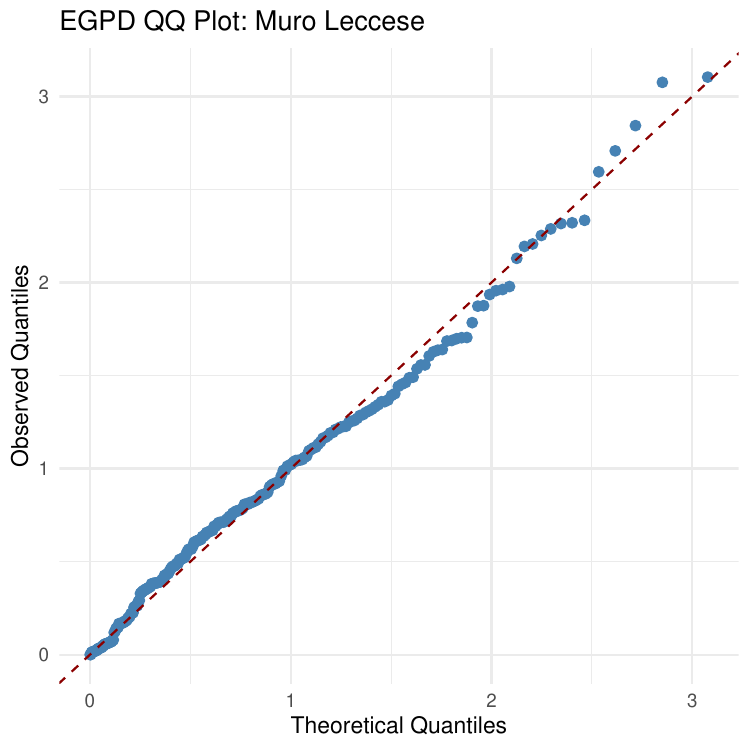}  
	\end{subfigure}
	\newline
    \begin{subfigure}{.32\textwidth}
		\centering
		\includegraphics[width=1\linewidth, height=0.15\textheight]{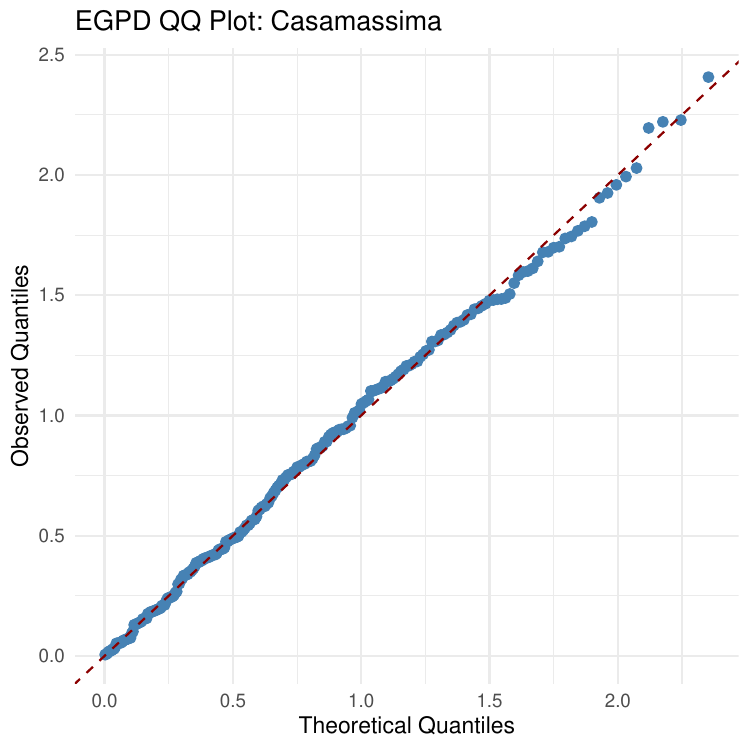}  
	\end{subfigure}
	\begin{subfigure}{.32\textwidth}
		\centering
		\includegraphics[width=1\linewidth, height=0.15\textheight]{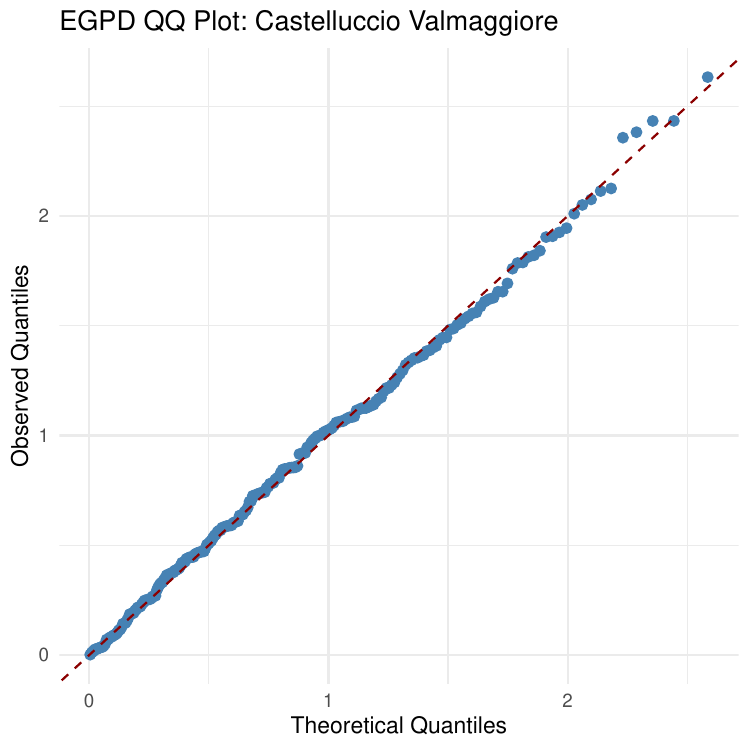}  
	\end{subfigure}
    \begin{subfigure}{.32\textwidth}
		\centering
		\includegraphics[width=1\linewidth, height=0.15\textheight]{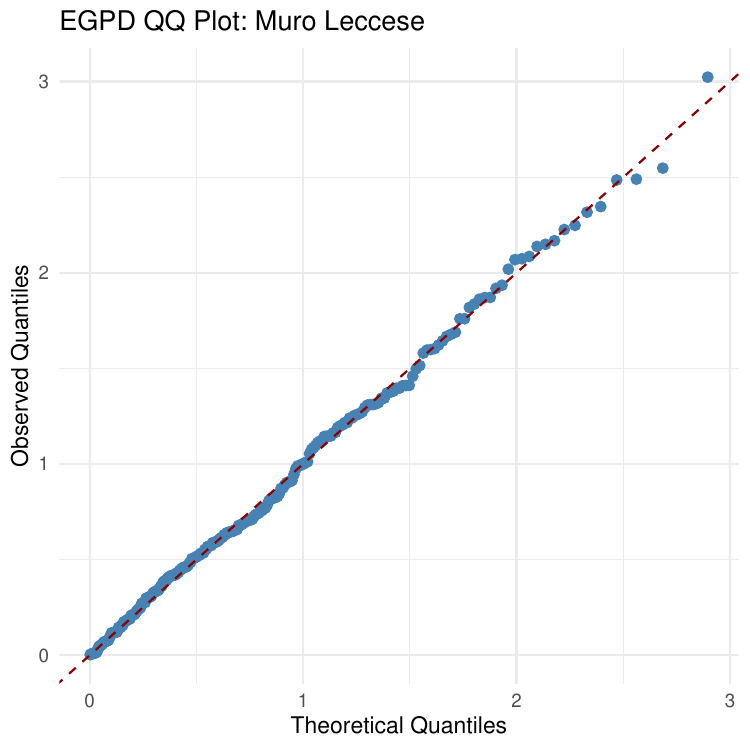}  
	\end{subfigure}
    \newline
    \begin{subfigure}{.32\textwidth}
		\centering
		\includegraphics[width=1\linewidth, height=0.15\textheight]{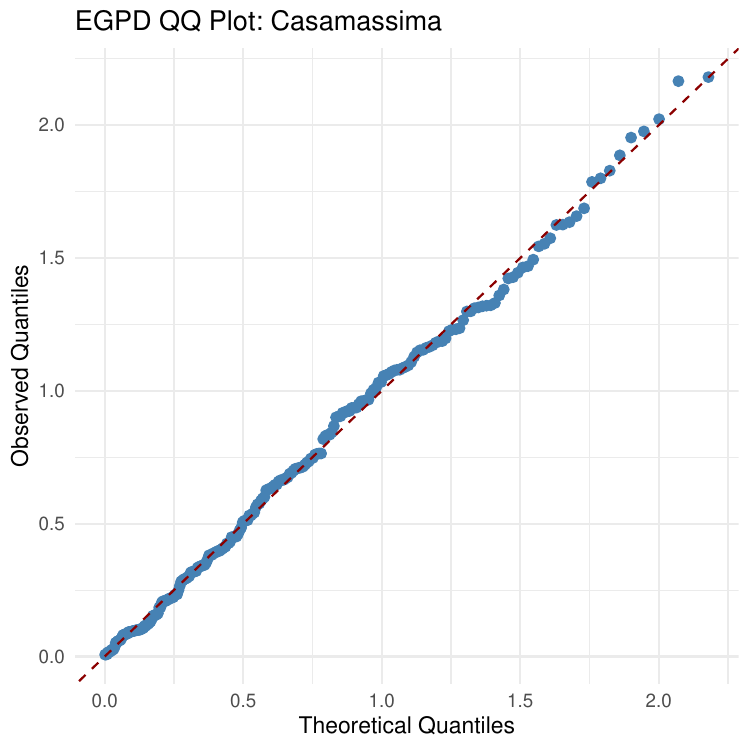}  
	\end{subfigure}
	\begin{subfigure}{.32\textwidth}
		\centering
		\includegraphics[width=1\linewidth, height=0.15\textheight]{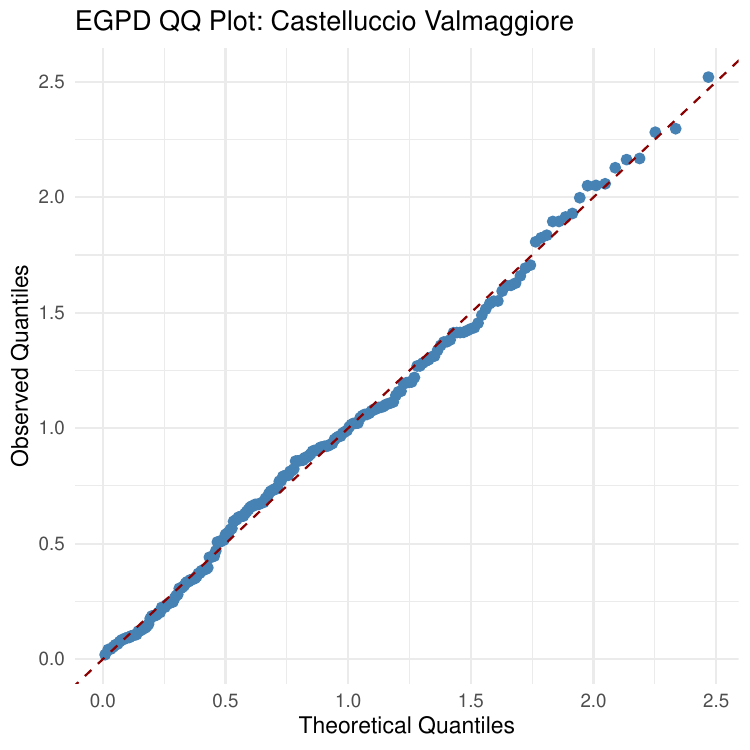}  
	\end{subfigure}
    \begin{subfigure}{.32\textwidth}
		\centering
		\includegraphics[width=1\linewidth, height=0.15\textheight]{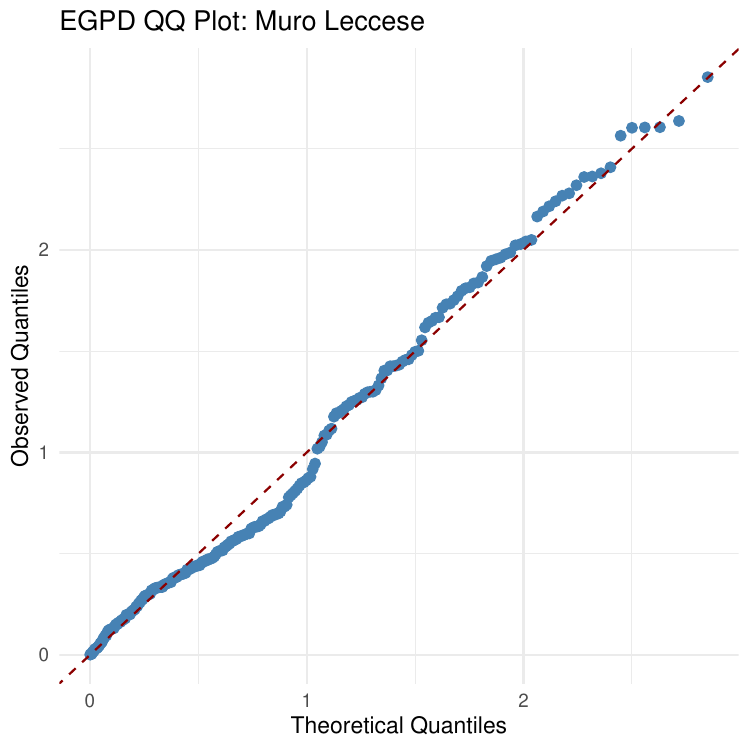}  
	\end{subfigure}
    
    \caption{QQ plots of fitted EGPD model to inverted dry periods at validation locations for $m=1, 3, 12$ (top to bottom rows), respectively. }
	\label{fig:qqplot-egpd-neg}
\end{figure}

\begin{figure}[t]
	\begin{subfigure}{.5\textwidth}
		\centering
\includegraphics[width=1\linewidth, height=0.25\textheight]{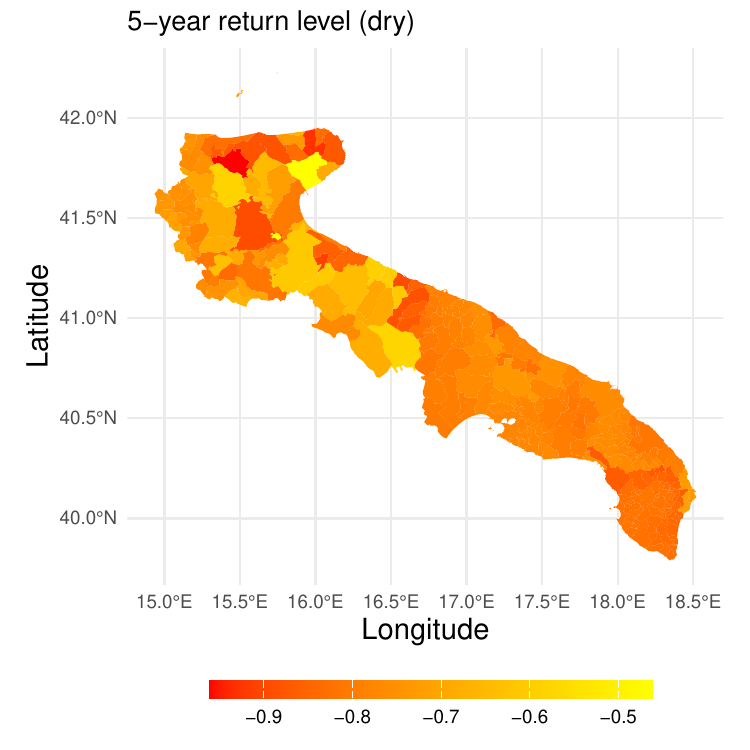}  
	\end{subfigure}
	\begin{subfigure}{.5\textwidth}
		\centering
\includegraphics[width=1\linewidth, height=0.25\textheight]{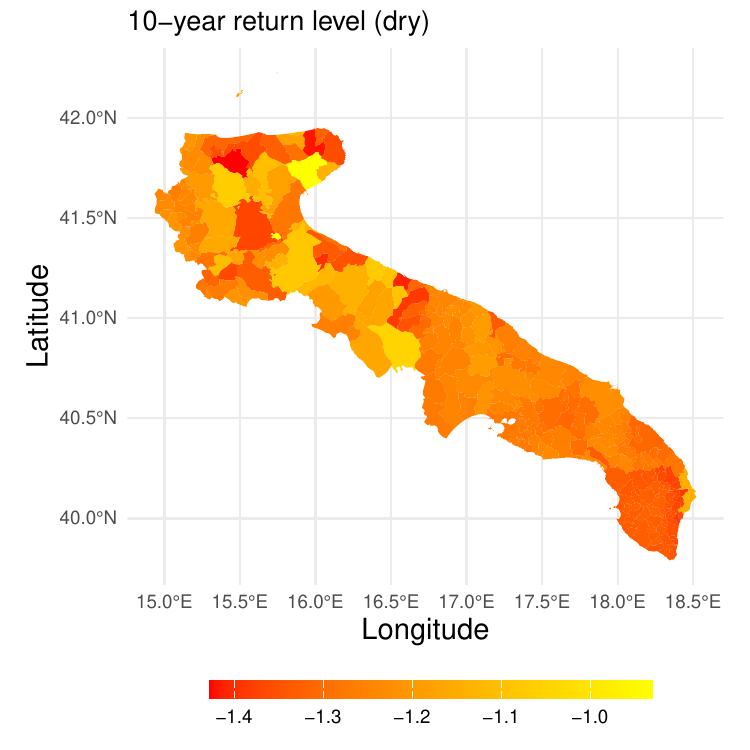}  
	\end{subfigure}
    \newline
    \begin{subfigure}{.5\textwidth}
		\centering
\includegraphics[width=1\linewidth, height=0.25\textheight]{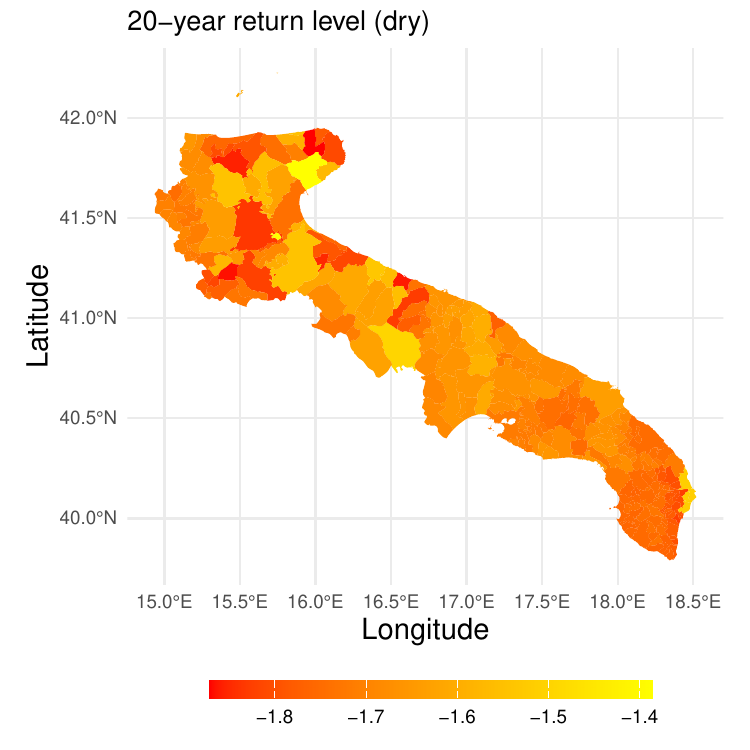}  
	\end{subfigure}
	\begin{subfigure}{.5\textwidth}
		\centering
\includegraphics[width=1\linewidth, height=0.25\textheight]{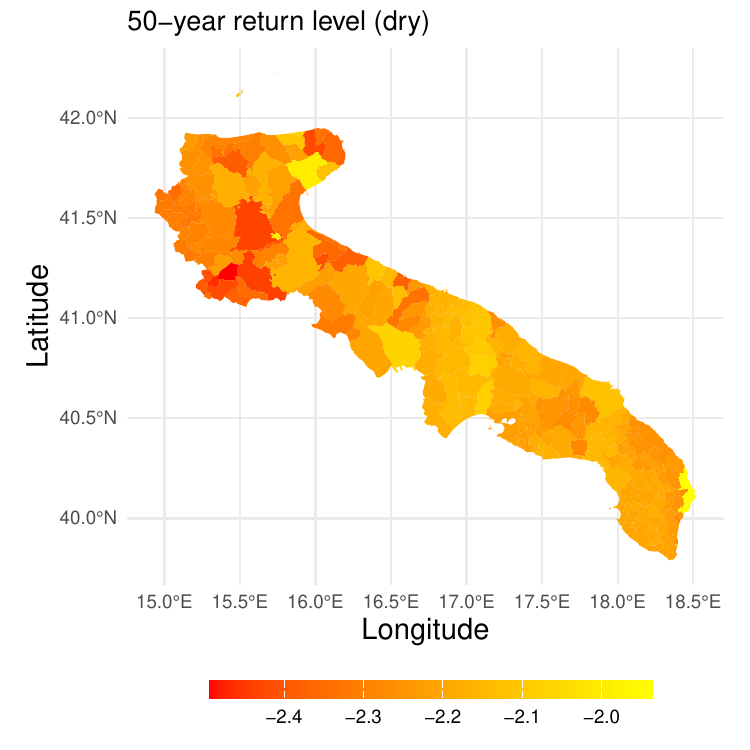}  
	\end{subfigure}
	\caption{Spatial maps of the estimated return levels for 5-, 10-, 20-, and 50-year dry return periods, based on an accumulation period of $m = 1$. }
	\label{fig:rl-dry-m1}
\end{figure}

\begin{figure}[t]
	\begin{subfigure}{.5\textwidth}
		\centering
\includegraphics[width=1\linewidth, height=0.25\textheight]{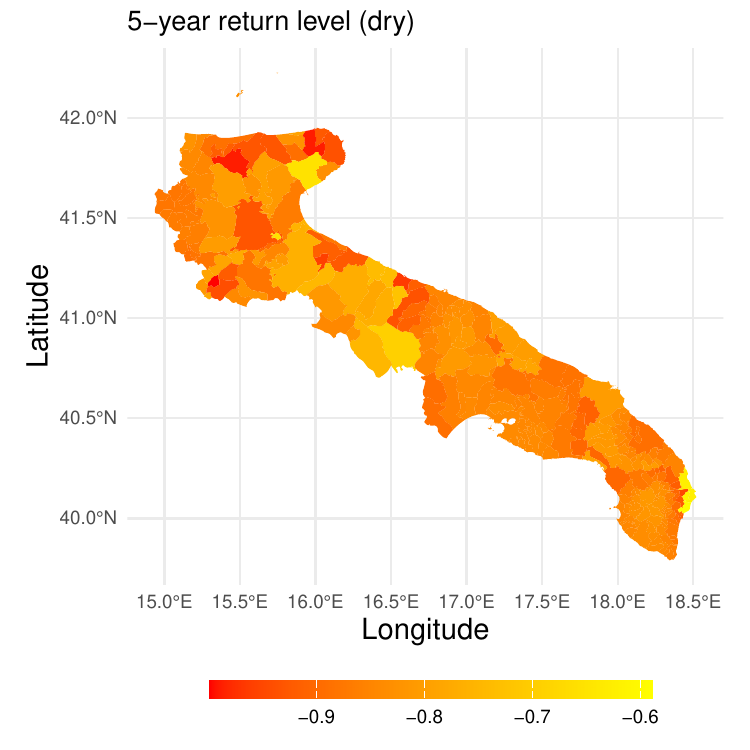}  
	\end{subfigure}
	\begin{subfigure}{.5\textwidth}
		\centering
\includegraphics[width=1\linewidth, height=0.25\textheight]{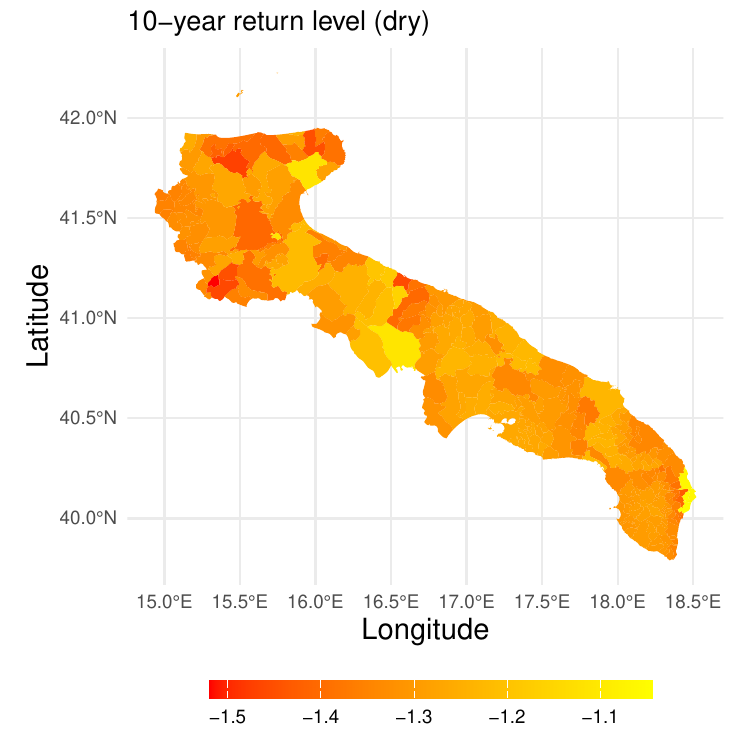}  
	\end{subfigure}
    \newline
    \begin{subfigure}{.5\textwidth}
		\centering
\includegraphics[width=1\linewidth, height=0.25\textheight]{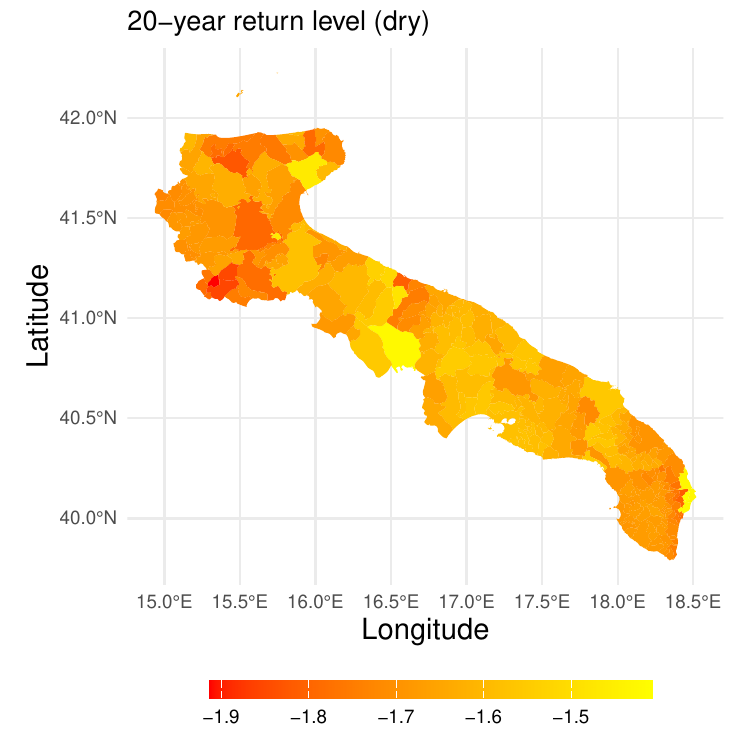}  
	\end{subfigure}
	\begin{subfigure}{.5\textwidth}
		\centering
\includegraphics[width=1\linewidth, height=0.25\textheight]{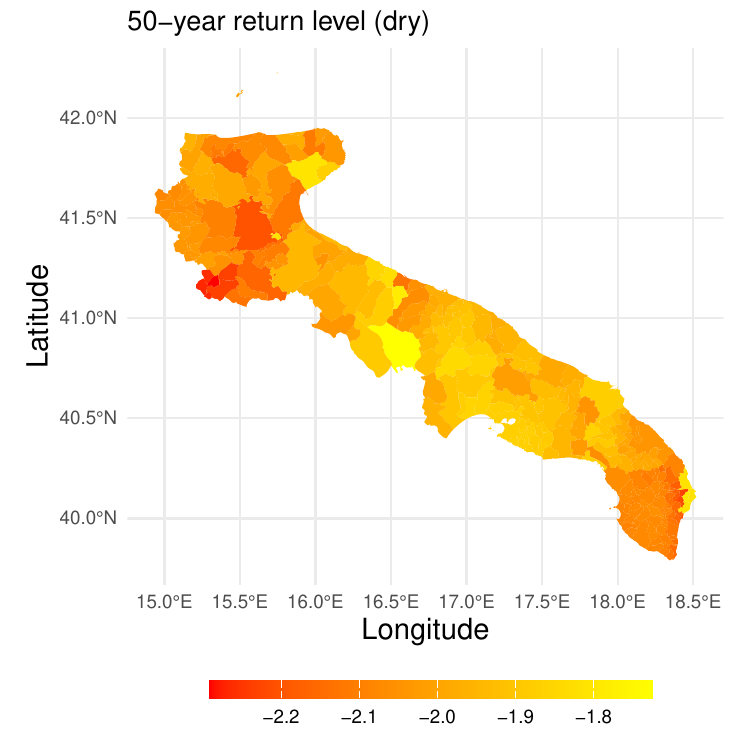}  
	\end{subfigure}
	\caption{Spatial maps of the estimated return levels for 5-, 10-, 20-, and 50-year dry return periods, based on an accumulation period of $m = 3$. }
	\label{fig:rl-dry-m3}
\end{figure}

\begin{figure}[t]
	\begin{subfigure}{.5\textwidth}
		\centering
\includegraphics[width=1\linewidth, height=0.25\textheight]{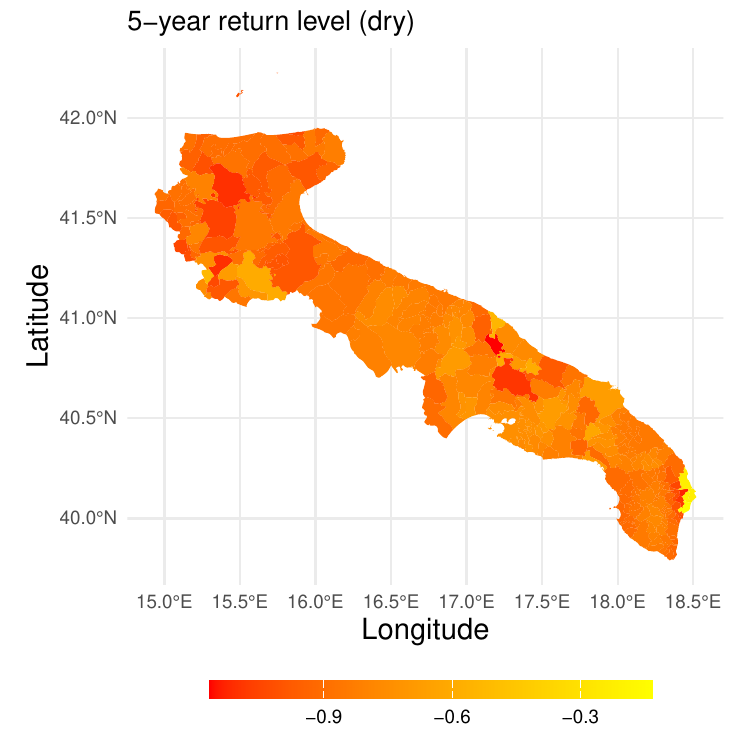}  
	\end{subfigure}
	\begin{subfigure}{.5\textwidth}
		\centering
\includegraphics[width=1\linewidth, height=0.25\textheight]{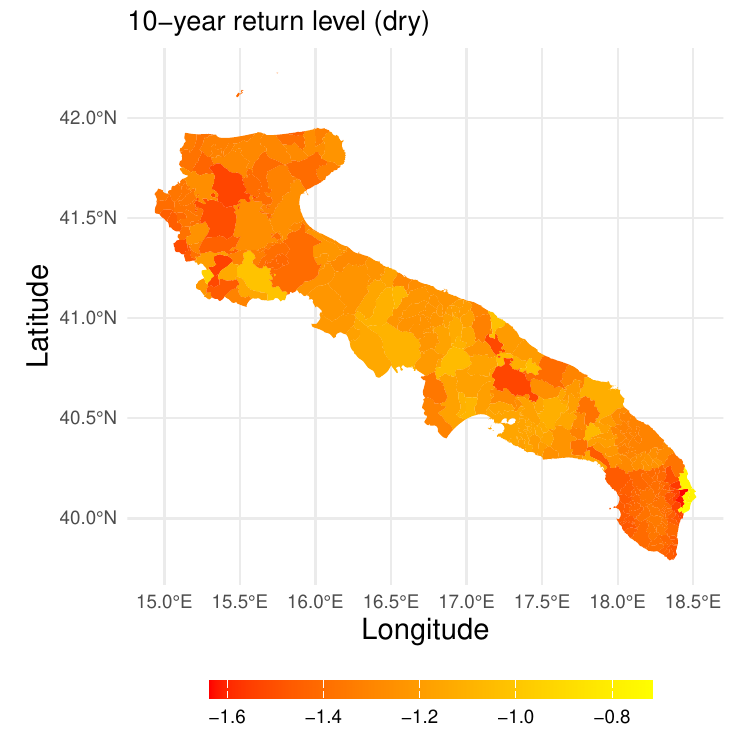}  
	\end{subfigure}
    \newline
    \begin{subfigure}{.5\textwidth}
		\centering
\includegraphics[width=1\linewidth, height=0.25\textheight]{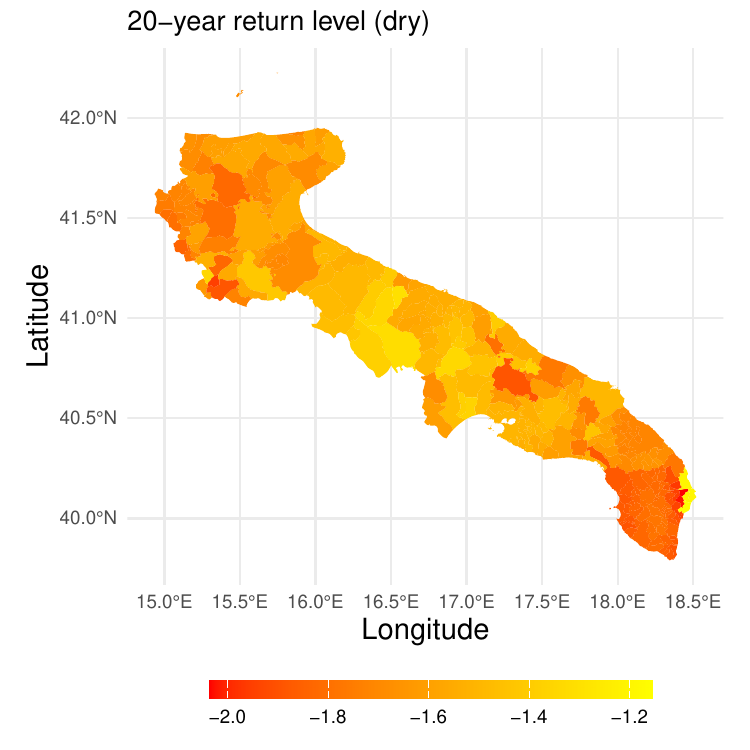}  
	\end{subfigure}
	\begin{subfigure}{.5\textwidth}
		\centering
\includegraphics[width=1\linewidth, height=0.25\textheight]{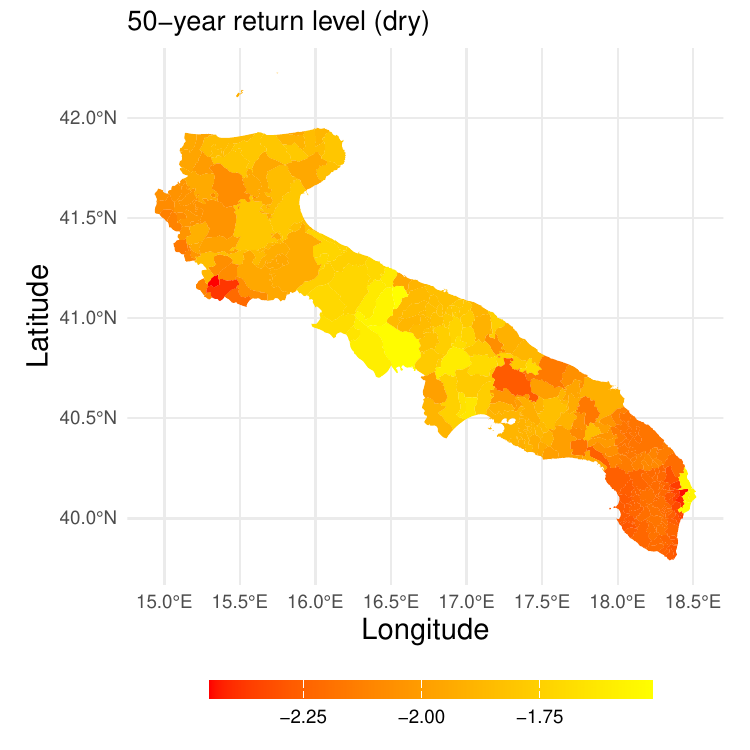}  
	\end{subfigure}
	\caption{Spatial maps of the estimated return levels for 5-, 10-, 20-, and 50-year dry return periods, based on an accumulation period of $m = 12$. }
	\label{fig:rl-dry-m12}
\end{figure}

\begin{figure}[t]
	\begin{subfigure}{.5\textwidth}
		\centering
\includegraphics[width=1\linewidth, height=0.25\textheight]{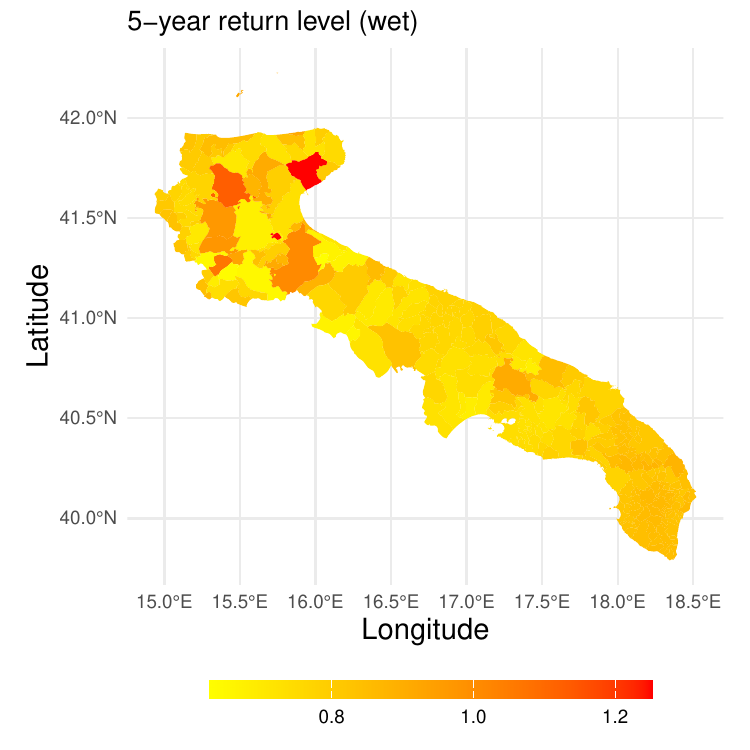}  
	\end{subfigure}
	\begin{subfigure}{.5\textwidth}
		\centering
\includegraphics[width=1\linewidth, height=0.25\textheight]{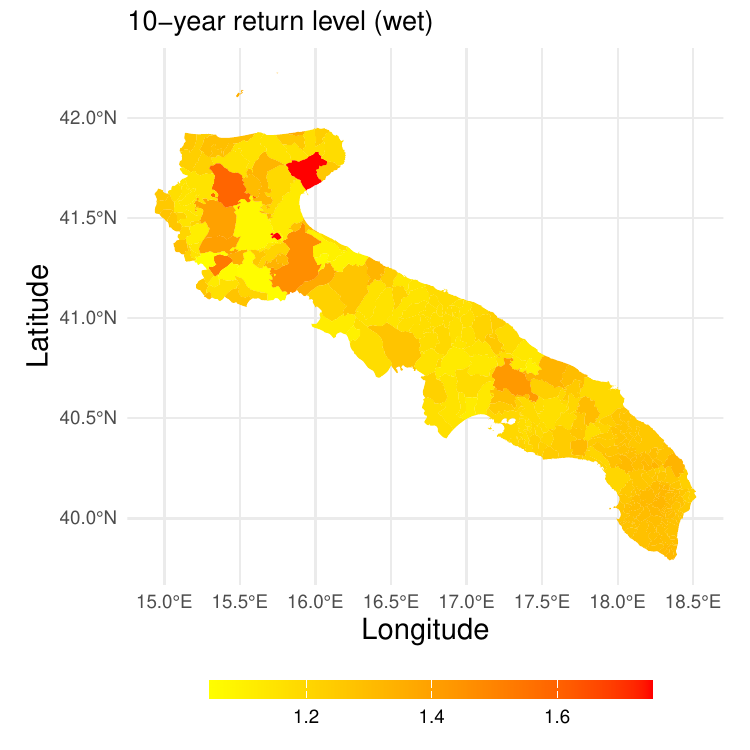}  
	\end{subfigure}
    \newline
    \begin{subfigure}{.5\textwidth}
		\centering
\includegraphics[width=1\linewidth, height=0.25\textheight]{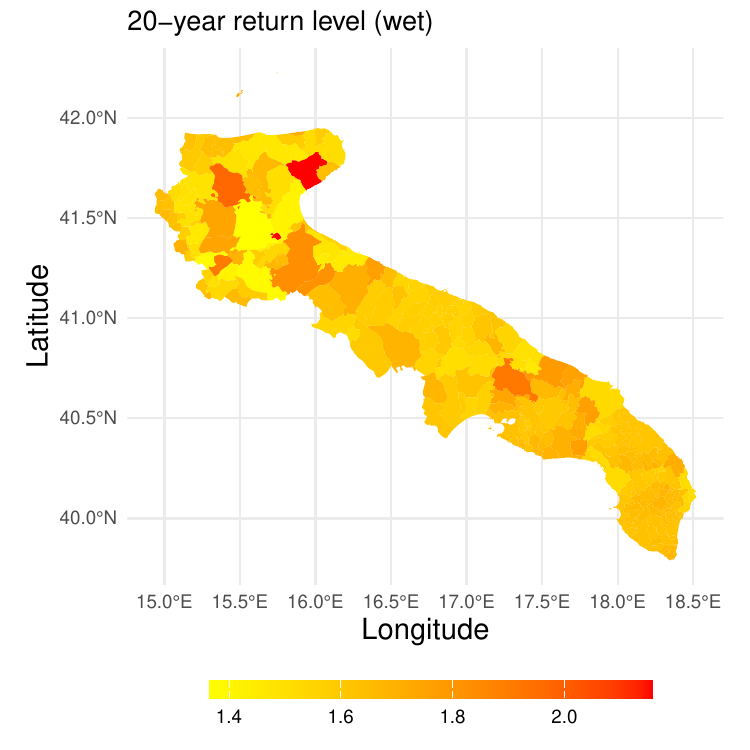}  
	\end{subfigure}
	\begin{subfigure}{.5\textwidth}
		\centering
\includegraphics[width=1\linewidth, height=0.25\textheight]{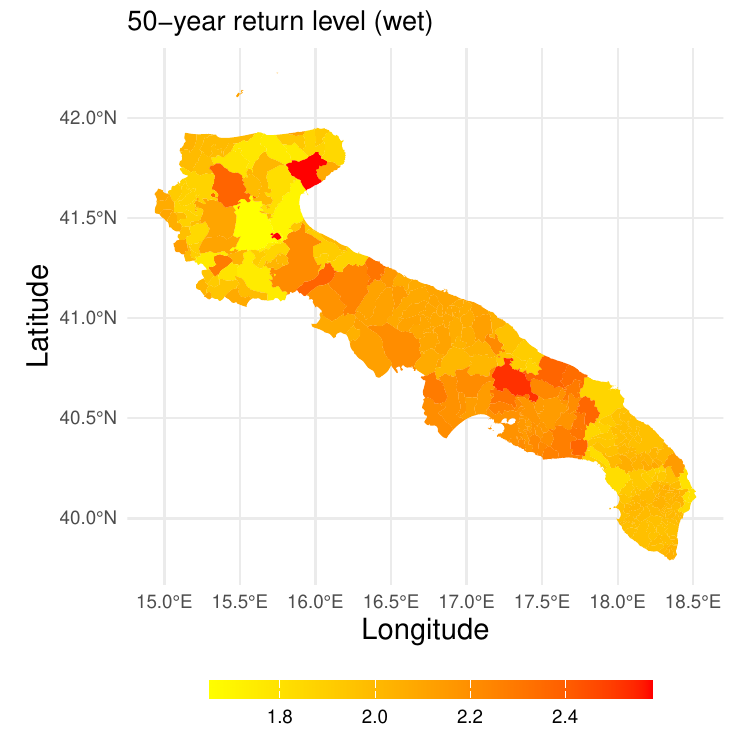}  
	\end{subfigure}
	\caption{Spatial maps of the estimated return levels for 5-, 10-, 20-, and 50-year wet return periods, based on an accumulation period of $m = 1$. }
	\label{fig:rl-wet-m1}
\end{figure}

\begin{figure}[t]
	\begin{subfigure}{.5\textwidth}
		\centering
\includegraphics[width=1\linewidth, height=0.25\textheight]{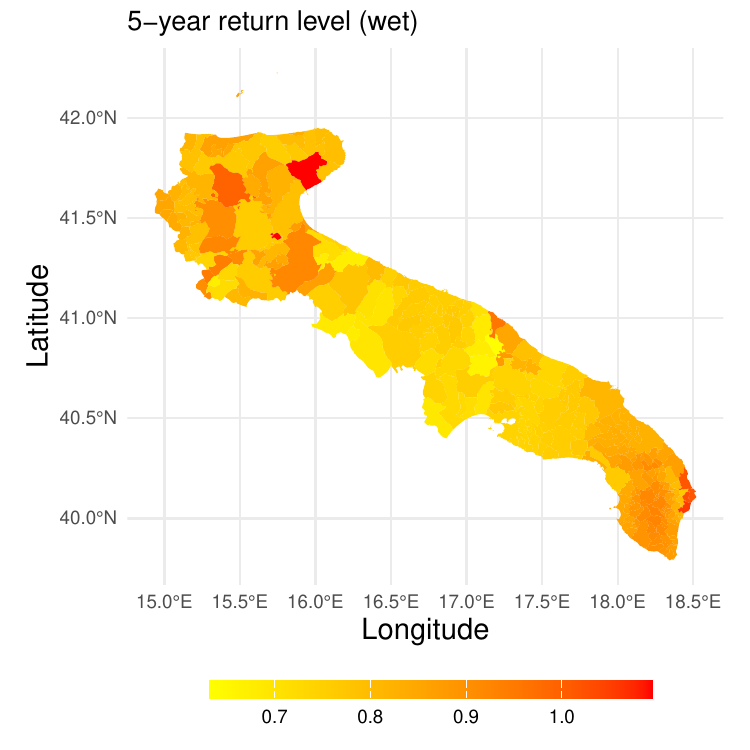}  
	\end{subfigure}
	\begin{subfigure}{.5\textwidth}
		\centering
\includegraphics[width=1\linewidth, height=0.25\textheight]{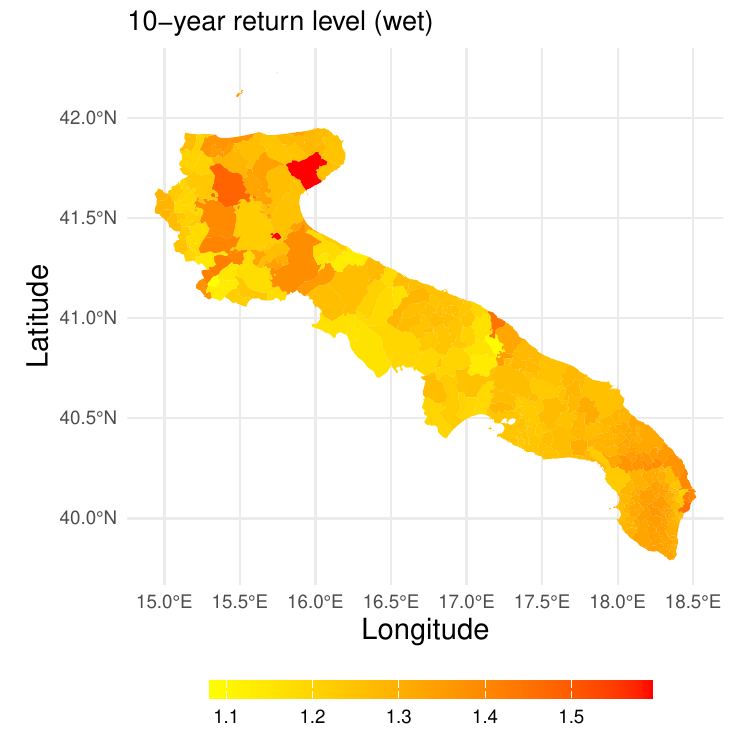}  
	\end{subfigure}
    \newline
    \begin{subfigure}{.5\textwidth}
		\centering
\includegraphics[width=1\linewidth, height=0.25\textheight]{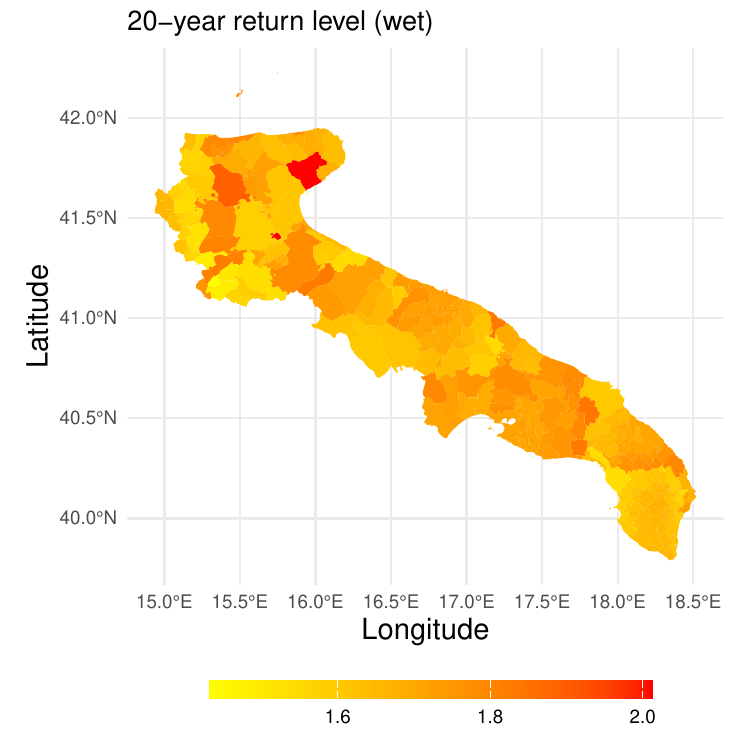}  
	\end{subfigure}
	\begin{subfigure}{.5\textwidth}
		\centering
\includegraphics[width=1\linewidth, height=0.25\textheight]{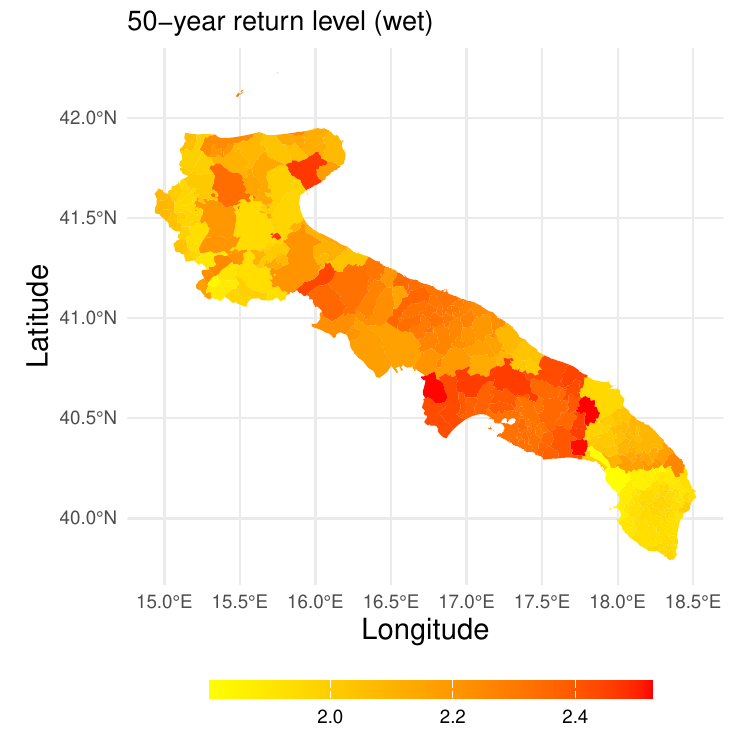}  
	\end{subfigure}
	\caption{Spatial maps of the estimated return levels for 5-, 10-, 20-, and 50-year wet return periods, based on an accumulation period of $m = 3$. }
	\label{fig:rl-wet-m3}
\end{figure}

\begin{figure}[t]
	\begin{subfigure}{.5\textwidth}
		\centering
\includegraphics[width=1\linewidth, height=0.25\textheight]{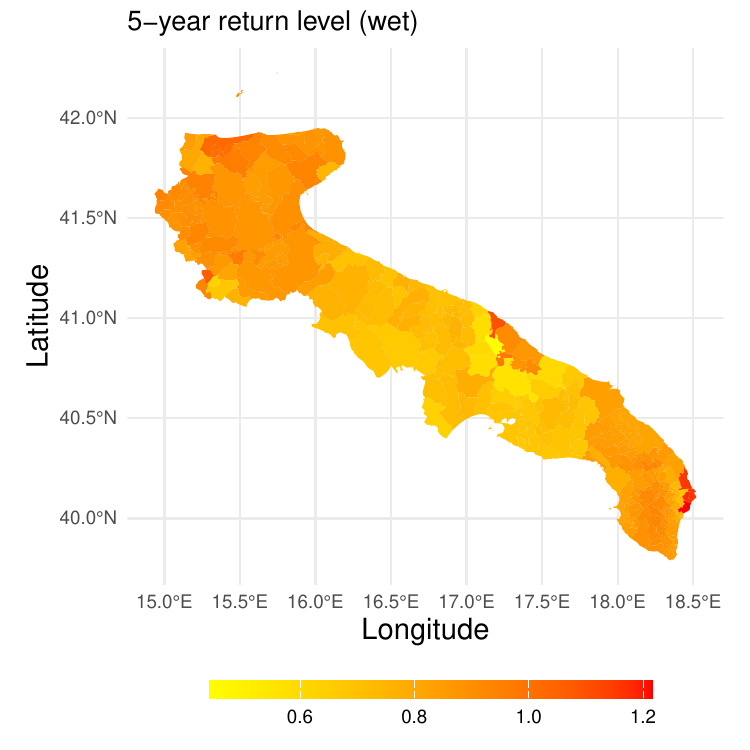}  
	\end{subfigure}
	\begin{subfigure}{.5\textwidth}
		\centering
\includegraphics[width=1\linewidth, height=0.25\textheight]{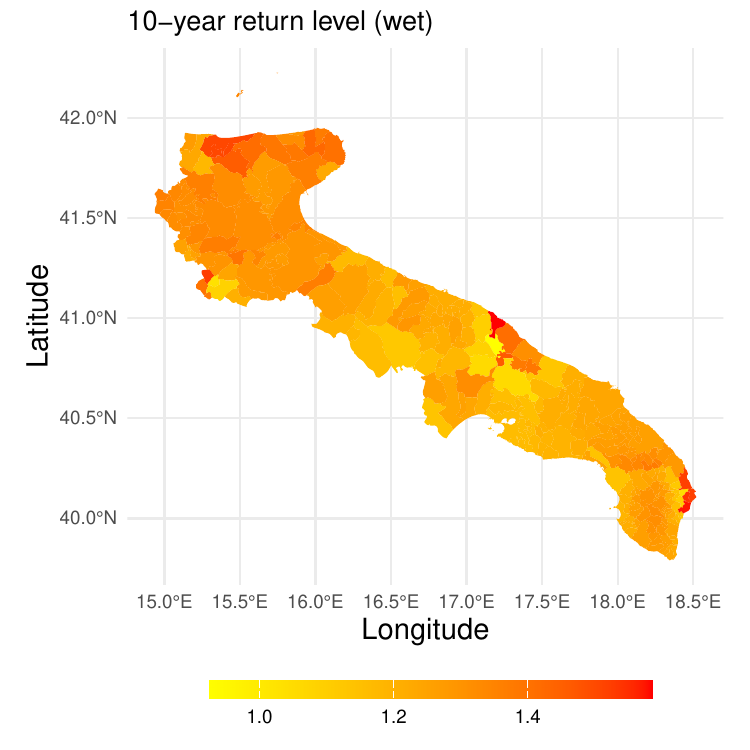}  
	\end{subfigure}
    \newline
    \begin{subfigure}{.5\textwidth}
		\centering
\includegraphics[width=1\linewidth, height=0.25\textheight]{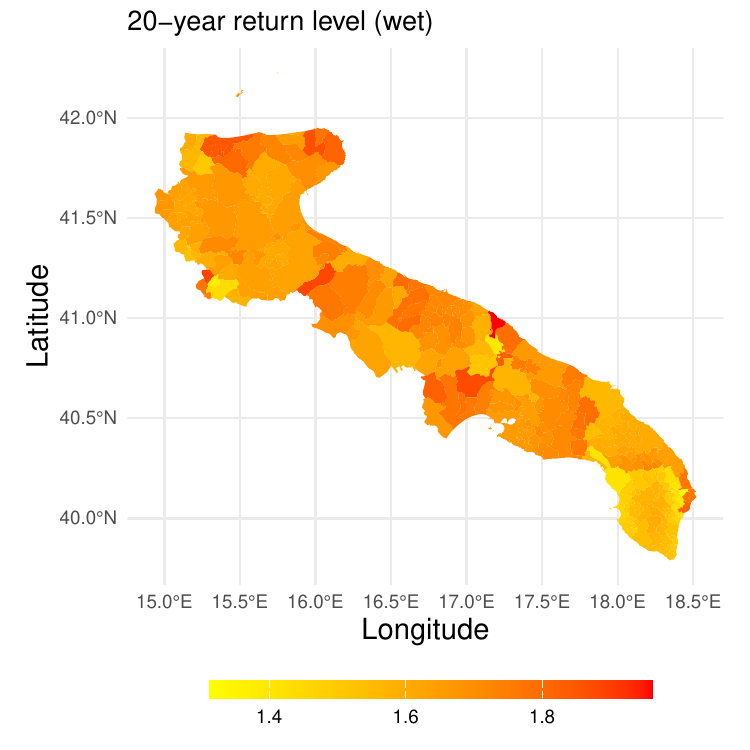}  
	\end{subfigure}
	\begin{subfigure}{.5\textwidth}
		\centering
\includegraphics[width=1\linewidth, height=0.25\textheight]{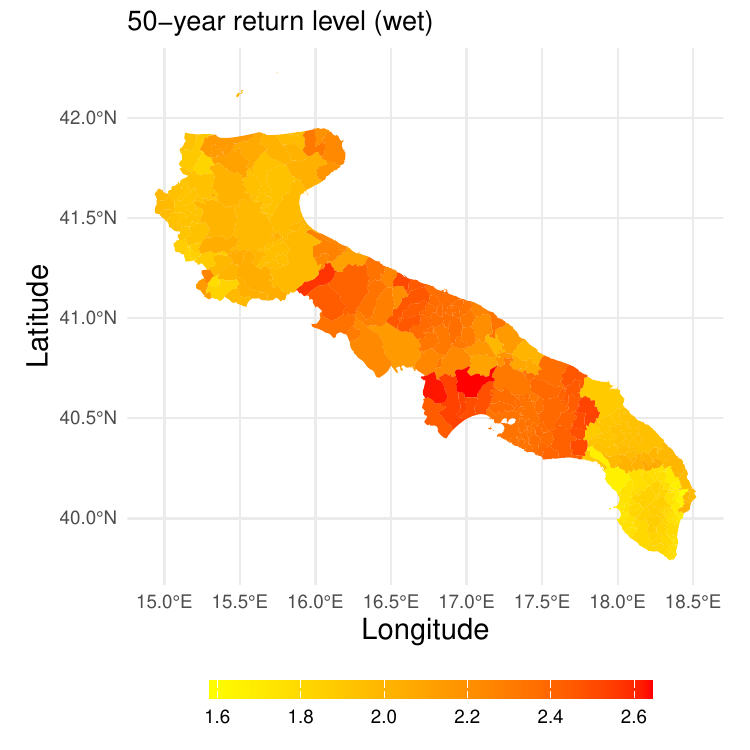}  
	\end{subfigure}
	\caption{Spatial maps of the estimated return levels for 5-, 10-, 20-, and 50-year wet return periods, based on an accumulation period of $m = 6$. }
	\label{fig:rl-wet-m6}
\end{figure}

\begin{figure}[t]
	\begin{subfigure}{.5\textwidth}
		\centering
\includegraphics[width=1\linewidth, height=0.25\textheight]{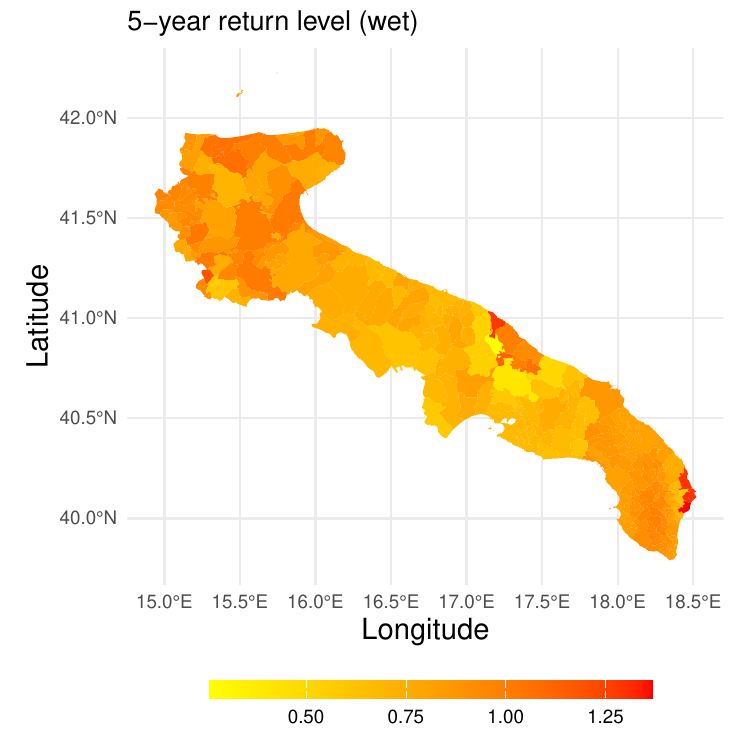}  
	\end{subfigure}
	\begin{subfigure}{.5\textwidth}
		\centering
\includegraphics[width=1\linewidth, height=0.25\textheight]{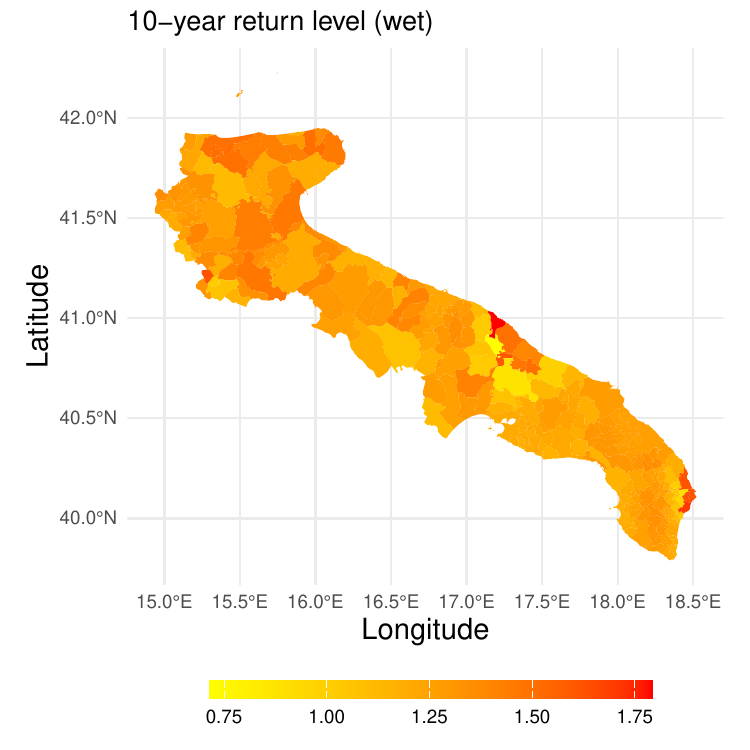}  
	\end{subfigure}
    \newline
    \begin{subfigure}{.5\textwidth}
		\centering
\includegraphics[width=1\linewidth, height=0.25\textheight]{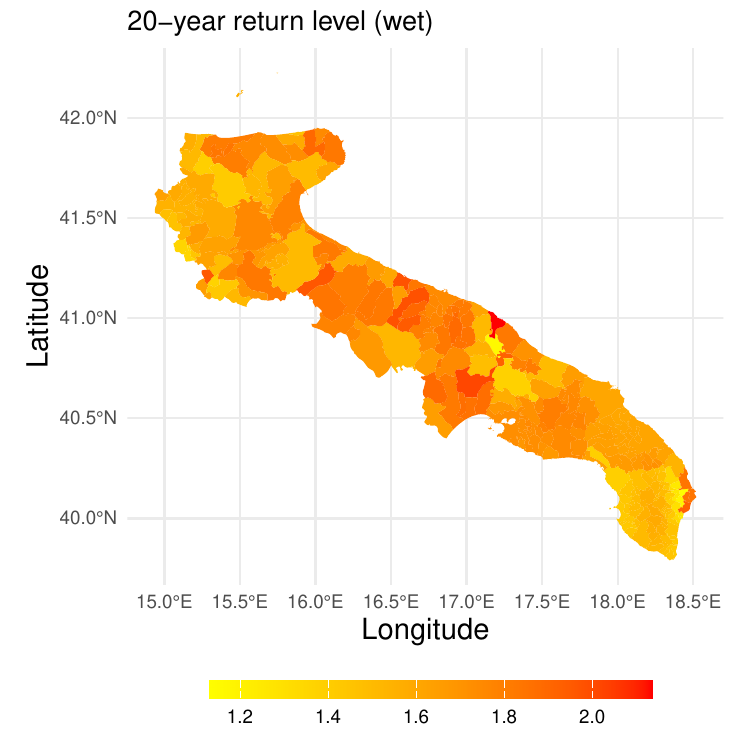}  
	\end{subfigure}
	\begin{subfigure}{.5\textwidth}
		\centering
\includegraphics[width=1\linewidth, height=0.25\textheight]{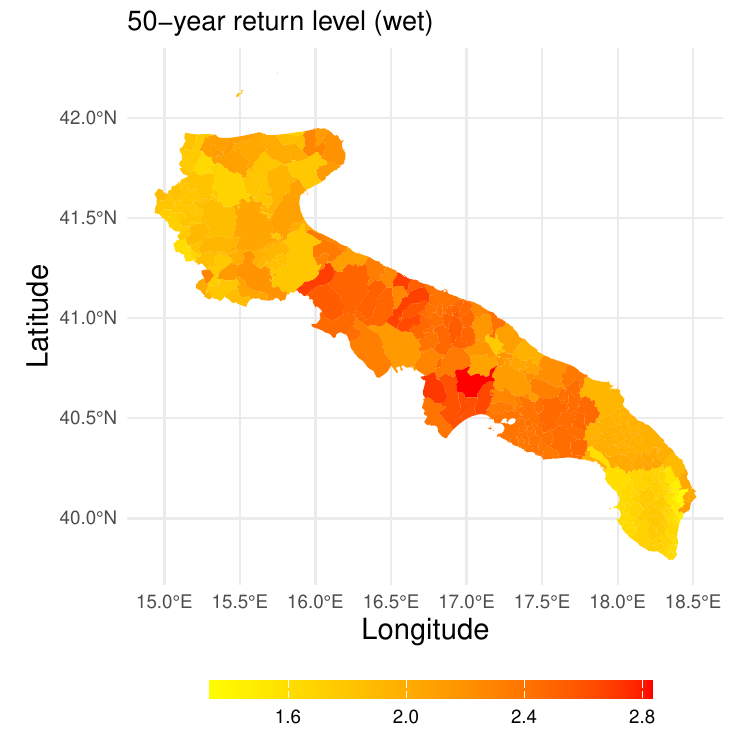}  
	\end{subfigure}
	\caption{Spatial maps of the estimated return levels for 5-, 10-, 20-, and 50-year wet return periods, based on an accumulation period of $m = 12$. }
	\label{fig:rl-wet-m12}
\end{figure}\end{document}